\documentclass[a4paper,fleqn,twocolumn,margin=1in]{aastex63}

\usepackage{float}
\usepackage{subfiles}
\usepackage{graphicx}
\usepackage{enumitem}
\usepackage{natbib}
\usepackage{amssymb}
\usepackage{amsmath}
\usepackage[lofdepth,lotdepth,caption=false]{subfig}
\def\muhz{\mu\mathrm{Hz}}
\def\numax{\nu_{\mathrm{max}}}

\def\dnu{\Delta \nu}

\def\teff{T_{\mathrm{eff}}}

\def\feh{\rm{[Fe/H]}}
\def\mgh{\rm{[Mg/H]}}

\def\teffsun{T_{\mathrm{eff,} \odot}}

\def\rsun{R_{\odot}}
\def\deg{^{\circ}}
\def\gyr{\rm{Gyr}}
\def\msun{M_{\odot}}
\def\numaxsun{\nu_{\mathrm{max,} \odot}}
\def\dnusun{\Delta\nu_{\odot}}

\def\Ks{K_{\mathrm{s}}}

\def\numaxmean{\langle \numax' \rangle}
\def\dnumean{\langle \dnu' \rangle}
\def\kapparmean{\langle \kappa_R \rangle}
\def\kappammean{\langle \kappa_M \rangle}
\def\kappar{\kappa_R}
\def\kappam{\kappa_M}
\newcommand{\fe}[1]{\rm{[#1/Fe]}}
\newcommand{\mas}{\rm{mas}}
\newcommand{\mg}[1]{\rm{[#1/Mg]}}

\def\highalpha{high-$\alpha$}
\def\lowalpha{low-$\alpha$}
\def\fdnu{f_{\Delta \nu}}
\def\fnumax{f_{\numax}}
\newcommand{\Kepler}{{Kepler}}
\newcommand{\Ktwo}{{K2}}
\newcommand{\Gaia}{{Gaia}}
\def\muas{\mu \mathrm{as}}
\usepackage{hyperref}
\setcitestyle{notesep={; }}
\graphicspath{{plots/}{../plots/}}
\begin{document}
\shorttitle{K2 GAP DR3}
\shortauthors{Zinn et al.}
\title{The K2 Galactic Archaeology Program Data Release 3: Age-abundance patterns in C1-C8, C10-C18}
\correspondingauthor{Joel C. Zinn}
\email{jzinn@amnh.org}

\author{Joel C. Zinn}
\altaffiliation{NSF Astronomy and Astrophysics Postdoctoral Fellow.}
\affiliation{Department of Astrophysics, American Museum of Natural History, Central Park West at 79th Street, New York, NY 10024, USA}
\affiliation{School of Physics, University of New South Wales, Barker Street, Sydney, NSW 2052, Australia}
\author{Dennis Stello}
\affiliation{School of Physics, University of New South Wales, Barker
  Street, Sydney, NSW 2052, Australia}
\affiliation{Sydney Institute for Astronomy, School of Physics, A28, The University of Sydney, NSW 2006, Australia}
\affiliation{Stellar Astrophysics Centre, Department of Physics and
Astronomy, Aarhus University, Ny Munkegade 120, DK-8000
Aarhus C, Denmark}
\affiliation{ARC Centre of Excellence for All Sky Astrophysics in 3 Dimensions (ASTRO 3D), Australia}
\author{Yvonne Elsworth}
\affiliation{School of Physics and Astronomy, University of Birmingham, Edgbaston, Birmingham, B15 2TT, UK}
\affiliation{Stellar Astrophysics Centre, Department of Physics and Astronomy, Aarhus University, Ny Munkegade 120, DK-8000 Aarhus C, Denmark}
\author{Rafael A. Garc{\'i}a}
\affiliation{AIM, CEA, CNRS, Universit{\'e} Paris-Saclay, Universit{\'e} Paris Diderot, Sorbonne Paris Cit{\'e}, F-91191 Gif-sur-Yvette, France}
\author{Thomas Kallinger}
\affiliation{Institute of Astrophysics, University of Vienna, T{\"u}rkenschanzstrasse 17, Vienna 1180, Austria}
\author{Savita Mathur}
\affiliation{Instituto de Astrof\'{\i}sica de Canarias, La Laguna, Tenerife, Spain}
\affiliation{Dpto. de Astrof\'{\i}sica, Universidad de La Laguna, La Laguna, Tenerife, Spain}
\author{Beno{\^i}t Mosser}
\affiliation{LESIA, Observatoire de Paris, PSL Research University, CNRS, Sorbonne Universit{\'e}, Universit{\'e} de Paris Diderot, 92195 Meudon, France}
\author{Marc Hon}
\affiliation{School of Physics, University of New South Wales, Barker
  Street, Sydney, NSW 2052, Australia}
\author{Lisa Bugnet}
\affiliation{AIM, CEA, CNRS, Universit{\'e} Paris-Saclay, Universit{\'e} Paris Diderot, Sorbonne Paris Cit{\'e}, F-91191 Gif-sur-Yvette, France}
\affiliation{Flatiron Institute, Simons Foundation, 162 Fifth Ave, New York, NY 10010, USA}
\author{Caitlin Jones}
\affiliation{School of Physics and Astronomy, University of Birmingham, Edgbaston, Birmingham, B15 2TT, UK}
\author{Claudia Reyes}
\affiliation{School of Physics, University of New South Wales, Barker
  Street, Sydney, NSW 2052, Australia}
\author{Sanjib Sharma}
\affiliation{Sydney Institute for Astronomy, School of Physics, A28, The University of Sydney, NSW 2006, Australia}
\affiliation{ARC Centre of Excellence for All Sky Astrophysics in 3 Dimensions (ASTRO 3D), Australia}
\author{Ralph Sch{\"o}nrich}
\affiliation{Mullard Space Science Laboratory, University College London, Holmbury St Mary, Dorking RH5 6NT, UK}
\author{Jack T. Warfield}
\affiliation{Department of Astronomy, The Ohio State University, 140 West
  18th Avenue, Columbus OH 43210, USA}
\affiliation{Department of Physics,
The Ohio State University, 191 West
Woodruff Avenue, Columbus OH 43210}
\author{Rodrigo Luger}
\affiliation{Center for Computational Astrophysics, Flatiron Institute, New York, NY, USA}
\affiliation{Virtual Planetary Laboratory, University of Washington, Seattle, WA, USA}
\author{Andrew Vanderburg}
\affiliation{Department of Astronomy, The University of Texas at Austin, Austin, TX 78712, USA}
\author{Chiaki Kobayashi}
\affiliation{Centre for Astrophysics Research, Department of Physics, Astronomy and Mathematics, University of Hertfordshire, Hatfield, AL10 9AB, UK}
\author{Marc H. Pinsonneault}
\affiliation{Department of Astronomy, The Ohio State University, 140 West
18th Avenue, Columbus OH 43210, USA}
\author{Jennifer A. Johnson}
\affiliation{Department of Astronomy, The Ohio State University, 140 West
18th Avenue, Columbus OH 43210, USA}
\author{Daniel Huber}
\affiliation{ Institute for Astronomy, University of Hawai`i, 2680 Woodlawn Drive, Honolulu, HI 96822, USA}
\author{Sven Buder}
\affiliation{Research School of Astronomy and Astrophysics, Australian National University, Canberra, ACT 2611, Australia}
\affiliation{ARC Centre of Excellence for All Sky Astrophysics in 3
  Dimensions (ASTRO 3D), Australia}
\author{Meridith Joyce}
\affiliation{Research School of Astronomy and Astrophysics, Australian National University, Canberra, ACT 2611, Australia}
\affiliation{ARC Centre of Excellence for All Sky Astrophysics in 3
  Dimensions (ASTRO 3D), Australia}
\author{Joss Bland-Hawthorn}
\affiliation{Sydney Institute for Astronomy, School of Physics, A28, The University of Sydney, NSW 2006, Australia}
\affiliation{ARC Centre of Excellence for All Sky Astrophysics in 3
  Dimensions (ASTRO 3D), Australia}
\author{Luca Casagrande}
\affiliation{Research School of Astronomy and Astrophysics, Mount
  Stromlo Observatory, The Australian National University, ACT 2611,
  Australia}
\author{Geraint F. Lewis}
\affiliation{Sydney Institute for Astronomy, School of Physics, A28,
  The University of Sydney, NSW 2006, Australia}
\author{Andrea Miglio}
\affiliation{Stellar Astrophysics Centre, Department of Physics and Astronomy, Aarhus University, Ny Munkegade 120, DK-8000 Aarhus C, Denmark}
\affiliation{School of Physics and Astronomy, University of Birmingham, Edgbaston, Birmingham, B15 2TT, UK}
\author{Thomas Nordlander}
\affiliation{Research School of Astronomy and Astrophysics, Australian National University, Canberra, ACT 2611, Australia}
\affiliation{ARC Centre of Excellence for All Sky Astrophysics in 3
  Dimensions (ASTRO 3D), Australia}
\author{Guy R. Davies}
\affiliation{School of Physics and Astronomy, University of Birmingham, Edgbaston, Birmingham, B15 2TT, UK}
\affiliation{Stellar Astrophysics Centre, Department of Physics and Astronomy, Aarhus University, Ny Munkegade 120, DK-8000 Aarhus C, Denmark}
\author{Gayandhi De Silva}
\affiliation{Australian Astronomical Optics, Faculty of Science and Engineering, Macquarie University, Macquarie Park, NSW 2113, Australia}
\affiliation{Macquarie University Research Centre for Astronomy, Astrophysics \& Astrophotonics, Sydney, NSW 2109, Australia}
\affiliation{ARC Centre of Excellence for All Sky Astrophysics in 3
  Dimensions (ASTRO 3D), Australia}
\author{William J. Chaplin}
\affiliation{School of Physics and Astronomy, University of Birmingham, Edgbaston, Birmingham, B15 2TT, UK}
\affiliation{Stellar Astrophysics Centre, Department of Physics and Astronomy, Aarhus University, Ny Munkegade 120, DK-8000 Aarhus C, Denmark}
\author{Victor Silva Aguirre}
\affiliation{Stellar Astrophysics Centre, Department of Physics and Astronomy, Aarhus University, Ny Munkegade 120, DK-8000 Aarhus C, Denmark}

\begin{abstract}
We present the third and final data release of the K2 Galactic
Archaeology Program (K2 GAP) for Campaigns C1-C8 and C10-C18. We
provide asteroseismic radius and mass coefficients, $\kappa_R$ and
$\kappa_M$, for $\sim 19,000$ red giant stars, which translate
directly to radius and mass given a temperature. As such, K2 GAP DR3
represents the largest asteroseismic sample in the literature to
date. K2 GAP DR3 stellar parameters are calibrated to be on an
absolute parallactic scale based on Gaia DR2, with red giant branch
and red clump evolutionary state classifications provided via a
machine-learning approach. Combining these stellar parameters with
GALAH DR3 spectroscopy, we determine asteroseismic ages with
precisions of $\sim 20-30\%$ and compare age-abundance relations to
Galactic chemical evolution models among both low- and high-$\alpha$
populations for $\alpha$, light, iron-peak, and neutron-capture
elements. We confirm recent indications in the literature of both
increased Ba production at late Galactic times, as well as significant
contribution to r-process enrichment from prompt sources associated
with, e.g., core-collapse supernovae. With an eye toward other
Galactic archaeology applications, we characterize K2 GAP DR3
uncertainties and completeness using injection tests, suggesting K2
GAP DR3 is largely unbiased in mass/age and with uncertainties of
$2.9\%\,(\rm{stat.})\,\pm0.1\%\,(\rm{syst.})$ \&
$6.7\%\,(\rm{stat.})\,\pm0.3\%\,(\rm{syst.})$ in $\kappar$ \&
$\kappam$ for red giant branch stars and $4.7\%\,(\rm{stat.})\,\pm0.3\%\,
(\rm{syst.})$ \& $11\%\,(\rm{stat.})\,\pm0.9\%\,(\rm{syst.})$ for
red clump stars. We also identify percent-level asteroseismic systematics, which are likely related to the time baseline of the underlying data, and which therefore should be considered in TESS asteroseismic analysis.
\end{abstract}

\section{Introduction}
\label{sec:intro}
Studies of Galactic chemical evolution have mostly focussed on targets in the solar neighborhood, where stars are relatively easy to observe, and which was the sole domain, historically, of precise parallaxes, and therefore, stellar ages \citep[e.g.,][]{nordstrom+2004}. Because stars for the most part maintain their birth abundances, stellar abundances and ages can be used to infer the chemical enrichment history of the Galaxy, providing information on the details of contributions to the insterstellar medium from nucleosynthetic channels like supernovae and stellar winds.

The local stellar population has been found to have a bimodal chemical distribution in alpha elements (e.g., O, Mg, Ca, Si) as seen in \fe{$\alpha$} versus \feh\footnote{Here and throughout the paper we use the standard notation \fe{X} $\equiv \log_{10}\left(\frac{\rm{X}}{\rm{Fe}}\right) - \log_{10}\left(\frac{\rm{X}_{\odot}}{\rm{Fe}_{\odot}}\right)$.}: there exists one population of \lowalpha\ stars with spatial distributions apparently more confined to the plane of the Galaxy and with intermediate to young ages, and there exists another population of \highalpha\ stars with hotter kinematics; centrally-concentrated spatial distributions in the disk; and older ages \citep[e.g.,][]{fuhrmann1998,prochaska+2000,gratton+2000,bensby_feltzing_lundstrom2003,haywood+2013}. With the understanding that $\alpha$ elements are produced primarily in core-collapse supernovae (CCSNe), with some contributions to the heavier nuclei from Type Ia supernovae (SNe Ia), whereas iron is mostly produced in SNe Ia, the low-$\alpha$ population has been interpreted as having mostly contributions from SNe Ia and the high-$\alpha$ population has been interpreted as having mostly contributions from CCSNe \citep{burbidge+1957,timmes_woosley_weaver1995}.

Studies of stellar populations beyond the solar neighborhood have shown that the \lowalpha\ and \highalpha\ populations maintain their chemical bimodality, and to a certain extent, their distinct radial spatial distributions, with the high-alpha stars more centrally concentrated \citep{nidever+2014,hayden+2015}, though not necessarily intrinsically having a different vertical spatial distribution \citep{hayden+2017}. There are also interesting chemical distinctions among these populations when looking at non--$\alpha$-element abundance ratios \citep[e.g.,][]{prochaska+2000,bensby_feltzing_lundstrom2003,adibekyan+2012,weinberg+2019,griffith_johnson_weinberg2019,nissen+2020}. Given expectations that the bimodality is ultimately related to different chemical enrichment histories, it is natural to ask how these populations evolved, and, in so doing, test understandings of the nucleosynthetic production sites' yields with time. Indeed, the underlying origin of these spatial and chemical distinctions is under debate \citep[e.g.,][]{chiappini_matteucci_gratton1997,schonrich_binney2009a,schonrich_binney2009,kobayashi_nakasato2011,minchev+2015,hayden+2017,mackereth+2018,spitoni+2019,clarke+2019}.

Thanks to an unprecedented collection of well-measured stellar kinematics from \Gaia\ \citep{gaia+2016,gaia2018}, and with large spectrocopic surveys like APOGEE \citep{majewski+2017}, GALAH \citep{de_silva+2015}, and LAMOST \citep{newberg+2012}  providing hundreds of thousands of detailed abundance measurements probing well beyond the solar vicinity, the bottleneck to progress on the origin and evolution of the elements in the Galaxy becomes stellar age.

Stellar composition has historically been used as a proxy for stellar age, based on the idea that Galactic enrichment increases with time, with input to the interstellar medium through asymptotic giant branch (AGB) stars, SNe Ia, and CCSNe continually injecting metals over time according to so-called delay-time distributions, in ratios that themselves depend on time due to the birth composition of the stars producing the elements changing  \citep{tinsley+1979,mcwilliam1997}. Kinematic information can also serve as an age proxy, with older stellar populations experiencing dynamical heating from discrete merger events \citep[e.g.,][]{grand+2016} as well as secular processes involving, e.g., spiral arms \citep{carlberg_sellwood1985,minchev_quillen2000}; the central bar \citep{saha_tseng_taam2010,grand+2016}; and giant molecular clouds \citep[e.g.,][]{spitzer_schwarzschild1951}.

To make precise statements about the chemical evolution of the Galaxy, however, requires genuine stellar age estimates independent of kinematics and abundances so as to not assume the age-abundance patterns under question.  
Work appealing to stellar ages to solve this conundrum has either made recourse to turnoff stars --- whose ages can be reliably determined through isochrone matching \citep{pont_eyer2004,jorgensen_lindegren2005,lin+2019}, but which are relatively dim and therefore probe predominantly local volumes --- or to spectroscopic ages \citep[e.g.,][]{sun_huang+2020,ness+2016}. Although machine learning--based spectroscopic ages have increased sample sizes to hundreds of thousands, they do not seem to yield reliable for ages $> 8$ Gyr \citep{ting_rix2019}. Another approach is to make reference to Galactic stellar population models, and match a predicted age-color relation to observed colors \citep[e.g.,][]{bland_hawthorn_2019}. As we will show here, such photometric ages --- with typical uncertainties of 30-50\% --- are significantly less precise than the 20-30\% K2 asteroseismic age uncertainties we present here.

Because asteroseismic ages do not assume either an age-kinematics
relation or age-metallicity relation, they can provide interesting
constraints on Galactic chemical evolution. Indeed, asteroseismic ages
have supported literature estimates that the two $\alpha$ populations
have different age distributions
\citep[e.g.,][]{silva-aguirre+2018a,miglio+2021}, while complicating the
understanding that the \highalpha\ population is uniformly old
\citep{chiappini+2015,anders+2017,warfield+2021}. These studies have mostly been
limited to asteroseismic data from the \Kepler\ mission
\citep{borucki+2008}, which probes a single, 100 sq. deg. field of
view at roughly fixed Galactic radius corresponding to that of the
Sun. With the extended
\Kepler\ mission, \Ktwo\ \citep{howell+2014}, has come access to
regions of the sky across the ecliptic, and exploratory work based on
data from four K2 campaigns has shown promise for better understanding
the $\alpha$ bimodality \citep{rendle+2019,warfield+2021}. The K2 Galatic Archaeology Program (GAP; \citealt{stello+2015}) takes advantage off this opportunity, targeting red giant stars with the express intent of investigating the chemical evolution of the Galaxy beyond the solar vicinity using asteroseismic ages.

In this paper, we describe the final data release of K2 GAP, which combines red giant asteroseismic data from Campaign 1 (C1) from K2 GAP DR1 \citep{stello+2017} and data from C4, C6, \& C7 from K2 GAP DR2 \citep{zinn+2020} with results from the remaining \Ktwo\  campaigns. In K2 GAP Data Release 3 (DR3), we improve upon K2 GAP DR2 by verifying the accuracy and precision of asteroseismology with an injection test exercise, and calibrate our results against \Gaia\ DR2 \citep{gaia2018} radii\footnote{Here and throughout the text, when we mention \Gaia\ radii, we refer to radii we calculate using APOGEE spectroscopy and \Gaia\ parallaxes, according to the Stefan-Boltzmann law. These radii are distinct from the \texttt{radius\_val} values provided as part of \Gaia\ DR2; the latter are not as accurate as we require here because they do not account for extinction, and assume inhomogeneous temperatures. See \S\ref{sec:calibratoin} for details.}. Finally, we derive ages based on these calibrated asteroseismic masses in order to compare abundance enrichment histories of low- and high-$\alpha$ populations with Galactic chemical evolution models from \cite{kobayashi_karakas_lugaro2020}.

\section{Data}
\label{sec:data}
\subsection{Asteroseismic data}
In this data release, we add asteroseismic data from C2, C3, C5, C8, and C10-C18 to results from C1 from K2 GAP DR1 \citep{stello+2017} and C4, C6, \& C7 from K2 GAP DR2 \citep{zinn+2020}. In what follows, we describe the procedure to derive the asteroseismic values for stars in these new campaigns, and we also describe how the results from all of the campaigns are combined together.

The majority of K2 GAP targets were chosen to satisfy simple color and magnitude cuts, with a minority chosen based on surface gravity selections from spectroscopic surveys (APOGEE [\citealt{majewski+2010}], SEGUE [\citealt{yanny+2009}]; or RAVE [\citealt{steinmetz+2006}]). Most of the campaigns have targets that are chosen based on a $J-\Ks > 0.5$ color cut and a magnitude cut of $9 \lesssim V \lesssim 15$, where the visual magnitude is computed from 2MASS photometry according to 
\begin{equation*}
    V \approx \Ks + 2((J-\Ks)+0.14)+0.256 e^{{2(J-\Ks)}},
\end{equation*}
which is a relation introduced by \citep{de_silva+2015}. The targets were prioritized for the most part by ranking targets in order of brightest to faintest visual magnitude, with higher priority given to targets selected based on spectroscopy.

\begin{figure*}
    \centering
    \includegraphics[width=0.95\textwidth]{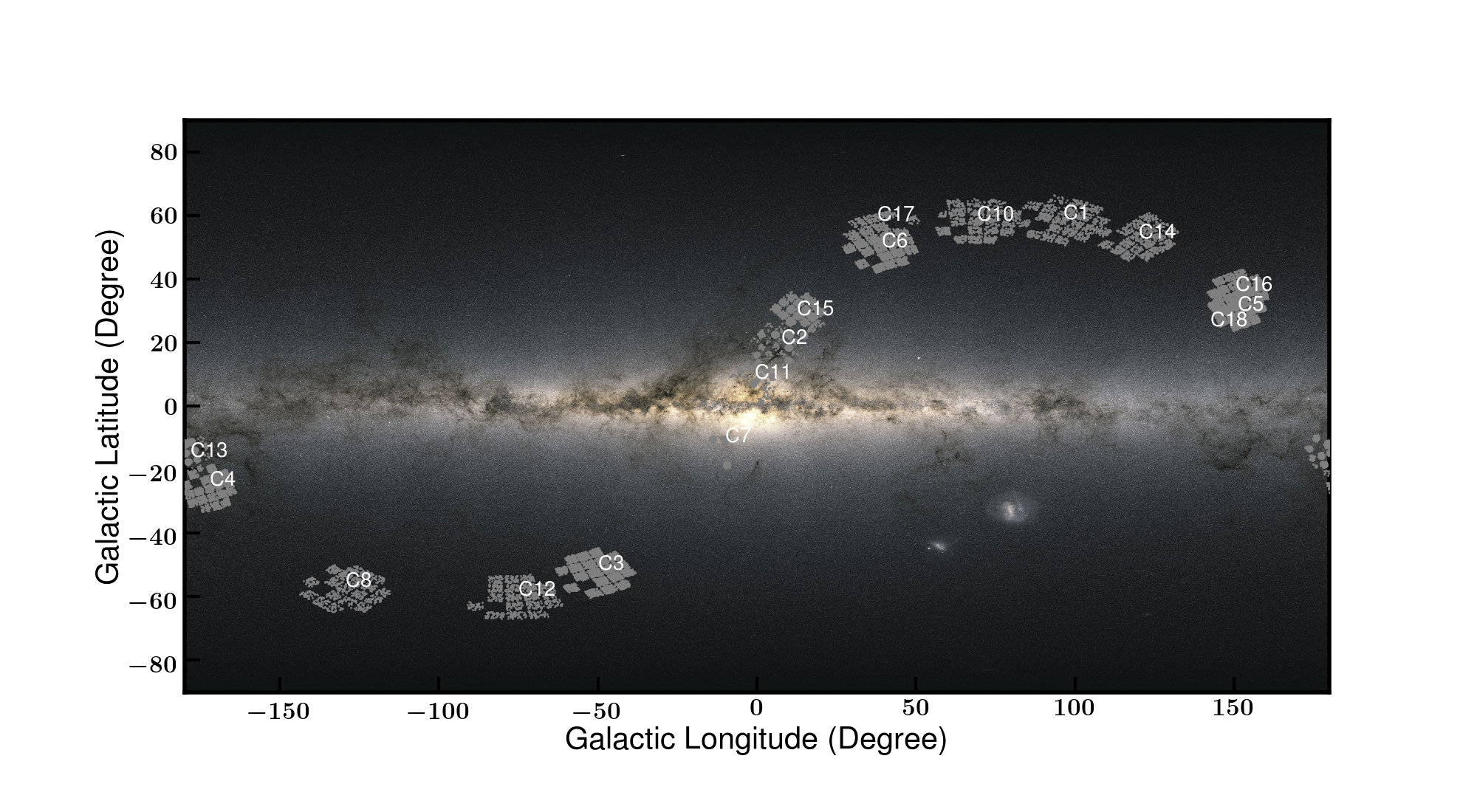}
    \caption{Distribution of the K2 GAP sample across the sky. Targets
      in dense fields close to the Galactic plane were selected only
      within 1-degree circles centered on each module rather than
      across entire modules. Also visible are the rectangular gaps
      corresponding to CCD modules 3 \& 7, which failed prior to the
      start of the K2 mission, and module 4, which failed during Campaign 10. Background image modified from ESA/\Gaia/DPAC.}
    \label{fig:radec}
\end{figure*}

The majority of the targeted stars were observed by \Ktwo, and follow
the target selection functions, with a few exceptions. Notably, the
priorities of C7 targets were mistakenly reversed during the
\Kepler\ office target list consolidation. For details of the effects
of this on the C7 selection function, see \cite{zinn+2020} and
\cite{sharma+2019}; the selection functions for all the campaigns are
described in S. Sharma et al., submitted. In addition, module 4 failed
while taking data in C10. We therefore exclude data from this module
in C10 because of the short duration of data collection before the
failure; modules 3 and 7 had already failed by the time the K2 mission
began, and so there are no data for these modules in K2 GAP. These missing modules can be seen in the K2 GAP DR3 footprint shown in Figure~\ref{fig:radec}.

There are known systematics in the \Ktwo\ light curves that require
special processing beyond the raw light curves produced by the
\Ktwo\ office. In particular, the \Ktwo\ satellite repositioned itself
every $\sim$6 hours to maintain pointing following the partial failure
of its gyroscope system. These thruster firings induce trends in the
light curves that would hinder asteroseismic analysis. The light
curves used in our analysis were therefore detrended from the raw
\Ktwo\ data using the EVEREST pipeline \citep{luger+2018} for all
observed K2 GAP target stars except C1, for which we used the K2SFF pipeline
\citep{vanderburg&johnson2014}, and targets classified as extended in the
Ecliptic Plane Input Catalog \citep[EPIC;][]{2016ApJS..224....2H},
which were not processed by EVEREST. C10 suffered a
failure of module 4 shortly into the start of the campaign, and so we
did not use data from targets on module 4. C11 was separated into two
parts due to a roll angle correction such that some stars had light
curves only for one part of the campaign; we combined the light curves of the two parts when available for the same target. C18 lasted about 50d
due to the spacecraft running low on fuel, and has correspondingly
reduced quality data. C19 only had about a week's worth of data with
pointing comparable to previous campaigns, and, as such, we do not
consider the data from C19 in this data release. 

Following this detrending, we removed non-asteroseismic variability using a boxcar high-pass filter with a width of 4 days, and performed sigma-clipping to reject flux values more than 4-$\sigma$ discrepant. For the campaigns new to this data release, we additionally regularized the spectral window function by inpainting any gaps in the light curves according to the algorithm of \cite{garcia+2014} and \cite{pires+2015}.

\subsection{Spectroscopic data}
APOGEE DR16 \citep{ahumada+2020} spectroscopic data are used for calibrating the asteroseismic data (\S\ref{sec:calibratoin}). APOGEE DR16 is part of SDSS-IV \citep{blanton+2017}, and which is described in \cite{ahumada+2020}. APOGEE observes in the $H$-band using the high-resolution ($R \sim 22,500$) APOGEE spectrograph \citep{wilson+2019} mounted on the Sloan Foundation 2.5-m telescope \citep{gunn+2006} at Apache Point Observatory. APOGEE observes about half of its targets in the disc, with Galactic latitude, $b \leq 16\deg$, with dedicated selection of the bulge, halo, and special programs comprising the rest of its observing allotment. Targets are selected according to color-magnitude cuts of $J-\Ks \geq 0.5$ and $7 \lesssim H \lesssim 14$, across the sky \citep{zasowski+2013,zasowski+2017}. The data are reduced according to \cite{nidever+2015}, using the APOGEE Stellar Parameters and Chemical Abundances Pipeline, \citep[ASPCAP;][]{holtzman+2015,garcia-perez+2016}. The final stellar parameter calibration and validation process is discussed by \cite{holtzman+2018}.

GALAH data are used for our analysis of age-abundance patterns (\S\ref{sec:ageabundance}). GALAH is an optical spectroscopic survey targeting stars in the Galactic disc with $12 < V < 14$ and $|b| > 10^{\circ}$ \citep{martell+2017}. The survey operates from the 3.9-m Anglo-Australian Telescope at Siding Spring Observatory in Australia, using the HERMES multi-object spectrograph \citep{sheinis+2014}. HERMES's high resolution ($R\sim 28,000$) spectra are reduced according to the procedure documented in \cite{kos+2017}. GALAH DR2 presented spectroscopic parameters from The Cannon \citep{ness+2015}, trained on a subset of $\sim 11,000$ stars \citep{buder+2018,heiter+2015} using Spectroscopy Made Easy \citep[SME;][]{valenti_piskunov1996,piskunov_valenti2017}. In this work, we use abundances from GALAH DR3, which improves upon GALAH DR2 by deriving stellar parameters and abundances for all stars directly through the spectroscopic analysis code SME, which performs on-the-fly spectrum synthesis calculations; this reduces potential bias from selection effects in the Cannon training process \citep[e.g.,][]{holtzman+2018}. The SME analysis code utilises grids of pre-computed non-LTE departure coefficients for thirteen chemical elements; these grids and the models they are based on are presented by \citet[and references therein]{amarsi+2020}, and are publicly available \citep{amarsi2020}.

\section{Methods}
\label{sec:methods}
\subsection{Asteroseismic radius and mass scaling relations}
Given the large sample size of the K2 GAP targets, it is not feasible to fit individual modes for each star to determine mass and radius. Instead, we condense the modes' information to two quantities, which can be measured relatively straightforwardly and which are related to the mass and radius of a star through so-called scaling relations. 

The first of these quantities, the frequency at maximum acoustic power, $\numax$, is thought to be related to the acoustic cutoff frequency, and therefore with the surface gravity of the star \citep{brown+1991,kjeldsen&bedding1995,chaplin+2008,belkacem+2011}. Assuming this relation holds homologously across evolutionary state, this implies a scaling relation of the form
\begin{equation}
\frac{\numax}{\numaxsun} \approx \frac{M/\msun}{(R/\rsun)^2\sqrt{(\teff/\teffsun)}}.
\label{eq:_scaling1}
\end{equation}

The second quantity of interest, the large frequency separation, $\dnu$, describes the frequency difference between modes of consecutive radial order that share the same degree. A second, independent scaling relation relates $\dnu$ to the average stellar density \citep{ulrich1986,kjeldsen&bedding1995}
\begin{equation}
\frac{\dnu}{\dnusun} \approx \sqrt{\frac{M/\msun}{(R/\rsun)^3}}.
\label{eq:_scaling2}
\end{equation}

The latter scaling relation is well-understood theoretically, and is valid, strictly speaking, in the limit of large radial order. However, given a stellar structure model, one can compute the expected $\dnu$ at the observed radial order as well as a $\dnu$ in the limit of large radial order, and therefore derive a correction factor, $\fdnu$, to translate the observed $\dnu$ to the large radial order $\dnu$ that enters into Equation~\ref{eq:_scaling2} \citep[e.g.,][]{white+2011,sharma+2016}. We therefore use a modified version of Equation~\ref{eq:_scaling2}: 
\begin{equation}
\frac{\dnu}{\fdnu \dnusun} \approx \sqrt{\frac{M/\msun}{(R/\rsun)^3}}.
\label{eq:scaling1}
\end{equation}

Note that these corrections do not take into account frequency shifts due to the approximations of adiabatic thermal structures and mixing length theory widely used in stellar evolution models \citep[e.g.,][]{jorgensen+2020,jorgensen+2021}. However, such considerations are secondary adjustments to $\fdnu$, given the empirical success of $\fdnu$ in producing agreement of asteroseismic radii and masses with independent estimates \citep[e.g.,][]{huber+2017,brogaard+2018a,zinn+2019rad}. We opt to use the $\fdnu$ corrections from \cite{sharma+2016}, which are computed on a star-by-star basis according to the star's properties (e.g., temperature, metallicity, etc.) by interpolation in a grid of theoretically computed $\fdnu$.  The \texttt{asfgrid} code to compute $\fdnu$ values is  publicly available \citep{sharma+2016,asfgrid}\footnote{\url{http://www.physics.usyd.edu.au/k2gap/Asfgrid/}}

In analogy with corrections to the $\dnu$ scaling relation, there are observational indications that the $\numax$ scaling relation of Equation~\ref{eq:_scaling1} should be modified to include a correction to the observed $\numax$, $\fnumax$ \citep{epstein+2014,yildiz_celik-orhan_kayhan2016,huber+2017,viani+2017,kallinger+2018}. For this reason, we use a modified $\numax$ scaling relation:
\begin{equation}
\frac{\numax}{\fnumax \numaxsun} \approx \frac{M/\msun}{(R/\rsun)^2\sqrt{(\teff/\teffsun)}}.
\label{eq:scaling2}
\end{equation}

Although progress is being made to make robust theoretical predictions of $\numax$ \citep[e.g.,][]{belkacem+2013,zhao+2016,zhou+2020}, it cannot yet be computed based on first principles to the precision required to be useful, as can be done for $\dnu$. We therefore make empirical estimates of $\fnumax$ in \S\ref{sec:calibratoin} for RGB and RC stars, which, in practice, are scalar values such that we can think of $\fnumax$ as indistinguishable from a modified $\numaxsun$.

\begin{deluxetable}{lcc}
  \tablecaption{Solar reference values for each pipeline contributing to K2 GAP DR3 \label{tab:solar_refs}}
\tablehead{Pipeline & $\numaxsun$ & $\dnusun$ }
\startdata
A2Z & 3097.33 & 134.92 \\
CAN & 3140 & 134.92 \\          
COR & 3050 & 134.92 \\          
SYD & 3090 & 135.1 \\           
BAM & 3094 & 134.84 \\              
BHM & 3050 & 134.92 \\
\enddata
\end{deluxetable}

The solar reference values in Equations~\ref{eq:scaling1}~\&~\ref{eq:scaling2} should, in theory, be measured using the same analysis as one would measure $\numax$ and $\dnu$. Therefore, each pipeline has different solar reference values, which are listed in Table~\ref{tab:solar_refs}. We assume here a solar temperature of $\teffsun = 5772
K$ \citep{mamajek+2015}.

By re-arranging Equations~\ref{eq:scaling1}~\&~\ref{eq:scaling2}, the radius scaling relation is found to be
\begin{align}
\frac{R}{\rsun} &\approx \left(\frac{\numax}{\fnumax \numaxsun}\right)
\left(\frac{\dnu}{\fdnu \dnusun}\right)^{-2}
\left(\frac{\teff}{\teffsun}\right)^{1/2}\\
&\equiv \kappa_R \left(\frac{\teff}{\teffsun}\right)^{1/2},
\label{eq:radius}
\end{align}
and the mass scaling relation expression is found to be
\begin{align}
\frac{M}{\msun} &\approx \left(\frac{\numax}{\fnumax \numaxsun}\right)^3
\left(\frac{\dnu}{\fdnu \dnusun}\right)^{-4}
\left(\frac{\teff}{\teffsun}\right)^{3/2}\\
&\equiv \kappa_M \left(\frac{\teff}{\teffsun}\right)^{3/2}.
\label{eq:mass}
\end{align}

Here, we have factored out the dependence on temperature. Since the majority of the K2 GAP DR3 stars do not have spectroscopic temperature estimates, we report, as we did in K2 GAP DR2, the radius and mass coefficients, $\kappar$ and $\kappam$. This allows the user to compute radii and masses using consistent temperature scales in the context of their work. We also provide the $\fdnu$ we compute according to \cite{sharma+2016} in Table~\ref{tab:derived}, though to maintain complete consistency, users should re-compute these values using the same temperature scale as they do to convert radius and mass coefficients into radii and masses. As a reference, should there be a $100K$ discrepancy in the EPIC temperatures used to compute $\fdnu$ as we do here and the user's temperatures, a $1\%$ systematic would be introduced into $\fdnu$. Users may generate their own $\fdnu$ values using the publicly-available \texttt{asfgrid} code.

\subsection{Derived asteroseismic parameters}
\label{sec:derived}
We make use of the same pipelines as the previous K2 GAP data releases
to extract these aforementioned asteroseimsic quantities, $\numax$ and
$\dnu$, from K2 light curves: A2Z \citep{mathur+2010}; BAM
\citep{zinn+2019bam}; BHM \citep{hekker+2010}; CAN
\citep{kallinger+2010,kallinger+2016}; COR
\citep{mosser&appourchaux2009,mosser+2010}; and SYD
\citep{huber+2009}. The generalized problem that each of these
pipelines addresses is to identify a regular pattern of solar-like
oscillations in the presence of red and white noise. The problem of
detecting solar-like oscillations in K2 data also involves systematic
noise that can mimic solar-like oscillations
\cite[see][]{stello+2017,zinn+2019bam}. Though their implementations
vary, the above asteroseismic pipelines share common approaches of 1)
fitting a model to the power spectrum to remove the stellar red noise;
2) fitting a Gaussian excess in power above the red noise, with a mean
corresponding to $\numax$ (A2Z, BAM, BHM, CAN, COR) or heavily
smoothing the excess to localize the frequency of its peak as
$\numax$ (SYD); and 3) identifying $\dnu$ using either individually fitted
modes (CAN) or some version of the autocorrelation function (A2Z, BAM,
BHM, CAN, COR, SYD). For more details on their implementation and
methodology in the context of K2, please see \cite{stello+2017} and
\cite{zinn+2020}.\footnote{The following changes were implemented in
  the SYD pipeline compared to its description and use in K2 GAP DR2:
  1) in addition to the nominal $\numax$ and $\dnu$ confidence cuts
  mentioned in \cite{zinn+2020}, stars are required to fall within the
  empirical $\dnu$--$\numax$ relation from \cite{stello+2009}, such
  that $0.75 (0.262 \numax^{0.772}) < \dnu < 1.5 (0.262
  \numax^{0.772})$; 2) stars for which $\dnu$ was deemed measurable
  were determined based on a machine learning approach from an
  independent analysis of K2 data \citep{reyes+2022}.}

We follow the procedure laid out in K2 GAP DR2 to derive average asteroseismic parameters for each star. This method is similar to the one adopted for the APOKASC-2 sample, which is described in \cite{pinsonneault+2018}. In short, we re-scale each of the pipeline $\numax$ and $\dnu$ values such that the average values for the entire sample across all pipelines is the same, which requires an iterative approach and results in averaged values for each star, denoted $\numaxmean$ and $\dnumean$. Three modifications have been implemented here compared to the methodology described in \cite{zinn+2020}. First, the A2Z $\dnu$ values are not incorporated into the $\dnumean$ due to a significant systematic offset from other pipeline values. Second, for stars that were observed in more than one campaign, variance-weighted averages for each pipeline are computed before proceeding, such that there is only one measurement per star. Third, whereas previously the sigma-clipping was done at the end of each iteration, we now allow the average $\numax$ to converge before performing a 3-$\sigma$ clipping and continuing the iteration process. For each star with at least two pipeline values returned, we then take the average $\numax$ value, $\numaxmean$, and adopt the scatter in those $\numax$ values as the uncertainty on $\numax$, $\sigma_{\numaxmean}$. The same exercise is performed for $\dnu$ to compute $\dnumean$ and $\sigma_{\dnumean}$. In so doing, we are assuming that the different pipelines have systematic differences in $\dnu$ and $\numax$ measurements that tend to cancel out when averaged together. This exercise is done for RGB and RC stars separately, based on evolutionary states computed using the machine-learning approach described in \cite{hon+2017a,hon+2018a}. In brief, the machine learning approach takes advantage of the fact that red giant branch and red clump stars exhibit differences in the observed mode structure \citep{bedding+2011}. These differences are detectable by visual inspection, and are therefore amenable to being learned by machine learning algorithms. The classifier developed by \cite{hon+2017a,hon+2018a} uses a convolutional neural network --- an architecture optimized for image processing --- to learn characteristic red giant and red clump mode features present in power spectra rendered as 2D images. In this work, evolutionary states are assigned arbitrarily at the initial iteration, and in subsequent iterations, for stars with a defined $\dnumean$ and $\numaxmean$,  machine-learning evolutionary states are assigned. The final iteration proceeds with only stars with defined $\dnumean$ and $\numaxmean$. As part of this process, each pipeline has assigned scale factors, $X_{\numax,\mathrm{\ RGB}}$, $X_{\dnu,\mathrm{\ RGB}}$, $X_{\numax,\mathrm{\ RC}}$, and $X_{\dnu,\mathrm{\ RC}}$ that describe by how much the pipeline-specific solar reference value (Table~\ref{tab:solar_refs}) should be multiplied to be put on the $\numaxmean$ and $\dnumean$ scale for RGB stars and RC stars, respectively. These modified solar reference values are provided in Table~\ref{tab:refs}. Here, we also indicate the analogous scaling factors from APOKASC-2 \citep{pinsonneault+2018}, where differences are a result of slightly different methodology and due to not working with the same pipelines: BAM was not a part of the APOKASC-2 analysis. It is also likely that there are significant differences introduced in the pipeline's asteroseismic scales due to the difference between the time baselines of \Kepler\ and \Ktwo, which we discuss in \S\ref{sec:calibratoin}.
\begin{figure}
    \centering
    \includegraphics[width=0.4\textwidth]{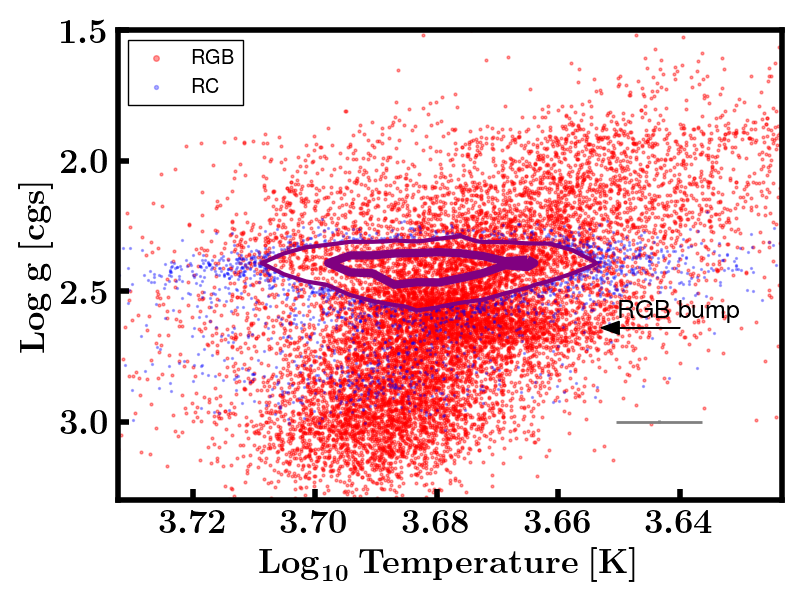}
    \caption{Kiel diagram for K2 GAP DR3, with stars colored red (blue) if classified as red giant branch stars (red clump stars). The purple curves delineate the 38\% and 68\% contours for red clump stars, for clarity in seeing the RGB bump (indicated by the arrow). The surface gravity is calculated using the EPIC temperature in combination with $\numaxmean$, according to Equation~\ref{eq:_scaling1}. The spread in temperature of the red clump is caused in part by its intrinsic width set by population-level variations in metallicity, mass, \& age, and also by the EPIC temperature uncertainty, which is indicated by the typical error bar indicated in the lower right.}
    \label{fig:kiel}
\end{figure}

\begin{figure}
    \centering
    \includegraphics[width=0.4\textwidth]{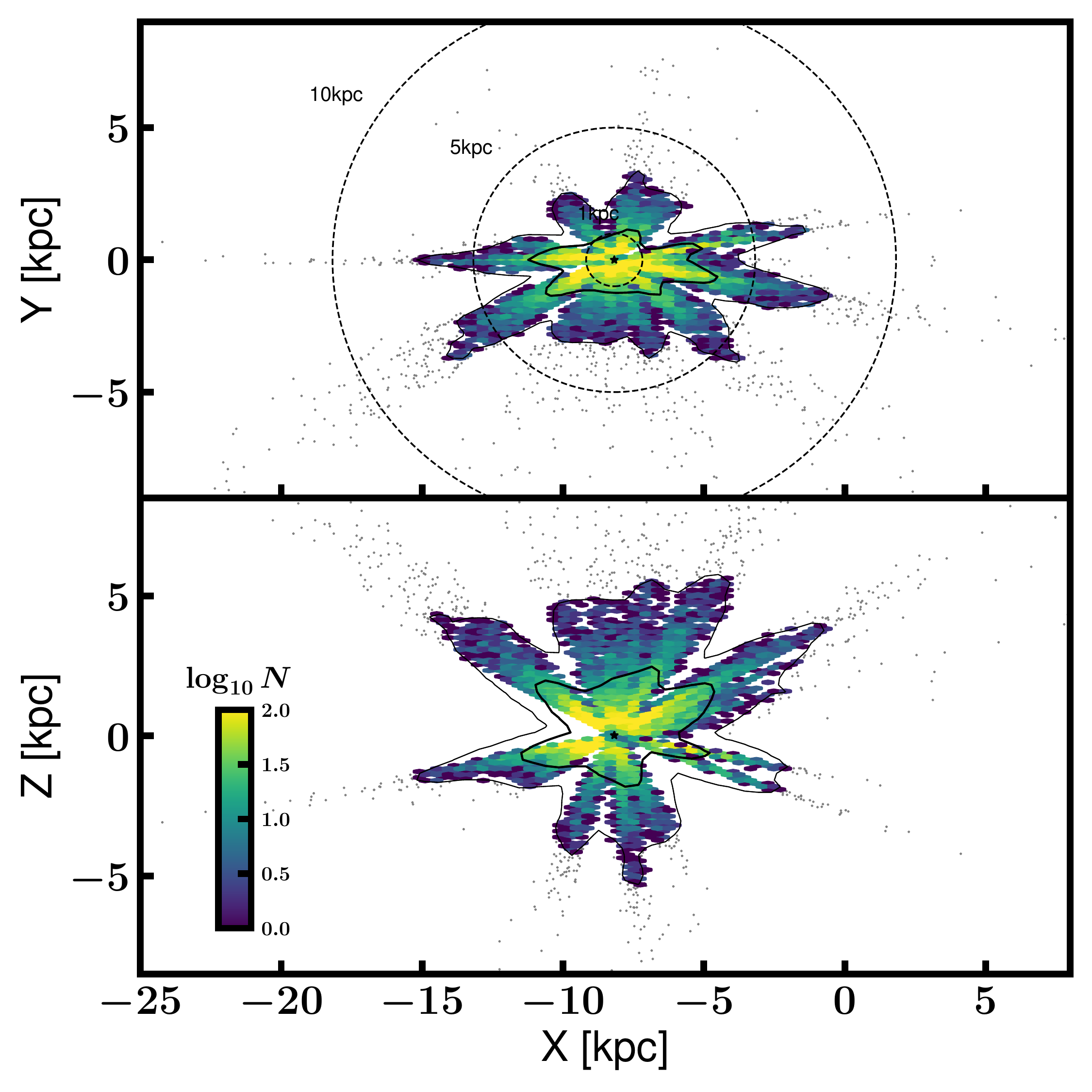}
    \caption{The distribution of K2 GAP DR3 stars (i.e., GAP targets with $\numaxmean$, as defined in \S\ref{sec:derived}) shown in Galactocentric coordinates. The Sun's position of (X, Y, Z) = (-8.18kpc, 0kpc, 0.021kpc) is marked as the black star, and is taken from a combination of the distance to Sgr A* \citep[][X]{gravity-collaboration+2019} and the Gaia DR2 Galactic disc velocity distribution symmetry analysis from \citet[Z]{bennett_bovy2019}. The inner (outer) contour represents the 68th (98th) percentile of the plotted stars. Within these contours, the logarithmic density of stars is indicated according to the color bar. Dashed circles indicate distances of 1kpc, 5kpc, and 10kpc.}
    \label{fig:pos}
\end{figure}
We list in Table~\ref{tab:derived} the individual, re-scaled pipeline values, $\numax' = X_{\numax} \numax$ and $\dnu' = X_{\dnu} \dnu$. As in K2 GAP DR2, we do not list $\numax'$ or $\dnu'$ if that pipeline value is sigma-clipped in the averaging procedure. We correct pipeline-specific $\dnu'$ as well as $\dnumean$ with the theoretical $\fdnu$ from \cite{sharma+2016} using the EPIC temperatures \& metallicities listed in Table~\ref{tab:derived}. We use these re-scaled $\numax'$ and $\dnu'$ values to compute re-scaled $\kappar'$ and $\kappam'$ values for each star and for each pipeline, using the solar reference values appropriate for each pipeline (see Table~\ref{tab:solar_refs}). Our recommended radius and mass coefficients, $\kapparmean$ and $\kappammean$, are those computed using the average parameters $\numaxmean$ and $\dnumean$ and APOKASC-2 solar reference values modified so that our radii are on the \Gaia\ parallactic scale (see \S\ref{sec:calibratoin}): $\nu_{\mathrm{max,}\odot\rm{,\ RGB}} = 3081\muhz$, $\nu_{\mathrm{max}\odot\rm{,\ RC}} = 3096\muhz$ and $\dnu_{\odot\rm{,\ RGB/RC}} = 135.146\muhz$ \citep{pinsonneault+2018}. The pipeline-specific and average radius \& mass coefficients are provided in Table~\ref{tab:kappa}, with their uncertainties calculated according to standard propagation of uncertainty.\footnote{Since A2Z $\dnu$ values do not contribute to $\dnumean$, there are no $\dnu_{\mathrm{A2Z}}'$ values populated in Table~\ref{tab:derived}, and the $\kappa_{R\mathrm{,A2Z}}$ and $\kappa_{M\mathrm{,A2Z}}$ values in Table~\ref{tab:kappa} are calculated using the raw $\dnu$ and re-scaled $\numax'$ values.}

In K2 GAP DR2, we established that the uncertainties from our
averaging process follow $\chi^2$ statistics, and can be described to
a good appproximation by fractional uncertainties that are mostly a
function of evolutionary state. We report in Table~\ref{tab:unc} the
median fractional uncertainties in $\numaxmean$, $\dnumean$,
$\kapparmean$, and $\kappammean$ for both RGB stars and RC stars, and
which may be considered typical of the uncertainties in our sample. We
also include typical fractional uncertainties in these parameters from
K2 GAP DR2, APOKASC-2 \citep{pinsonneault+2018}, and another,
independent analysis of \Kepler\ data \citep{yu+2018}. The typical
$\dnu$ uncertainty for K2 GAP DR3 is somewhat larger than it was in K2
GAP DR2 due to the previously mentioned difference in how sigma
clipping is performed in the averaging procedure used in the two data releases. The resulting precisions in RGB masses, which are determinative in asteroseismic age precisions, are about a factor of two larger than those of \Kepler, corresponding to uncertainties of about $20-30\%$ in age.

We provide all the results returned by every pipeline in
Table~\ref{tab:raw}. Included in this table are machine-learning
evolutionary states based on $\dnumean$ and $\numaxmean$, as well as
evolutionary states based on individual pipeline values, which are
taken to be $\dnu'$ and $\numax'$.\footnote{The A2Z evolutionary
  states are based on raw $\dnu$ and re-scaled $\numax'$. Also, in the
  small number of cases where there were multiple observations of the
  same star across different campaigns, we adopted the evolutionary state from the campaign with the smallest evolutionary state uncertainty according to the machine learning approach.} We also include the EPIC IDs for stars that had no measured asteroseismic parameters by any pipeline but that were targeted as part of K2 GAP so that users may investigate asteroseismic selection functions as needed; we quantify K2 GAP DR3 completeness as a function of mass and radius in \S\ref{sec:injection}. The K2 GAP DR3 sample that we refer to in what follows is a subset of the totality of targeted stars, and consists only of the stars with a valid $\numaxmean$. There are 19417 such stars, 18821 of which also have a valid $\dnumean$ and therefore $\kapparmean$ and $\kappammean$. Stars with both $\numaxmean$ and $\dnumean$ are assigned an evolutionary state, resulting in 12978 RGB stars and 5843 RC stars. The numbers of stars with asteroseismic detections broken down by campaign and pipeline are listed in Table~\ref{tab:numbers}. The Kiel diagram for the K2 GAP DR3 sample is shown in Figure~\ref{fig:kiel} and its distribution on the sky is shown in Figure~\ref{fig:radec}; the sample is also shown in Galactocentric coordinates in Fig.~\ref{fig:pos}.

\section{Validation of asteroseismic values in K2 GAP DR3}
\label{sec:calibratoin}
\subsection{Injection tests}
\label{sec:injection}
In the previous section, we detailed the dependence of asteroseismic results across pipelines. However, there are likely additional systematics due to the length of the K2 light curves compared to, e.g., Kepler light curves. Indeed, \cite{hekker+2012} revealed non-negligible variations in completeness, precision, and accuracy in red giant asteroseismic parameters due to the length of the time series (i.e., the time baseline). In order to test the completeness, precision, and accuracy of the different asteroseismic modelling pipelines for K2-like data, we generated synthetic data for which we knew the ``true" $\numax$ and $\dnu$ from Kepler and performed blind injection recovery tests.

 We first created a grid in magnitude-$\numax$ space from the
 distribution of Kepler stars from APOKASC \citep{pinsonneault+2018};
 the faint giant sample of \cite{mathur+2016}; and the M-giant sample
 of \cite{stello+2014}, in order to select Kepler stars evenly across
 this parameter space. From each bin, when possible, we generated
 K2-like light curves based on 80d segments of Kepler light curves via
 two methods. First, we attempted to select from each bin three Kepler
 stars with at least five quarters of data each, from which we created
 15 synthetic K2 light curves (selecting five different 80d sections
 from three stars). Second, we attempted to generate 15 synthetic K2
 light curves from 15 different Kepler stars from a single 80d section
 of each of their light curves; in practice, not each bin had enough
 stars to create 30 synthetic K2 light curves from these two methods. Each of these synthetic K2 light curves was created using KASOC v1
Q1-Q14 light curves \citep{handberg_lund2014}, linearly interpolating
the Kepler flux onto a cadence of a star in \Ktwo\ C3 to mimic the
spectral window of actual C3 data and the frequency resolution of
\Ktwo. We then increased the white noise level for each of the synthetic K2 light curves according to the following procedure. First, the white noise as a function of magnitude was computed for the entire grid of Kepler stars as well as the 10291 non-GAP C3 targets with EVEREST long-cadence light curves. The white noise for each star was computed by taking the standard deviation of its light curve filtered to remove variability slower than $\sim 150\muhz$. For both of these samples, the 20th percentile of the white noise levels as a function of magnitude were fitted using third degree polynomials. Each synthetic K2 light curve's white noise level was increased by the ratio of the \Kepler-to-\Ktwo\ white noise if that ratio was less than unity at the \Kepler\ star's magnitude. In practice, this resulted in increasing the white noise level for stars fainter than $Kp = 14$, by on average 10\% and by no more than 20\%.

\begin{figure*}[htp]
\centering
  \subfloat{\includegraphics[width=0.4\textwidth]{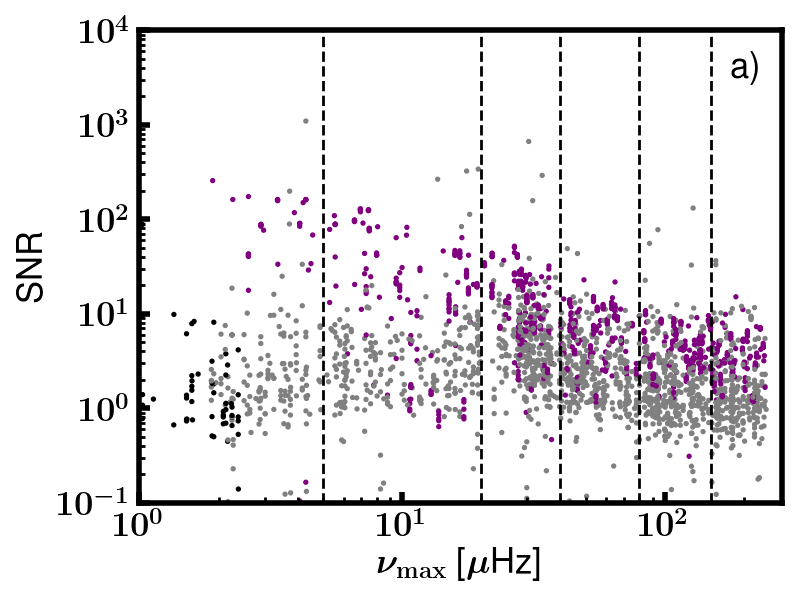}}
  \subfloat{\includegraphics[width=0.4\textwidth]{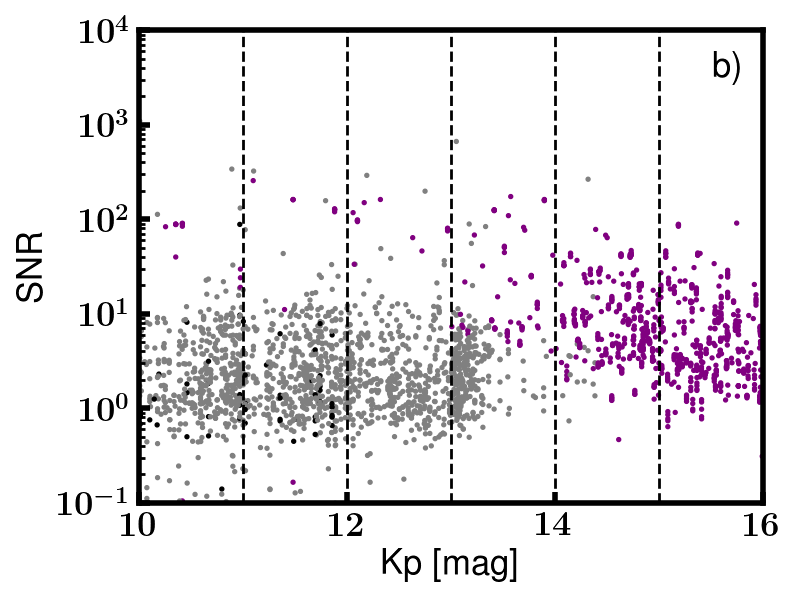}}
\caption{Distribution of the expected signal-to-noise ratio (SNR) for synthetic \Ktwo\ stars as a function of the \Kepler\ $\numax$ (a) and \Kepler\ magnitude (b). Synthetic stars are colored according to their provenance: \cite{pinsonneault+2018} (grey); \cite{mathur+2016} (purple); or \cite{stello+2014} (black). The synthetic population was drawn from a range of SNR and $\numax$ in order to test the accuracy and precision of the asteroseismic pipelines in K2 GAP DR3.}
\label{fig:snr}
  \end{figure*}

We show in Figure~\ref{fig:snr}a the SNR of the synthetic sample.\footnote{Note that this is the SNR in power, not amplitude.} We compute the SNR of the synthetic K2 data in a way that takes into account both the expected maximum mode amplitude and the granulation background level at $\numax$. To do so, we adopt the approach from \cite{campante+2016a}, assuming 3 modes per order, ignoring observation integration time effects, and assuming a noise level according to the observed star-to-star white noise level at high frequencies in the spectra. For the mode maximum amplitude, we adopt the model $\mathcal{M}_{4,\beta}$ from \cite{corsaro+2013}. The points are colored by the provenance of the \Kepler\ data, where there are potentially multiple synthetic stars per KIC ID because of the division of the \Kepler\ light curves into 80d sections. In total, there are 57 synthetic stars from the M-giant catalogue \citep{stello+2014}; 891 synthetic stars from the faint giant catalogue \citep{mathur+2016}; and 1691 synthetic stars from the APOKASC catalogue \citep{pinsonneault+2018}. The dashed lines demarcate the boundaries of our grid used to draw the synthetic light curves in $\numax$ space. We also show the distribution in magnitude space in Figure~\ref{fig:snr}b, with vertical lines demarcating the magnitude bins used to populate the synthetic sample.

The pipelines' analysis of these synthetic K2 data proceeded blindly  (i.e., the synthetic data were treated as real data), and the resulting asteroseismic parameters were processed using an iteration of the averaging procedure described in \S\ref{sec:derived}. The average results are denoted in the following figures as `ALL', and any pipeline-specific results for synthetic K2 data are only shown if they pass the same muster as the real data (i.e., having at least two pipelines return results).

\begin{figure}
    \centering
\subfloat{\includegraphics[width=0.4\textwidth]{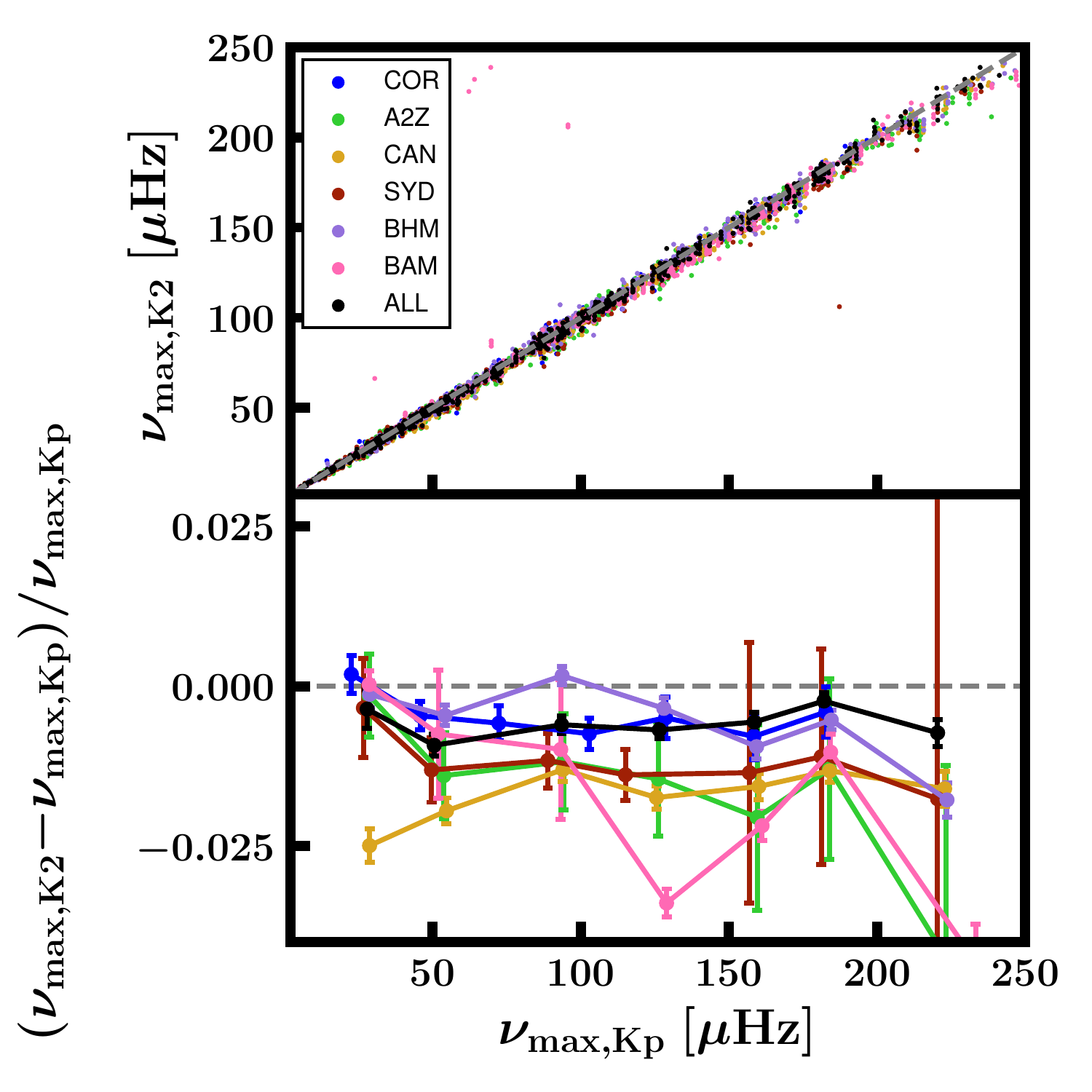}}\\
\subfloat{\includegraphics[width=0.4\textwidth]{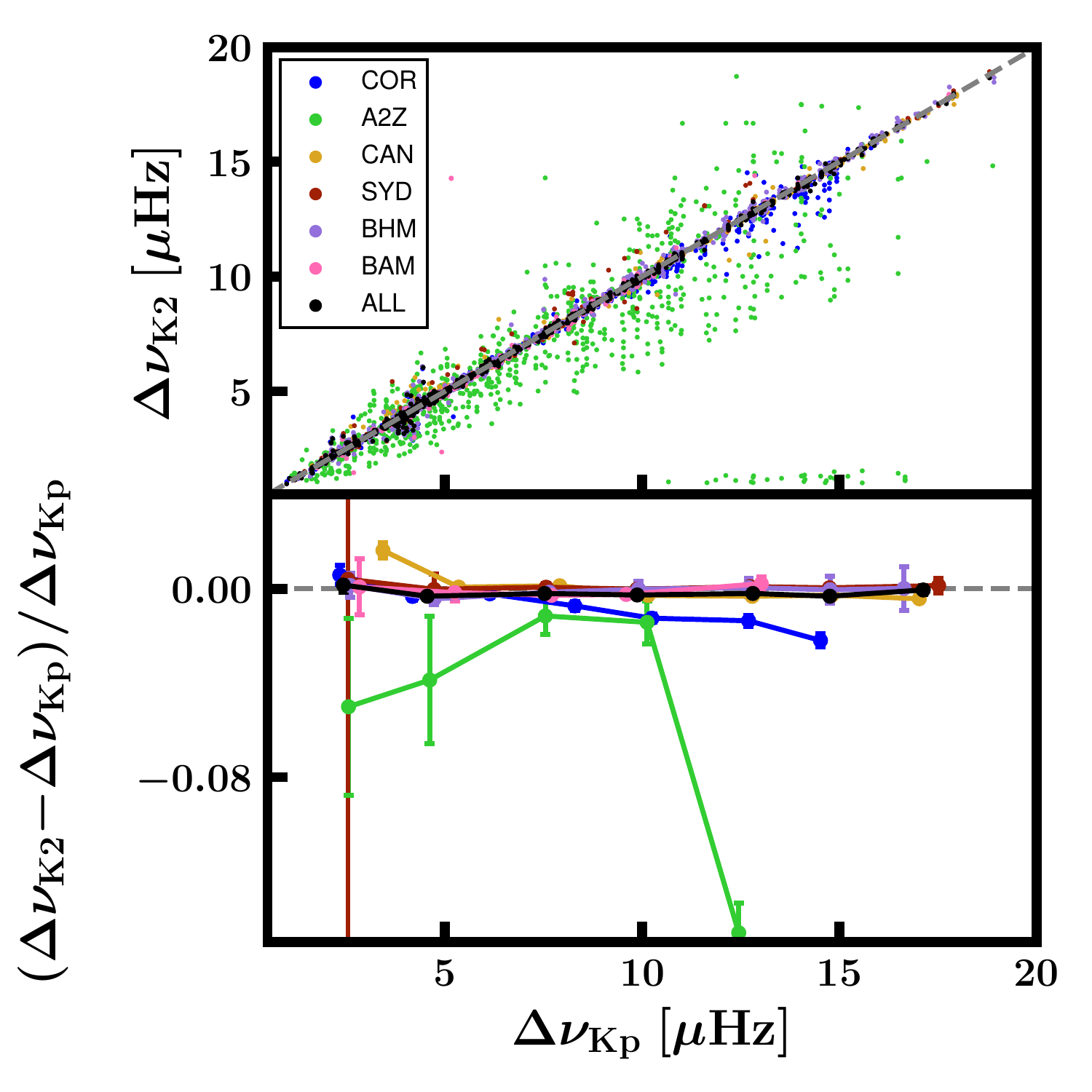}}
    \caption{Binned medians and uncertainties on the median of the fractional difference between \Kepler\ and synthetic \Ktwo\ $\numax$ (top) \& $\dnu$ (bottom) values for each pipeline contributing results to K2 GAP DR3, according to the legend. Deviations from the dashed line indicate that the pipeline returns \Ktwo\ values that are on a different scale than the pipeline's \Kepler\ results (labelled as $\nu_{\mathrm{max,K2}}$ \& $\nu_{\mathrm{max,Kp}}$ [top] and $\dnu_{\mathrm{K2}}$ \& $\dnu_{\mathrm{Kp}}$ [bottom], respectively).}
    \label{fig:truth}
\end{figure}

\begin{figure}
    \centering
\subfloat{\includegraphics[width=0.4\textwidth]{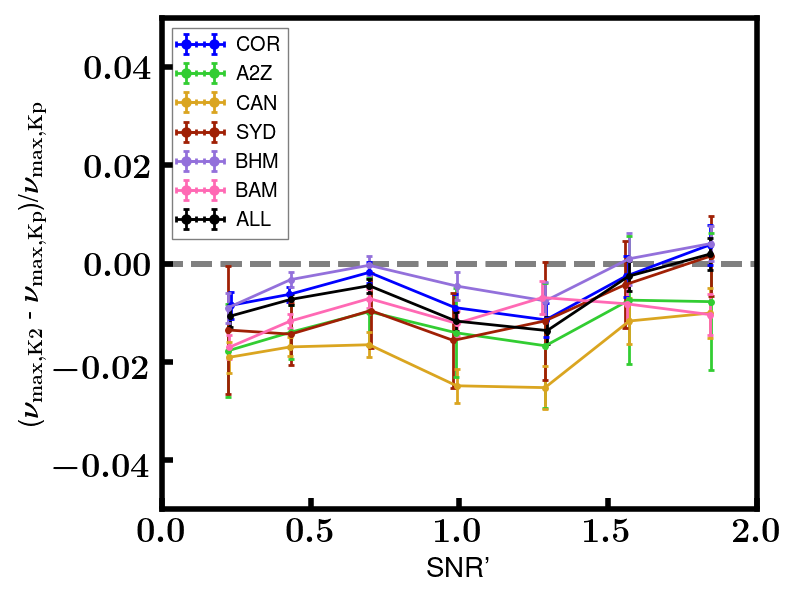}}\\
\subfloat{\includegraphics[width=0.4\textwidth]{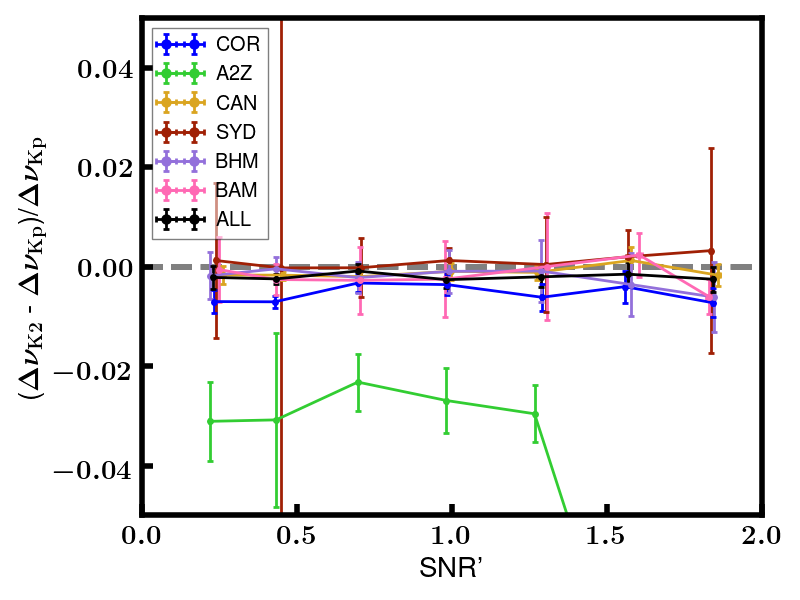}}
    \caption{Fractional difference between \Kepler\ and synthetic \Ktwo\ values, for each pipeline according to the legend, as a function of the synthetic \Ktwo\ signal-to-noise ratio (SNR). The differences are shown as binned medians and uncertainties on the medians. Trends as a function of SNR would indicate a pipeline's asteroseismic values are noise-dependent.}
    \label{fig:truth_snr}
\end{figure}

In Figure~\ref{fig:truth}, we show the accuracy of the recovery for each of the asteroseismic pipelines, based on the ground-truth \Kepler\ asteroseismic values. For this exercise, each pipeline analyzed the Kepler light curves to generate ground-truth labels. For the purposes of this plot and those that follow, uncertainties on the binned median are computed by inflating the standard uncertainty on the binned mean by a factor of $\sqrt{\frac{\pi}{2}}$ \citep{kenney_keeping1962}.

We show the trends in the \Ktwo\ asteroseismic values as a function of
both $\numax$ and $\dnu$, and which are evident at the percent level
as a function of $\numax$ and $\dnu$. There are also biases when
averaged over all of $\numax$ and $\dnu$, which can be seen by the
fact that the trends for some pipelines in Figure~\ref{fig:truth} are
systematically offset below the one-to-one line. This suggests that
there are non-negligible systematics in asteroseismic pipeline
recovery that are a function of baseline, and which would result in
too-small radii and masses compared to Kepler asteroseismology (see
the below comparison between mass distributions in K2 and
Kepler). Time baseline seems to have the smallest impact on $\dnu$,
since several pipelines report nearly identical $\dnu$ with
\Kepler\ as with \Ktwo\ data (though some pipelines show substantial
disagreement). $\numax$, however, suffers from significant biases
based on the time baseline: excursions of a $2-3\%$ and zero-point
biases of $1-2\%$ are observed. There are also indications that some
pipelines may have SNR-dependent biases, which manifest as trends in
the fractional agreement between \Kepler\ and synthetic \Ktwo\ values
as a function of SNR in Figure~\ref{fig:truth_snr}. Note that the SNR
shown in this figure is not the same SNR as shown in
Figure~\ref{fig:snr}a: the SNR in Figure~\ref{fig:truth_snr}
represents the relative SNR at fixed $\numax$, and is computed by
dividing out the median trend from Figure~\ref{fig:snr}a.

Although we will be calibrating our \Ktwo\ data based on independent estimates of radius in \S\ref{sec:calibratoin}, these biases are important to note, and are being investigated in the context of TESS \citep{stello+2021}. It should also be noted that there could be additional biases introduced in asteroseismic analysis based on the preparation of the pixel-level data and the details of processing the light curves into power spectra (e.g., choices in frequency filter). Based on internal consistency checks against K2SFF light curves \citep{vanderburg&johnson2014}, such effects are smaller than time baseline biases shown here ($< 1\%$).

\begin{figure*}
    \centering
    \includegraphics[width=0.7\textwidth]{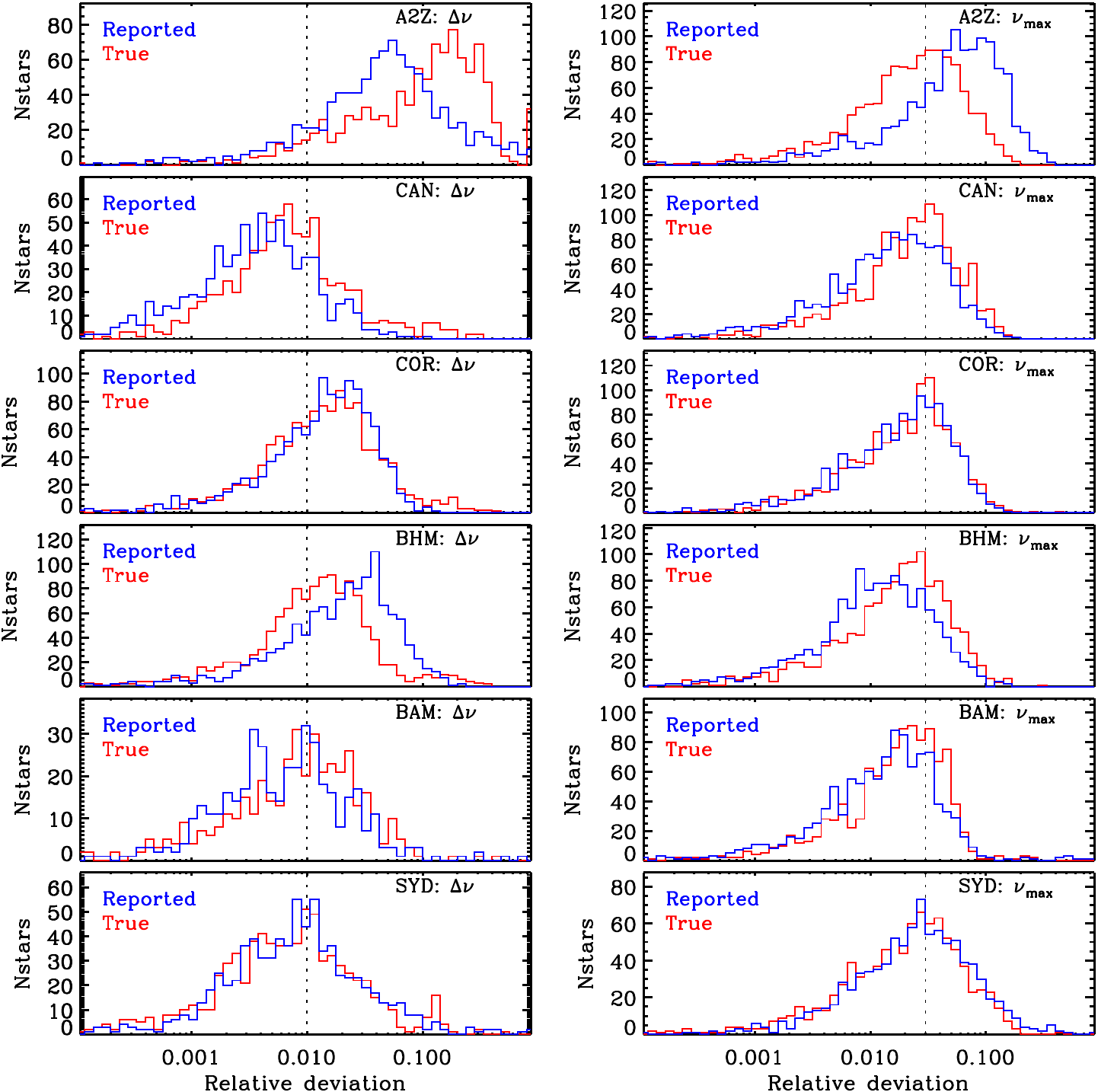}
    \caption{Reported pipeline fractional uncertainty distributions on \Ktwo\ asteroseismic values (`reported') and the inferred fractional uncertainty distributions by comparing \Kepler\ and \Ktwo\ values (`true'). The dashed lines are shown for reference, and correspond to fractional uncertainties of $1\%$ for $\dnu$ and $3\%$ for $\numax$. }
    \label{fig:dennis}
\end{figure*}

Apart from testing the $\numax$-, $\dnu$- SNR- and time baseline--dependent biases in pipeline results, we can also test the internal consistency of the uncertainties using the synthetic \Ktwo\ data. Since we have results from precise \Kepler\ data, we can compare it to the less precise, simulated \Ktwo\ data for the same stars, and evaluate if pipeline results are internally consistent to within their reported uncertainties. To do so, the observed distribution of the fractional deviation between the \Ktwo\ and the \Kepler\ measurements (`true' in Fig.~\ref{fig:dennis}) is compared to the expected distribution (`reported' in Figure~\ref{fig:dennis}), created by drawing Gaussian random variables assuming the reported \Ktwo\ uncertainty for each simulated \Ktwo\ star. If the reported uncertainties were self-consistent, then the two distributions would be identical. If the pipeline tends to over-estimate uncertainties, the `reported' distribution would be skewed toward higher uncertainties compared to the `true' distribution, and vice versa. The internal consistency is globally good for most pipelines. This plot also indicates the relative precision of the pipelines, with the dashed line indicating $\sigma_{\dnu} = 0.01$ and $\sigma_{\nu_{ \mathrm{max}}} = 0.03$, which are representative values for the internal uncertainties for the pipelines. For $\dnu$, there is perhaps a tendency that the pipelines that provide results for fewer stars (and hence maybe are more strict in accepting which measurements are valid) show smaller deviations between `true' and `reported' values. By the same token, the more values a pipeline accepts as valid, the more results deviating strongly from the truth are reported.

\begin{figure*}[htp]
\centering
\subfloat{\includegraphics[width=0.5\textwidth]{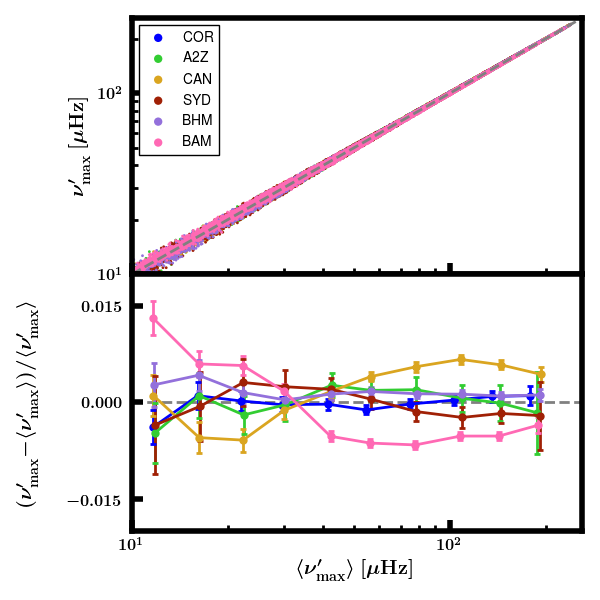}}
\subfloat{\includegraphics[width=0.5\textwidth]{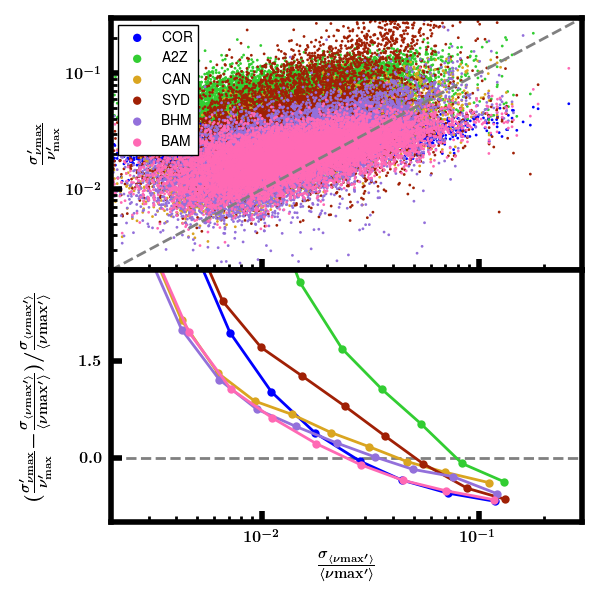}}
\caption{Left: comparison of RGB $\numax$ among pipelines, showing the re-scaled $\numax'$
  from each pipeline versus the mean $\numaxmean$ across pipelines in the top panel,
  and the fractional difference between $\numaxmean$ and $\numax'$ in the bottom panel, with error bars showing
  binned errors on the median fractional difference, assuming the uncertainty
  on $\numaxmean$ to be the standard deviation among the
  re-scaled pipeline $\numax'$ values, $\sigma_{\numaxmean}$. Right: the fractional scatter across
  $\numax'$ for a given star with multiple pipeline values is plotted
  against the reported fractional uncertainty on $\numax'$ for each pipeline. The
  fractional difference between the two fractional uncertainties is shown in the bottom panel.}
\label{fig:numax_rgb_mean}
\end{figure*}

\begin{figure*}[htp]
\centering
  \subfloat{\includegraphics[width=0.5\textwidth]{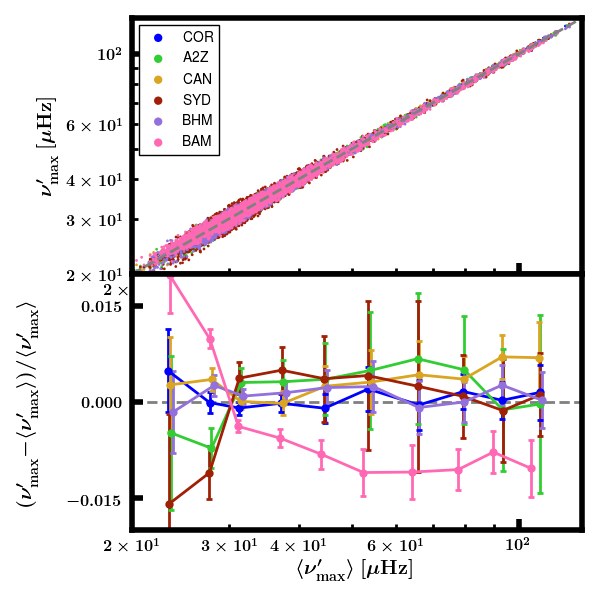}}
  \subfloat{\includegraphics[width=0.5\textwidth]{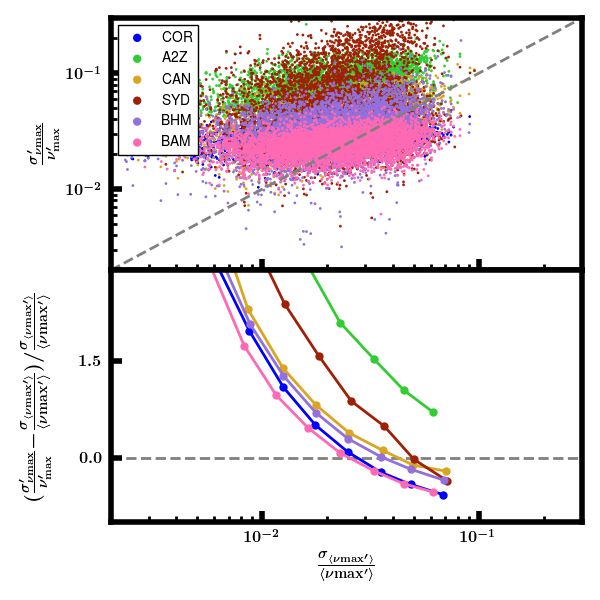} }
\caption{Same as Figure~\ref{fig:numax_rgb_mean}, but for RC stars.}
\label{fig:numax_rc_mean}
\end{figure*}

\begin{figure*}[htp]
\centering
  \subfloat{\includegraphics[width=0.5\textwidth]{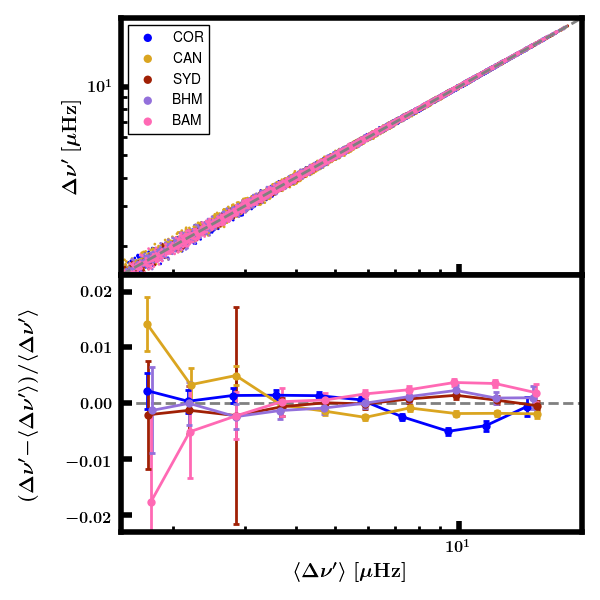}
  \subfloat{\includegraphics[width=0.5\textwidth]{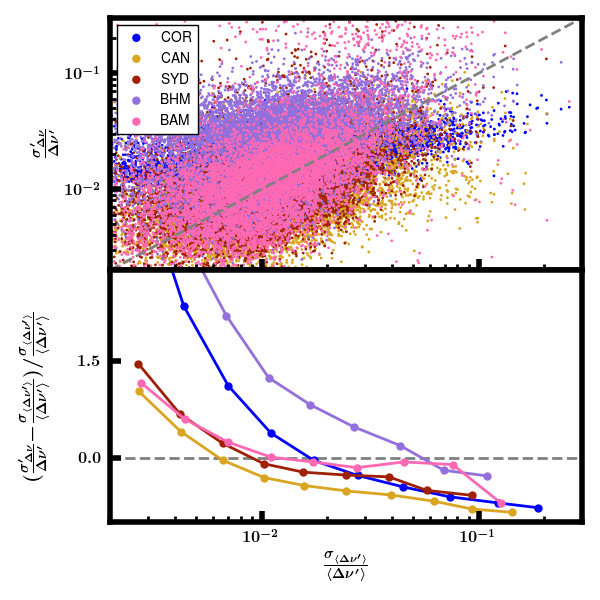})}}
\caption{Same as Figure~\protect\ref{fig:numax_rgb_mean}, but for $\dnu$.}
\label{fig:dnu_rgb_mean}
  \end{figure*}

\begin{figure*}[htp]
\centering
  \subfloat{\includegraphics[width=0.5\textwidth]{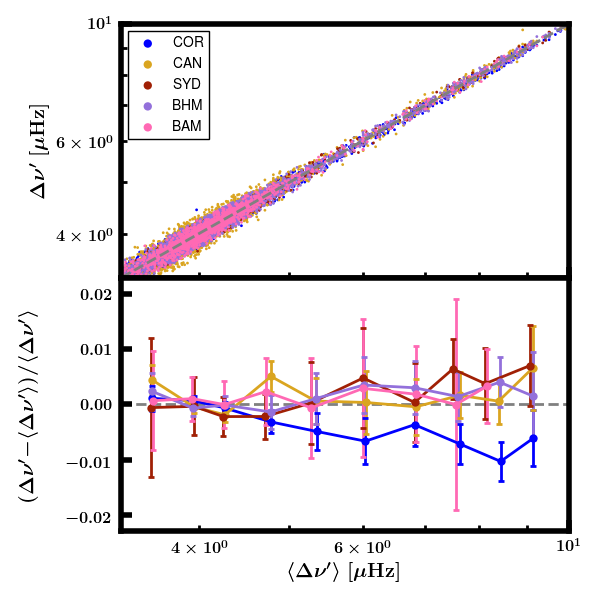}}
  \subfloat{\includegraphics[width=0.5\textwidth]{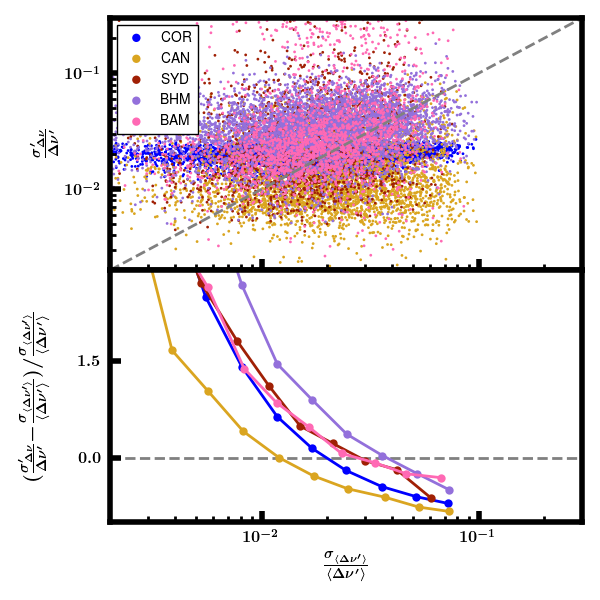}}
\caption{Same as Figure~\protect\ref{fig:dnu_rgb_mean}, but for RC stars.}
\label{fig:dnu_rc_mean}
  \end{figure*}

The above exercise tests the internal consistency of the uncertainties reported by each pipeline, but, by comparing the reported uncertainties to the scatter in the pipeline values for each star, $\sigma_{\numaxmean}$ and $\sigma_{\dnumean}$ described in \S\ref{sec:methods}, we can better establish the accuracy of the pipeline uncertainties. Indeed, even if a pipeline consistently assigns uncertainties to their parameters, it does not necessarily correspond to the true uncertainty --- i.e., including systematic uncertainties --- in the physical parameter: each pipeline's methodology is on its own system and measures $\numax$ and $\dnu$ in slightly different ways. This can be seen to the extent that the scaling factors for each pipeline, $X_{\numax}$ and $X_{\dnu}$, differ from unity, indicating that the pipelines measure asteroseismic values on scales that differ by up to $1\%$. Even after correction to the mean scale, the top panels of Figures~\ref{fig:numax_rgb_mean}-\ref{fig:dnu_rc_mean} show that there are residual fractional deviations between re-scaled pipeline values and the mean values across pipelines, $\numax'$ and $\numaxmean$ as a function of $\numax$ and $\dnu$. By adopting the scatter across pipelines in asteroseismic values as our uncertainties in K2 GAP DR3, we take into account the uncertainties due to these differences in pipeline methodologies. We show comparisons between the internal uncertainties for each pipeline and the K2 GAP DR3 uncertainties in the right panels of Figures~\ref{fig:numax_rgb_mean}-\ref{fig:dnu_rc_mean}. The region above (below) the dotted lines is a regime of where the pipeline-reported uncertainties are larger (smaller) than the K2 GAP DR3 adopted uncertainties. As found in DR2, the pipelines often agree on $\dnu$ and $\numax$ better than would be expected from their internal uncertainties. 

The uncertainties $\sigma_{\numaxmean}$ and $\sigma_{\dnumean}$ do not
explicitly take into account the reported measurement/statistical
uncertainties of the pipelines, but, by virtue of
$\sigma_{\numaxmean}$ and $\sigma_{\dnumean}$ being defined based on
the pipeline-to-pipeline scatter, they capture both systematic
uncertainties in the pipeline methods and statistical measurement
uncertainties: large bias in the pipeline results will tend to
increase the pipeline-to-pipeline scatter, as would large measurement
uncertainty. Even if we assume the reported pipeline measurement
uncertainties represent the true uncertainties, which is to varying
degrees an inaccurate assumption (cf., Fig.~\ref{fig:dennis}), it is
not clear how the statistical uncertainties in the pipeline
measurements should be combined to yield a purely statistical uncertainty in $\numaxmean$ and $\dnumean$,
which are averages of the pipeline measurements. This is because the
pipelines will have some degree of correlation in their measurements
owing to all the pipelines analyzing the same power spectrum for a
given star (i.e., there is only one realization of the data). In order
to estimate a purely statistical uncertainty on $\numaxmean$ and
$\dnumean$, we conservatively assume that all the pipeline
measurements are completely correlated, and compute uncertainties on
$\numaxmean$ and $\dnumean$, which we report in Table~\ref{tab:derived} as
$\varsigma_{\numaxmean}$ and $\varsigma_{\dnumean}$. These latter
uncertainties are larger than our adopted empirical uncertainties in
this work,  $\sigma_{\numaxmean}$ and $\sigma_{\dnumean}$, by factors
of $\sim 2.2$ and $\sim 1.5$, respectively. Assuming a correlation of 0.1 among all pipelines reduces the differences to $\sim 1.2$ and $0.95$. Because the reported pipeline uncertainties are to varying degrees unreliable (Fig.~\ref{fig:dennis}) and because of the unknown correlation among different pipeline measurements, these uncertainties are not used in this analysis, but are rather provided as a conservative indication of a purely statistical uncertainty compared to our adopted empirical uncertainties, $\sigma_{\numaxmean}$ and $\sigma_{\dnumean}$. 

  \begin{figure}[htp]
\centering
\includegraphics[width=0.4\textwidth]{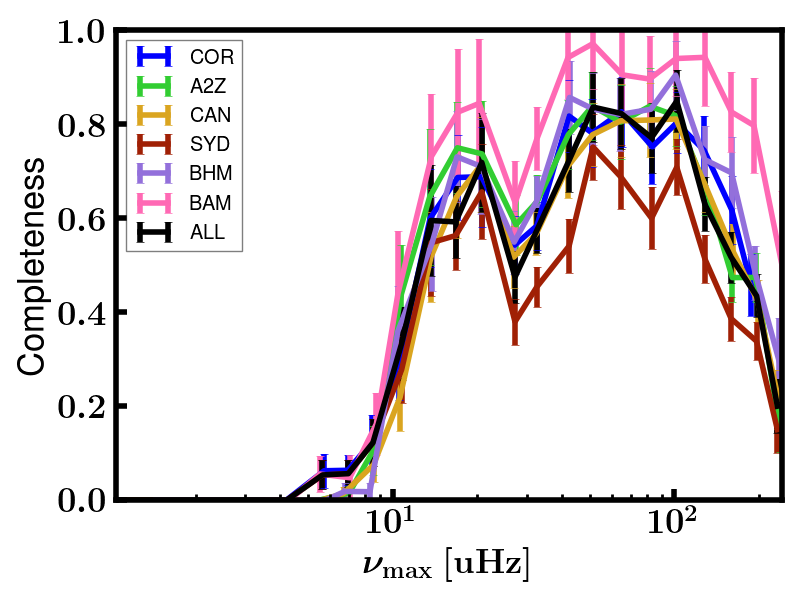}
\caption{The recovery rate as a function of $\numax$ for the pipelines contributing to K2 GAP DR3. A completeness of 1 would indicate that the pipelines recover all of the injected synthetic \Ktwo\ stars generated from \Kepler\ data, and gives confidence that the pipeline is recovering all the detectable stars in real \Ktwo\ data.}
\label{fig:compnumax}
  \end{figure}

  \begin{figure}[htp]
\centering
\includegraphics[width=0.4\textwidth]{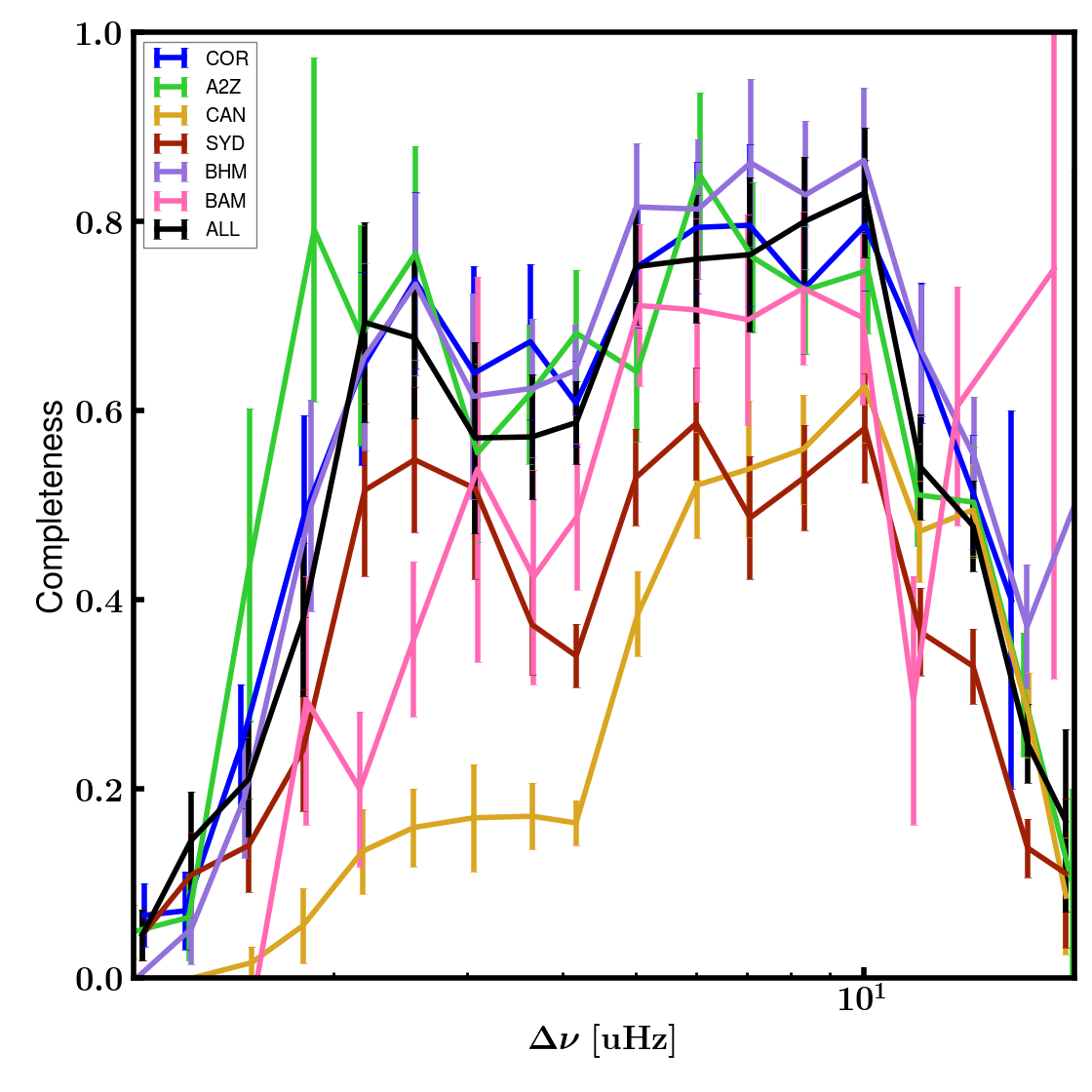}
\caption{The same as Figure~\ref{fig:compnumax}, but for $\dnu$.}
\label{fig:compdnu}
  \end{figure}
  
   \begin{figure}[htp]
\centering
\includegraphics[width=0.4\textwidth]{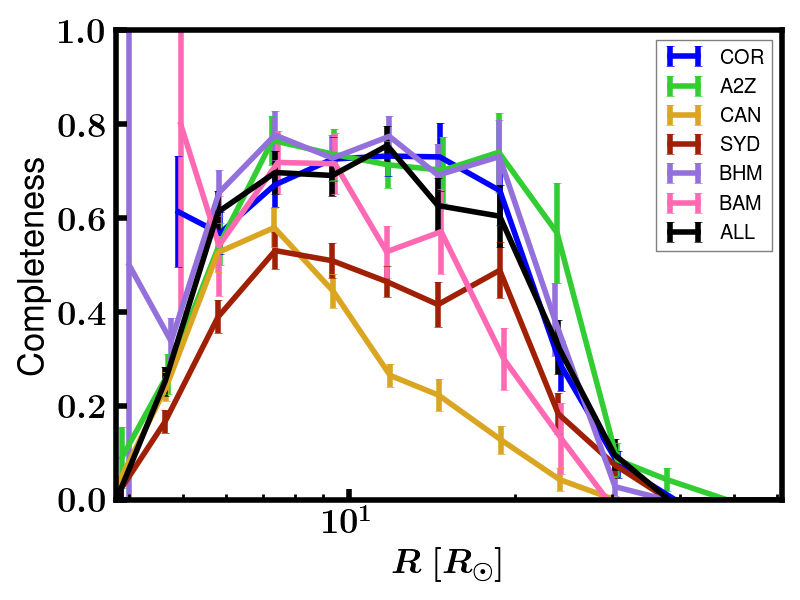}
\caption{The same as Figure~\ref{fig:compnumax}, but for radius.}
\label{fig:compradius}
  \end{figure} 
  
  \begin{figure}[htp]
\centering
\includegraphics[width=0.4\textwidth]{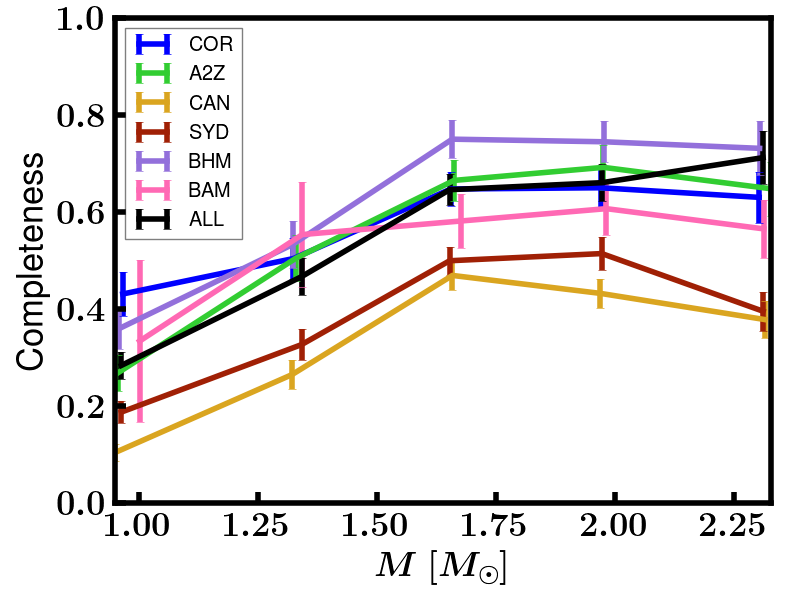}
\caption{The same as Figure~\ref{fig:compnumax}, but for mass.}
\label{fig:compmass}
  \end{figure}
  
We next estimate the completeness of each pipeline's results by comparing the number of recovered stars to the total number of synthetic stars. The completeness fraction, where $1.0$ indicates a perfect recovery rate, is shown as a function of $\numax$, $\dnu$, radius, and mass in Figures~\ref{fig:compnumax}-\ref{fig:compmass}. The synthetic sample was created with a range of SNRs, and with magnitude-dependent noise consistent with \Ktwo\ data, but the distribution of the synthetic sample is not, in detail, representative of the K2 GAP DR3 sample. For this reason, the completeness curves plotted in Figures~\ref{fig:compnumax}-\ref{fig:compmass} are indicative but not determinative of the completeness of the respective parameters in K2 GAP DR3. Note also that the completeness is defined with respect to Kepler results, so this completeness is, strictly speaking, an estimate of the completeness of recovering K2 data with respect to Kepler and not necessarily an absolute completeness estimate, which must await a future analysis using Gaia as a reference \cite[e.g., following the Kepler observation completeness analysis from][]{wolniewicz_berger_huber2021}. 

We see that the completeness curves are peaked in the middle of parameter space for $\numax$ and $\dnu$, with lower completeness at high and low values of $\numax$ and $\dnu$. This is understood to be related to the frequency resolution: both $\dnu$ and $\numax$ detection is limited on the lower end by the time baseline, and on the upper end by the sampling rate of the \Ktwo\ observations. It is even more difficult to recover $\numax$ and $\dnu$ at low values because of another effect: there are fewer modes that are excited at low $\numax$, and they can be difficult to distinguish from noise, especially at the frequency resolution of \Ktwo. This latter effect is the reason why there is a marked decrease in completeness for $\numax \lesssim 10\muhz$. This incompleteness has been noted in previous data releases \citep{stello+2017,zinn+2020} but we are able to robustly quantify it here for the first time: although it varies by pipeline, at least $\approx 20\%$ of stars with $\numax \lesssim 10 \muhz$ are not detected.

 The completeness fractions in radius and mass space are not one-to-one mappings from $\numax$ and $\dnu$, since, for a given surface gravity ($\numax$), there is a spread in mass ($\dnu$). For this reason, we consider the radius and mass completeness curves separately from the $\numax$ and $\dnu$ cases. The completeness in radius suffers from a drop-off in recovery with increasing radius, due to incompleteness in $\numax$ and $\dnu$ at lower frequencies. Given the lack of a strong correlation between radius and age on the giant branch (since the majority of a red giant's lifetime is spent on the main sequence as opposed to climbing the giant branch), the drop-off in recovery with increasing radius does not require a selection function correction in age space, but does have implications for a selection function correction as a function of distance.
The completeness curves are much less peaked in mass space. This is of particular interest for Galactic archaeology applications of K2 GAP DR3: were completeness a strong function of mass, it would require special treatment in the selection function. There is a tendency for low-mass stars to be under-represented among some pipelines, for $M \lesssim 1.2 \msun$. This may be relevant for detailed studies since this will map onto an under-representation of older stars. Regarding the completeness of the underlying K2 GAP sample itself, typically 97\% of the proposed targets in any given campaign were observed, with the targets following simple color-magnitude cuts (S. Sharma, in prep.).

\begin{figure*}
    \centering
    \includegraphics[width=0.7\textwidth]{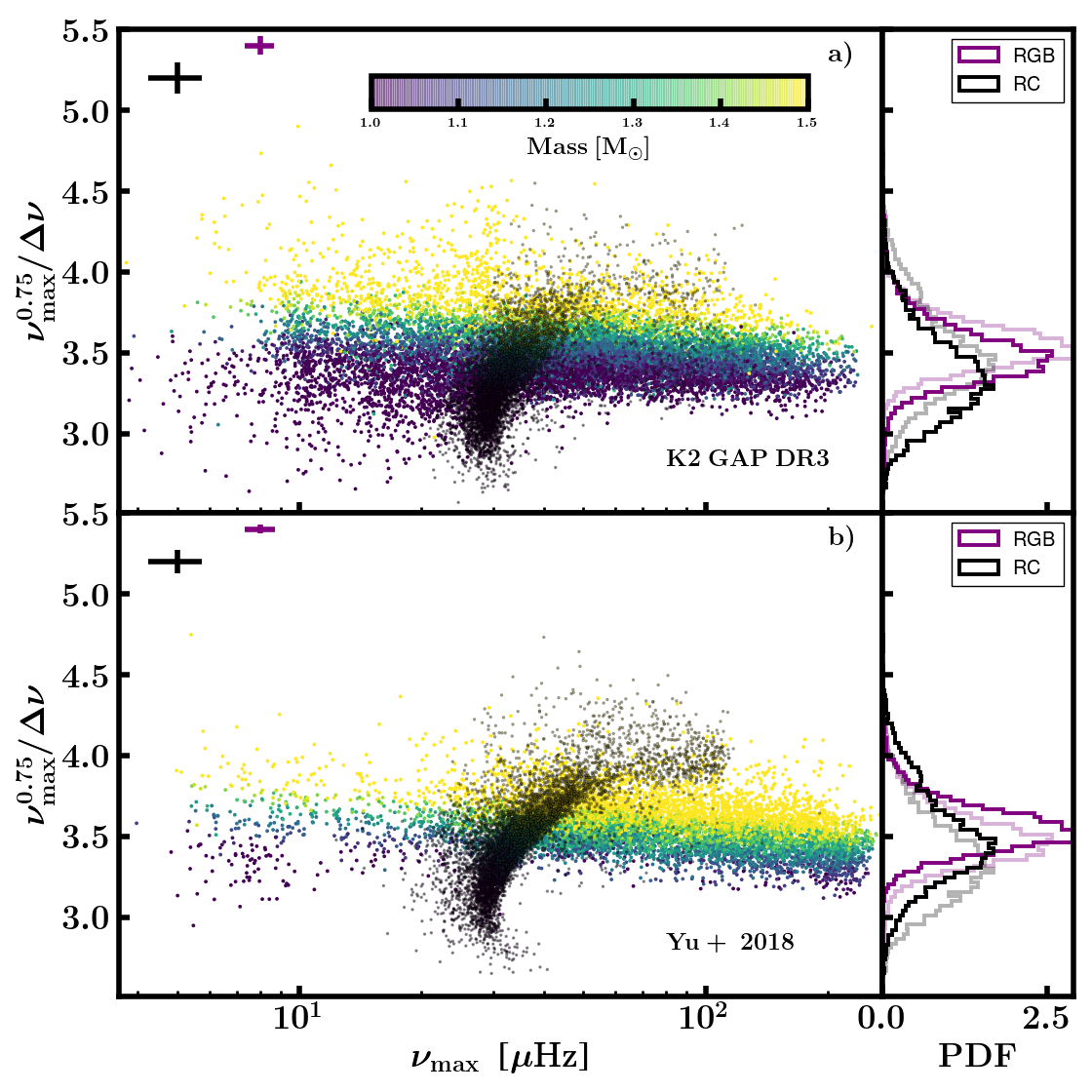}
    \caption{Asteroseismic mass diagram for K2 GAP DR3 (a) and the \cite{yu+2018} Kepler sample (b), where the ordinate is a proxy for mass. Stars are colored by their asteroseismic mass, which is computed according to Equation~\ref{eq:mass}, using EPIC temperatures. Red clump stars are indicated by black symbols. Stars are classified as first-ascent red giant or not based on a machine-learning approach (see \S\ref{sec:derived}). Typical uncertainties for red giant branch stars and red clump stars are indicated by the purple and black error bars. The probability density functions (PDFs) of red giant branch and red clump mass proxies are shown on the panels on the right, where the lighter curves in each panel show the bold curves from the other panel, for comparison.}
    \label{fig:nike}
\end{figure*}

Figure~\ref{fig:nike}a is indicative of the mass distribution for
those stars in the K2 GAP DR3 sample with both $\numaxmean$ and
$\dnumean$, where the ordinate is an asteroseismic proxy for mass
proposed by \cite{huber+2010} that scales like $M^{0.25}$, given the asteroseismic scaling relations (Eqs.~\ref{eq:scaling1} \&~\ref{eq:scaling2}). For reference, Figure~\ref{fig:nike}b shows the Kepler sample from \cite{yu+2018}. Comparing the Kepler and K2 samples, we find a good correspondence, with a couple of differences worth noting. First, the right edge of the clump is better defined in Kepler data by having better precision and more high-mass secondary red clump stars. The Kepler sample also extends to higher frequencies than does K2 GAP DR3, presumably due to better noise properties in Kepler compared to K2. However, K2 has double the fraction of low-frequency ($< 20\muhz$) oscillators than does Kepler, in spite of the tendency to not recover stars in this frequency regime with K2-like time baselines (Fig.~\ref{fig:compnumax}). Note that the overall shift in mass between the \cite{yu+2018} and K2 GAP DR3 samples is consistent with the time baseline systematics in $\numax$ (Fig.~\ref{fig:truth}), such that the SYD Kepler $\numax$ values would be expected to be larger by $\sim 1\%$ than K2 $\numax$ values.

\begin{figure*}
    \centering
    \includegraphics[width=0.7\textwidth]{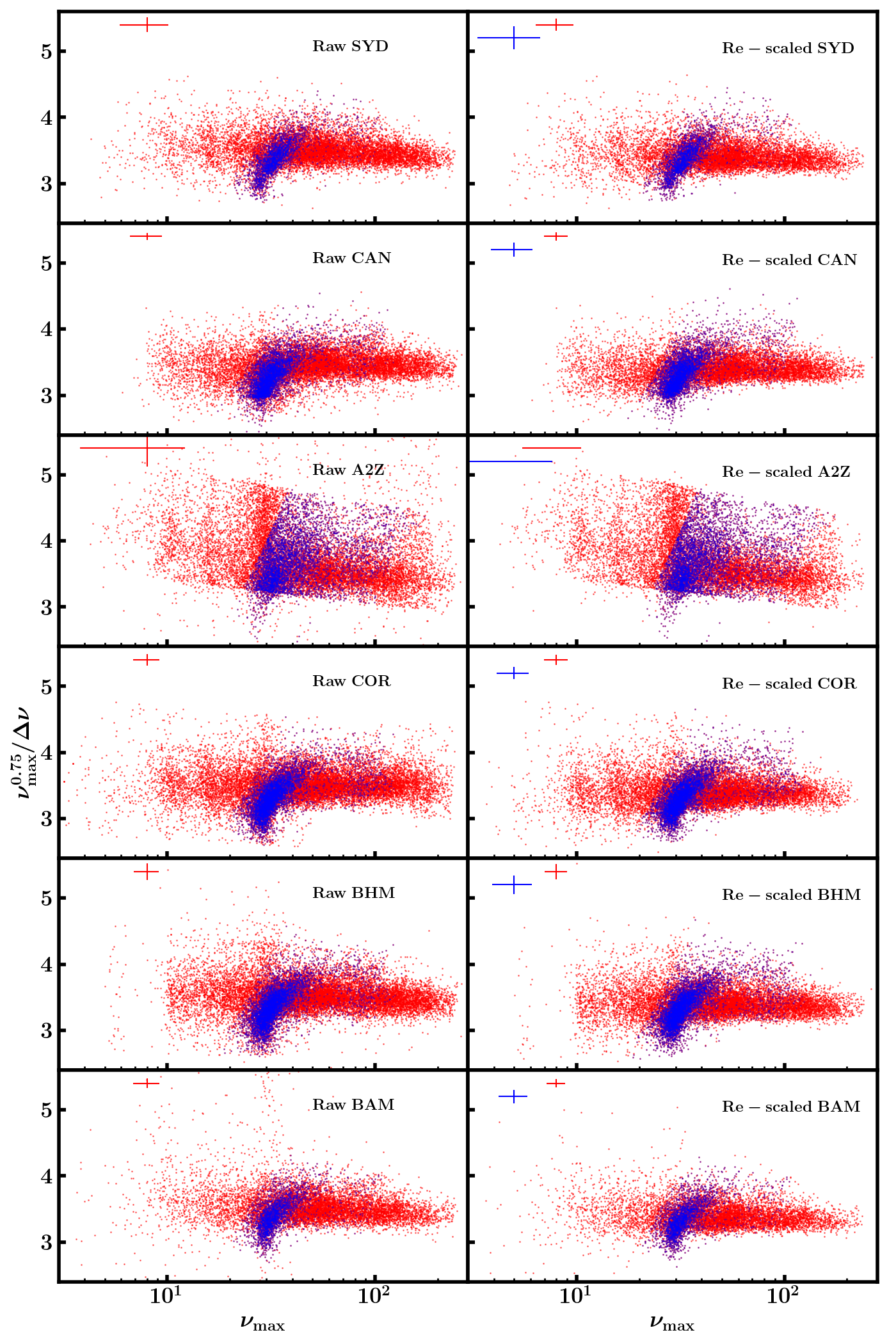}
    \caption{Asteroseismic mass diagram for each pipeline, where the ordinate is a proxy for mass. Red giant branch stars are shown as red points, and red clump stars are shown as blue points. The left column shows the raw pipeline-specific values ($\dnu$, $\numax$) and the right column shows the re-scaled pipeline-specific values ($\dnu'$, $\numax'$), which are only given for stars with results from at least two pipelines (see \S\ref{sec:derived}). Stars are classified as first-ascent red giant or not based on a machine-learning approach using pipeline-specific asteroseismic values (see \S\ref{sec:derived}). Typical uncertainties for RGB and RC stars are indicated by the red and blue error bars.}
    \label{fig:nike_pip}
\end{figure*}

As with Figure~\ref{fig:nike}a, Figure~\ref{fig:nike_pip} shows the K2 GAP DR3 stars in the mass proxy v. $\numax$ space, but for each pipeline and separately for raw pipeline results ($\numax$, $\dnu$; left panels) and re-scaled pipeline values ($\numax'$, $\dnu'$; right panels). The structures of the distributions in this space are generally similar across pipelines, though there are differences in detail. For instance, we see that there are some pipeline-dependent differences in the recovery of low-mass red clump stars and the recovery of low-frequency stars. There are also differences between the raw and re-scaled values, the most salient of which are that 1) raw values have more scatter in the ordinate (due to requiring more than one pipeline returning results to define the re-scaled values, which will tend to select stars with more precise asteroseimsic values) and 2) there tend to be fewer low-frequency and high-frequency re-scaled values (a selection effect of it being less likely for multiple pipelines to return values for stars affected by K2's white noise and time baseline). The diagonal ridge on the left side of the red clump distribution is due to requiring that stars with $\dnu < 3.2$ be assigned a red giant branch evolutionary state (see \S\ref{sec:derived}). However, we see that this choice does not cut out true red clump stars, which are found in the locus where the density of the blue points saturates.

\subsection{Asteroseismic calibration with \Gaia}
\label{sec:abs}
In \S\ref{sec:methods}, we indicated that it is important to use appropriate solar reference values, according to the asteroseismic pipeline being used. The K2 GAP DR3 values are averages across pipelines, so the question arises as to what solar reference value scale is appropriate. One proposal would be to adopt the solar reference values from APOKASC-2 \citep{pinsonneault+2018},  $\numaxsun = 3076\muhz$ and $\dnusun = 135.146\muhz$, given that APOKASC-2 values are also averages across pipelines. Although we follow a very similar methodology to place the pipeline values on a common scale, it differs in some regards (e.g., sigma-clipping and not weighting pipeline values by uncertainties during the averaging process). As well, we include results from BAM, which was not a pipeline considered in \cite{pinsonneault+2018}. For this reason, we cannot assume that $\numaxmean$ and $\dnumean$ are on the same scale as defined by \cite{pinsonneault+2018} just because we use the solar reference values from the cluster calibration procedure in \cite{pinsonneault+2018}. It is also possible that the difference in Kepler versus K2 observation duration results in systematically different parameter measurements (see \S\ref{sec:injection}). 

With this in mind, in what follows, we calibrate the K2 GAP DR3 $\numaxmean$ values using a non-unity, scalar $\fnumax$ (\S\ref{sec:methods}), or, equivalently, by re-scaling the APOKASC-2 $\numaxsun$ value. 
Our \Gaia\ calibration sample is the subset of stars in the K2 GAP DR3 sample with $\numaxmean$ and $\dnumean$ that have APOGEE DR16 \citep{ahumada+2020} temperatures \& metallicities and \Gaia\ parallaxes \& proper motions from  \textit{Gaia} Data Release 2 \citep{gaia2018,lindegren+2018}. 

With the known zero-point offset in \Gaia\ parallax \citep[e.g.,][]{lindegren+2018,khan+2019,zinn+2019zp} in mind, we appeal to the methodology described in \cite{schoenrich+2019a}, which infers distances in \Gaia-based bulk stellar motions. This method can be sensitive to knowing the selection function of the stellar population, and so we take care to model the selection function of GAP targets according to \cite{schoenrich_aumer2017}. The resulting parallax zero-points show a scatter of $\sim 10\muas$ across the campaigns, comparable to the positional variation found by \cite{chan_bovy2020} and \cite{khan+2019}.

We perform the calibration using a subset of the \Gaia-APOGEE-K2 overlap, knowing that there are certain known systematics that could bias the calibration. First, we limit the impact of parallax zero-point by only working with stars with raw Gaia parallaxes of $\pi > 0.4$\mas, parallax uncertainties less than $10\%$, and \Gaia\ $G$-band magnitude $< 13$mag, out of an abundance of caution, in light of indications of parallax- and magnitude dependent offsets \citep{schoenrich+2019a,zinn+2019zp}. We also reject metal-poor stars ($\feh < -1$) from subsequent analysis, since there are indications that asteroseismic scaling relation systematics could exist in the metal-poor regime \citep[][though see \citealt{kallinger+2018}]{epstein+2014,zinn+2019rad}. We further reject stars that are highly evolved ($R > 30\rsun$), in order to avoid potential systematics in the asteroseismic scale in the luminous regime \citep{mosser+2013a,stello+2014,kallinger+2018,zinn+2019rad}. Finally, we reject from consideration 12 RGB and 2 RC stars that have asteroseismic and Gaia radius disagreement by more than $3\sigma$, leaving 841 RGB and 214 RC stars for calibration. Since this sample has APOGEE spectroscopic abundances, we also modify the $\fdnu$ for our calibration sample by adjusting the metallicity that goes into computing $\fdnu$ to account for non-solar $\alpha$ abundances according to the \cite{salaris+1993} prescription. 

The \Gaia\ radii are computed following the procedure from \cite{zinn+2017plxtgas}, wherein a bolometric flux, \Gaia\ parallax, and APOGEE effective temperature are combined using the Stefan-Boltzmann law. We use a $K_{\mathrm{s}}$-band bolometric correction \citep{ghb09} to minimize extinction effects, and employ the three-dimensional dust map of \protect\cite{green+2015}, as
implemented in \texttt{mwdust}\footnote{\url{https://github.com/jobovy/mwdust}} \citep{bovy+2016}.

\begin{figure}
    \centering
    \includegraphics[width=0.5\textwidth]{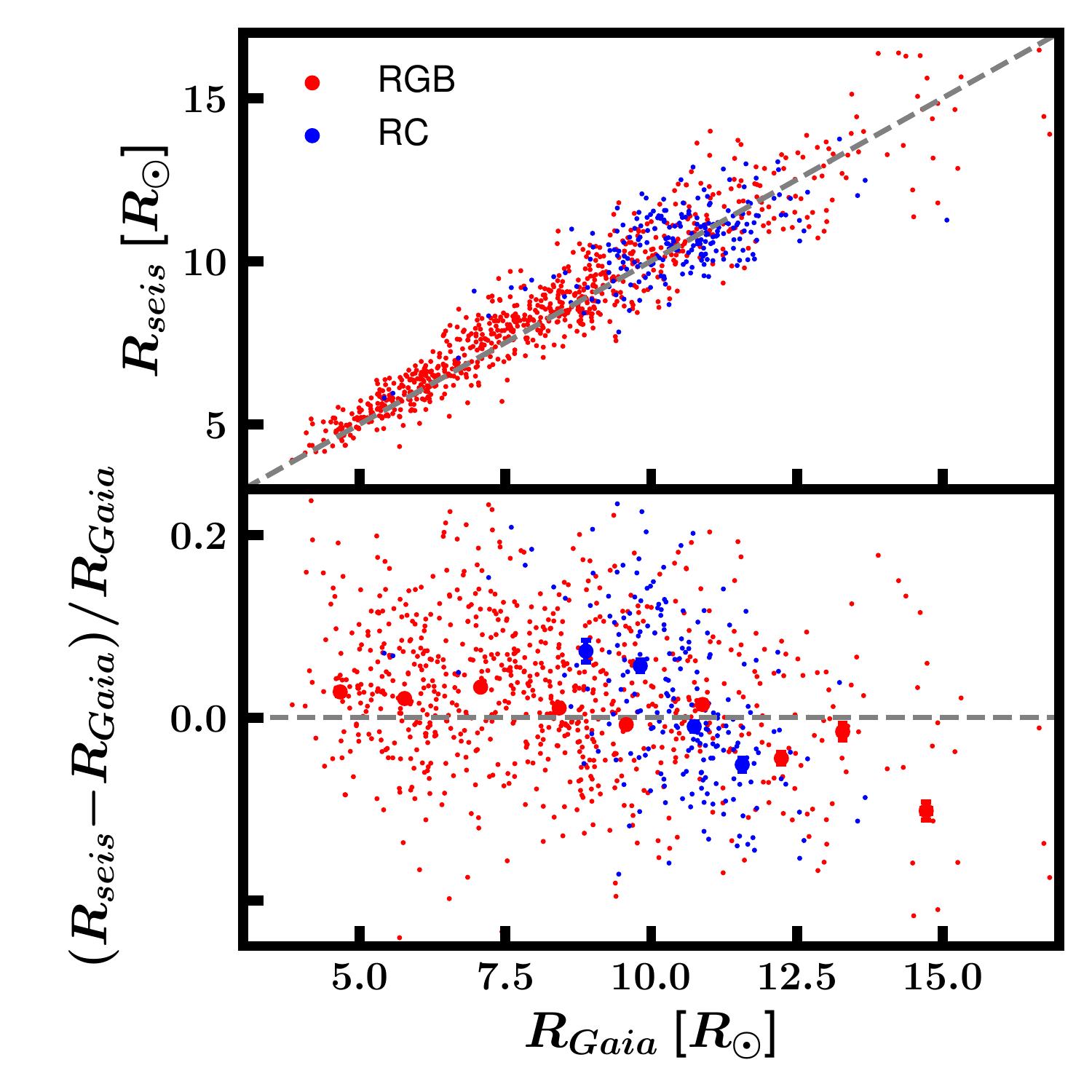}
    \caption{Comparison between \Gaia\ radius and asteroseismic radius for the \Gaia-APOGEE-K2 sample used for the calibration of K2 GAP DR3. Binned medians and uncertainties on the medians of the fractional difference in asteroseismic and \Gaia\ radii are plotted as blue (red) error bars for red clump (red giant branch) stars in the bottom panel, while the radii are plotted versus each other in the top panel. Both panels have a grey dashed line to indicate perfect agreement between the two radii. We re-define $\numaxsun$ for the RGB stars and RC stars separately in K2 GAP DR3 such that their radii agree on average with \Gaia\ radii (see Table~\ref{tab:refs}).}
    \label{fig:radius_scales}
\end{figure}

We see in Figure~\ref{fig:radius_scales} similar trends as we did in K2 GAP DR2 \citep{zinn+2020}: there is an over-estimation in the asteroseismic radii compared to \Gaia\ at and below $R \approx 8\rsun$ among red giant branch stars.

The strong trend in radius agreement for the RC is of astrophysical interest, particularly given constraints on mass-loss \citep[e.g.,][]{miglio+2012,kallinger+2018} that rely on the accuracy of asteroseismic scaling relations for the RC. However, as we noted in \citealt{zinn+2020}, the trend seems to be mostly a function of $\dnu$, and therefore may be related to inadequacies in the red clump stellar structure models that underpin theoretical $\fdnu$ calculations \citep{an+2019a}. It is beyond the scope of the present work to further examine the cause of the discrepancy, but developments in better understanding this behavior in the RC are in preparation.

We calibrate our K2 GAP DR3 asteroseismic values to be on the \Gaia\ parallactic scale by adopting the following:
\begin{equation*}
\fnumax \equiv \langle R_{\rm{seis}}/R_{Gaia} \rangle = \frac{\sum \frac{R_{\rm{seis}}}{R_{Gaia}} / \sigma^2_R}{\sum{1/\sigma^2_R}},
\end{equation*}
 where $\sigma_R = \frac{R_{\rm{seis}}}{R_{Gaia}} \sqrt{ \left( \frac{\sigma_{R, Gaia}}{R_{Gaia}} \right )^2 + \left( \frac{\sigma_{R,\rm{seis}}}{R_{\rm{seis}}}\right)^2}$. We do this separately for RGB and RC stars, finding $f_{\nu_{\mathrm{max}}\rm{,RGB}} = 1.017 \pm 0.001$ and $f_{\nu_{\mathrm{max}}\rm{,RC}} = 1.008 \pm 0.003$. This can be thought of as a re-scaling of the solar reference value, $\numaxsun$, though for convenience, we apply this correction directly to the $\numaxmean$, $\kapparmean$, and $\kappammean$ values provided in Table~\ref{tab:derived}, and thus when working with $\numaxmean$, the K2 GAP DR3 $\numaxsun$ value given in Table~\ref{tab:solar_refs} should be used, which is the same as that from \citep{pinsonneault+2018}. Even after accounting for this $\fnumax$, the uncertainty in the $\fnumax$ becomes a systematic uncertainty in the $\kapparmean$, and $\kappammean$ scales, viz., $0.1\%$ \& $0.3\%$ in $\kapparmean$ \& $\kappammean$ for RGB stars and $0.3\%$ \& $0.9\%$ for RC stars. Note that it is possible there is a scalar correction required of $\dnumean$, as well. We therefore conservatively treat the uncertainty in $\fnumax$ as an uncertainty in a scalar contribution to $\fdnu^2$, given that our calibration of the asteroseismic radius, which scales as $\fnumax^{-1} \fdnu^{2}$ (Eq.~\ref{eq:radius}), is formally a calibration of the quantity $\fnumax^{-1} \fdnu^{2}$. This implies a systematic uncertainty in $\dnumean$ of $0.05\%$ and $0.15\%$ for RGB stars and RC stars, respectively. As discussed in \cite{zinn+2019rad}, there are additional systematics in the asteroseismology-\Gaia\ radius comparison that could amount to about $\pm2\%$ in $\fnumax$, and which are due to intrinsic uncertainties in the bolometric correction scale, the temperature scale, and the spatial correlations in \Gaia\ parallaxes.

On balance, the modest corrections required to bring the asteroseismic data onto the \Gaia\ parallactic scale support previous findings that the asteroseismic scaling relations are accurate to within a few to several percent on the lower giant branch \citep[e.g.,][]{silva_aguirre+2012,huber+2012,hall+2019a,khan+2019,zinn+2019rad}. With the assurance that the K2 GAP DR3 asteroseismic masses are well-calibrated, we turn to applications of those data to age-abundance patterns.

\section{Age-abundance patterns in K2 GAP DR3}
\label{sec:ageabundance}
\subsection{Notes on GALAH abundances}
Our examination of age-abundance patterns makes use of GALAH (GALactic Archaeology with HERMES) DR3 \citep{buder+2021} abundances for stars targeted as part of the K2-HERMES
\citep{wittenmyer+2018} program. Although our asteroseismic calibration uses APOGEE temperatures and metallicities for deriving asteroseismic radii (\S\ref{sec:abs}), we note that calibration using GALAH spectroscopic parameters instead results in an equivalent $\fnumax$ to within uncertainties. We opt to use GALAH abundances in what follows because 1) there are neutron-capture element lines in the optical unavailable to APOGEE's infrared bandpass, and 2) GALAH abundances are corrected for non-LTE effects for the elements H, Li, C, O, Na, Mg, Al, Si, K, Ca, Mn, Fe, and Ba. On the latter point, non-LTE spectral analysis seems especially important for bringing into agreement dwarf and giant abundances at fixed metallicitiy within $\sim 0.05$ dex \citep{amarsi+2020}, though some systematics at the 0.1-0.2dex level may remain for Al, Ba, and $\alpha$-elements, which are mentioned below.

We note that APOGEE DR16 temperatures and GALAH DR3 temperatures for RGB stars differ by $\approx 30K$, in the sense that APOGEE temperatures tend to be hotter. This difference is at the same level as the intrinsic uncertainty in the APOGEE temperature scale, which is set by the accuracy of the infrared flux method (IRFM) temperature scale for red giants \citep[e.g.,][]{alonso+1999a,ghb09}. The metallicity scales of the two systems differ by $\approx -0.05$dex, in the sense that APOGEE is more metal-rich. The combined effect of these small offsets means that the asteroseismic parameter calibration performed with APOGEE temperatures in \S\ref{sec:calibratoin} is consistent to within systematic uncertainties of $\fnumax$ and thus the calibrated parameters are suitable for the following analysis using GALAH temperatures.
 
It should also be noted that scattering on background opacities was not included in the GALAH DR3 non-LTE calculations. Background scattering may affect giant abundances at the 0.01dex level for elements other than C, Mg, Ca, and Mn, which can have larger effects due to background scattering at lower metallicities \citep[e.g.,][]{hayek+2011}. Among metal-poor giants, Mg, Ca, and Mn may thus be under-estimated by up to 0.05 dex for stars with $\feh < -2$ \citep{amarsi+2020}.

\subsection{Benchmark Galactic chemical evolution model}
We compare our age-abundance patterns to the fiducial abundance models of \citet[K20]{kobayashi_karakas_lugaro2020}. The models use nucleosynthetic yields from CCSNe, SNe Ia, AGB stars, and neutron stars mergers, which are discussed as relevant in the discussion that follows. The K20 models assume a one-zone enrichment model, wherein mixing of the interstellar medium is instantaneous, and there is pristine gas inflow. The infall rate and star formation efficiency are chosen to match the metallicity distribution function of the solar neighborhood. For the solar neighborhood model considered here, there is assumed to be no gas outflow.
K20 assume single-degenerate SNe Ia, where the total number of SNe is
determined from the O/Fe slope. The fraction of main sequence+white
dwarf to RGB+white dwarf progenitors is fit to reproduce the observed
Galactic metallicity distribution function (see, e.g., Figure A2 of \citealt{kobayashi_leung_nomoto2020}). We note that the \feh\ at which SNe Ia begin to go off in the K20 models is not simply determined by the delay-time distribution of SNe Ia, but rather the metallicity-dependence of Fe production in SNe Ia \citep{kobayashi+1998}. This is because the K20 models' SNe Ia single-degenerate scenario assumes that white dwarfs surpass the Chandrasekhar mass limit by metallicity-dependent white dwarf winds that prevent common envelope production and encourage stable mass transfer \citep[see][]{hachisu_kato_nomoto1996,kobayashi+1998,kobayashi_nomoto2009}.

The K20 models we use are representative of the solar neighborhood,
and so we restrict our analysis to K2 GAP DR3 stars with Galactocentric distances between 7 and 9kpc. 

Because the K20 models are calibrated to observations purely in abundance space and not with reference to stellar age measurements, comparing the observed K2 GAP DR3 age-abundance patterns to the models provides an independent check on the success of the assumed global (star formation rate, infall rate) and local (nucleosynthetic yields) model choices. Of particular interest in what follows are the implications for the nucleosynthetic site of production and yields. For comparisons of GALAH DR3 abundance ratios to K20 models, we refer the reader to \cite{amarsi+2020}. 

\subsection{Ages}
We derive ages from K2 GAP DR3 asteroseismic masses computed according to Equation~\ref{eq:scaling2} with $\kappammean$ and GALAH DR3 temperatures. The age inference is performed in a Bayesian framework  using BSTEP \citep{sharma+2018}, a Bayesian stellar parameter estimator that may incorporate asteroseismic parameters, $\numax$ and $\dnu$, which essentially constrain the mass of the star and therefore its main sequence lifetime. Further details regarding the BSTEP ages used in this work are available  in \citet{sharma2021} (see also \citet{buder+2021}). In what follows, we only use stars that BSTEP classifies with high confidence as RGB, given uncertainties on RC ages due to mass loss \citep[e.g.,][]{casagrande+2016}.

\subsection{\mgh\ versus \feh\ space}
We begin by dividing our sample into high- and low-$\alpha$ samples, following the high-low boundary from \citet[W19]{weinberg+2019}:

\begin{align}
\label{eq:mgfe}
\mathrm{for\ \feh} &< 0: \mathrm{[Fe/Mg]} > 0.12 - 0.13\mathrm{[Fe/H]},\\
\mathrm{for\ \feh} &> 0: \mathrm{[Fe/Mg]} > 0.12.
\end{align}

The above division was initially used for stars with APOGEE abundances, though it has subsequently been used successfully to divide GALAH DR2 \citep{buder+2018} abundances into high- and low-$\alpha$ populations by \citet[GJW19]{griffith_johnson_weinberg2019}, who recently interpreted both APOGEE and GALAH DR2 abundance ratios in the context of Galactic chemical evolution. Following the example of GJW19, we also restrict analysis to those stars with effective temperatures between $4500$K and $6200$K, which avoids blending in cool stars from molecular lines and highly broadened lines in fully radiative stars. 

\begin{figure}
    \includegraphics[width=0.4\textwidth]{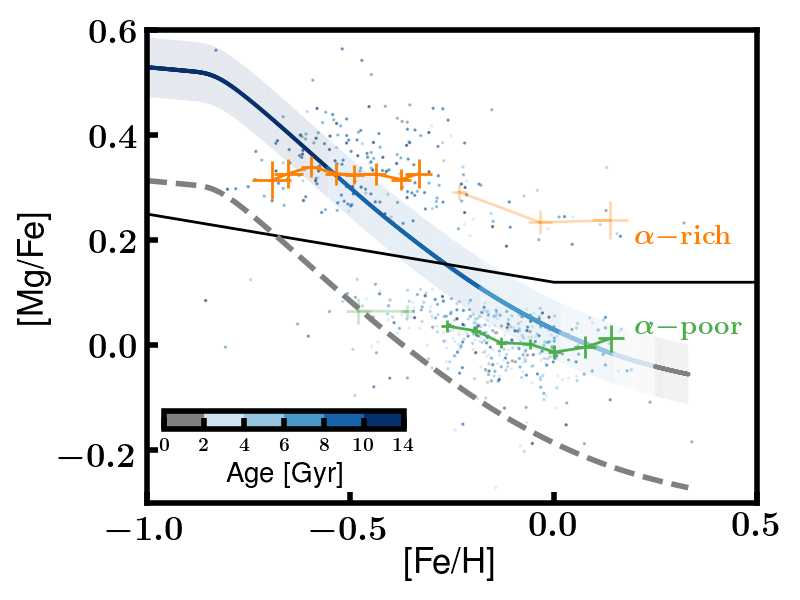}
    \caption{We divide our GALAH DR3 + K2 GAP DR3 sample into high-
      and low-$\alpha$ samples, following the proposed division from
      \citet[Eq.~\ref{eq:mgfe}]{weinberg+2019}. The stars indicated by
      orange points are accordingly classified as high-$\alpha$ stars
      and the stars indicated by green points as low-$\alpha$ stars;
      the green and orange error bars and curves indicate the
      $2\sigma$ uncertainties on the binned weighted mean of the
      low-$\alpha$ and high-$\alpha$ stars. The Galactic chemical
      evolution model from \citet[K20]{kobayashi_karakas_lugaro2020}
      is shown before (grey dotted curve) and after (blue segmented
      curve) an additive correction to [Mg/Fe] to enforce agreement with
      the median \fe{Mg} of the observed stars. The
      region around the blue segmented curve reflects the $1\sigma$ uncertainty in the K20 abundance normalization, taken to be the uncertainties in the solar abundances from \cite{asplund+2009}. The transparency of the binned weighted means of the data emphasizes where the K20 models track the high-$\alpha$ stars ($\feh \lesssim -0.3$ and $\tau \gtrsim 8 \mathrm{Gyr}$) and the low-$\alpha$ stars ($\feh \gtrsim -0.3$ and $\tau \lesssim 8 \mathrm{Gyr}$).} 
    \label{fig:alpha_fe}
\end{figure}

We believe there is some contamination from genuinely $\alpha$-poor stars
that, by virtue of their abundance uncertainties, scatter into the
high-$\alpha$ selection (and vice versa). For this reason, we require
that each star’s 2D uncertainty ellipse have more than 95\% of its
density on one side or the other of the high-/low-$\alpha$ division
line. In order to construct the 2D uncertainty ellipse, we assume a
uniform correlation between [Fe/H] and [Mg/Fe]. The Pearson
correlation coefficient between [Fe/H] and [Mg/Fe] is observed to be
$\sim -0.4$, though the precise value adopted does not significantly
affect our results. We also require stars to have [Fe/H] $> -1$ at
95\% confidence, since the metal-poor stellar population is likely
populated by accretion \citep[e.g.,][]{belokurov+2018,haywood+2018}
rather than in situ formation, as the K20 models assume.  The
resulting division of the GALAH abundances is demonstrated in
Figure~\ref{fig:alpha_fe}, where each star is colored by its age. The
high-/low-$\alpha$ division line is shown in black. The grey curve represents the raw K20
[Fe/H]-[Mg/Fe] trend, which has been shifted by a scalar offset
in [Fe/H] and a scalar offset in [Mg/Fe] to reflect the same solar abundance scale used by GALAH DR3
\citep[see Table A2 of][]{buder+2021}.\footnote{Where possible, we
  adopt the `composite' abundance normalizations listed in Table A2 of
  \cite{buder+2021} and, otherwise, the average of a given elements'
  line-by-line normalizations.} The segmented blue curve represents
the K20 [Fe/H]-[Mg/Fe] trend, re-scaled by an additive offset in Mg such that the median predicted
[Mg/Fe] agrees with the median observed [Mg/Fe]. The band around the curve corresponds to a $1\sigma$ uncertainty in the \cite{asplund+2009} solar abundances, which are used in the K20 models for abundance normalization.\footnote{The exception is O, whose solar abundance is taken to be $A_{\odot}(O)=8.76 \pm 0.02$ \citep{steffen+2015}.}

The sample consists of 396 high-$\alpha$ stars and 
208 low-$\alpha$ stars, with typical uncertainties of 20-30\% in age.\footnote{These are the number of stars with Mg and Fe measurements, which are necessary to define the high-$\alpha$ and low-$\alpha$ stars. Note that not all of these stars have abundance measurements for every element we consider in what follows.} The ages for this sample, as well as their GALAH spectroscopic information, and high-/low-$\alpha$ classification are provided in Table~\ref{tab:age}.

We follow the example of W19 and GJW19 in considering abundance ratio and age-abundance patterns in [X/Mg] space instead of only [X/Fe] space due to the expectation that nearly all Mg production occurs in CCSNe, whereas Fe is produced by both CCSNe and SNe Ia. Therefore, the low-$\alpha$ population can be interpreted as having SNe Ia contributions, and the high-$\alpha$ population as enriched by CCSNe. Normalizing by Mg means that elements produced only by CCSNe have the same trends in both the low- and high-$\alpha$ regimes. Elements with contributions from SNe Ia, however, will show a separation in [X/Mg] v. [Mg/H] that depends on the relative contribution of CCSNe and SNe Ia. We note that other enrichment channels like AGB winds that do not behave precisely like CCSNe or SNe Ia in their Mg production and delay-time distribution may complicate interpretations of the \mg{X} versus \mgh\ trends. Ultimately, showing age-abundance patterns and abundance ratios in \mgh\ space in addition to \feh\ space can offer complementary information to the asteroseismic age information. For instance, CCSNe elements would be expected to 1) have constant \mg{X} as a function of stellar age; 2) decreasing \fe{X} as a function of stellar age; and 3) show similar \mg{X} trends as a function of \mgh\ for both high- and low-$\alpha$ populations.

\begin{figure}
    \includegraphics[width=0.4\textwidth]{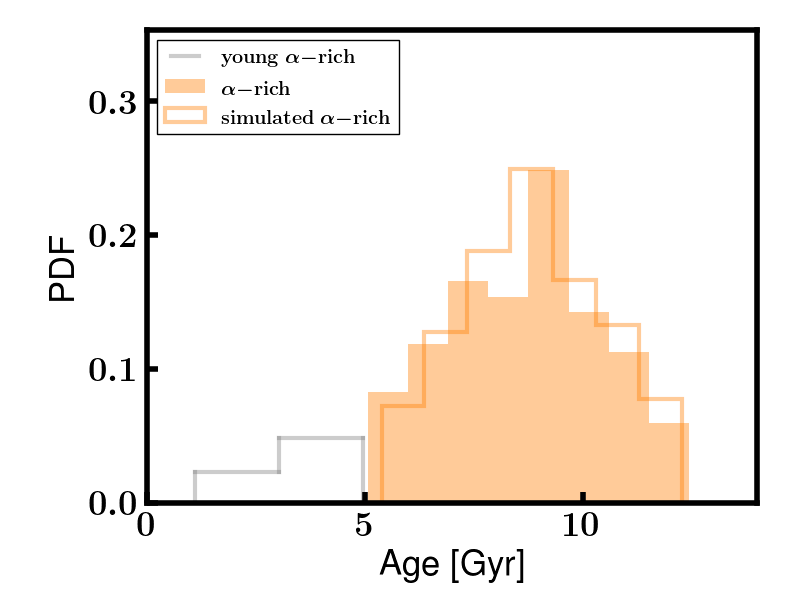}
    \caption{The distribution of high-$\alpha$ ages (filled orange) is consistent with being drawn from a uniform age of 9 Gyr (as simulated in the orange lines). A separate population of young high-$\alpha$ stars with ages $\lesssim 5$ Gyr is shown in grey, and is consistent with previous identifications of a young high-$\alpha$ population in the literature.} 
    \label{fig:hi_ages}
\end{figure}

 We show the distribution of \highalpha\ ages in
 Figure~\ref{fig:hi_ages}. The filled orange histogram shows the
 distribution of high-$\alpha$ ages larger than 5 Gyr. The orange line
 indicates the distribution of a simulated population of these stars
 assuming a mean age of 9 Gyr and with uncertainties taken from the fractional
 uncertainties of the data. The grey line indicates a
 population consistent with being `young' high-$\alpha$ stars, which
 were originally identified in CoRoT \citep{baglin+2006} \& APOGEE
 data \citep{chiappini+2015}, and have since been seen in Kepler and
 K2 data
 \citep[e.g.,][]{martig+2015,silva-aguirre+2018a,warfield+2021}. These
 stars may be genuinely old and appear young due to having gained mass
 through stellar mergers \citep[e.g.,][]{martig+2015,jofre+2016,izzard+2018,sun_huang+2020},
 or perhaps are genuinely young and have formed in gas relatively
 unenriched by SNe Ia \citep{chiappini+2015,johnson+2021}. We therefore draw a
 distinction between this population and the rest of the high-$\alpha$
 population, which are consistent with having a uniform age
 of $\sim 9$ Gyr according to a Kolmogorov-Smirnov
 test.

\begin{figure*}
    \includegraphics[width=0.95\textwidth]{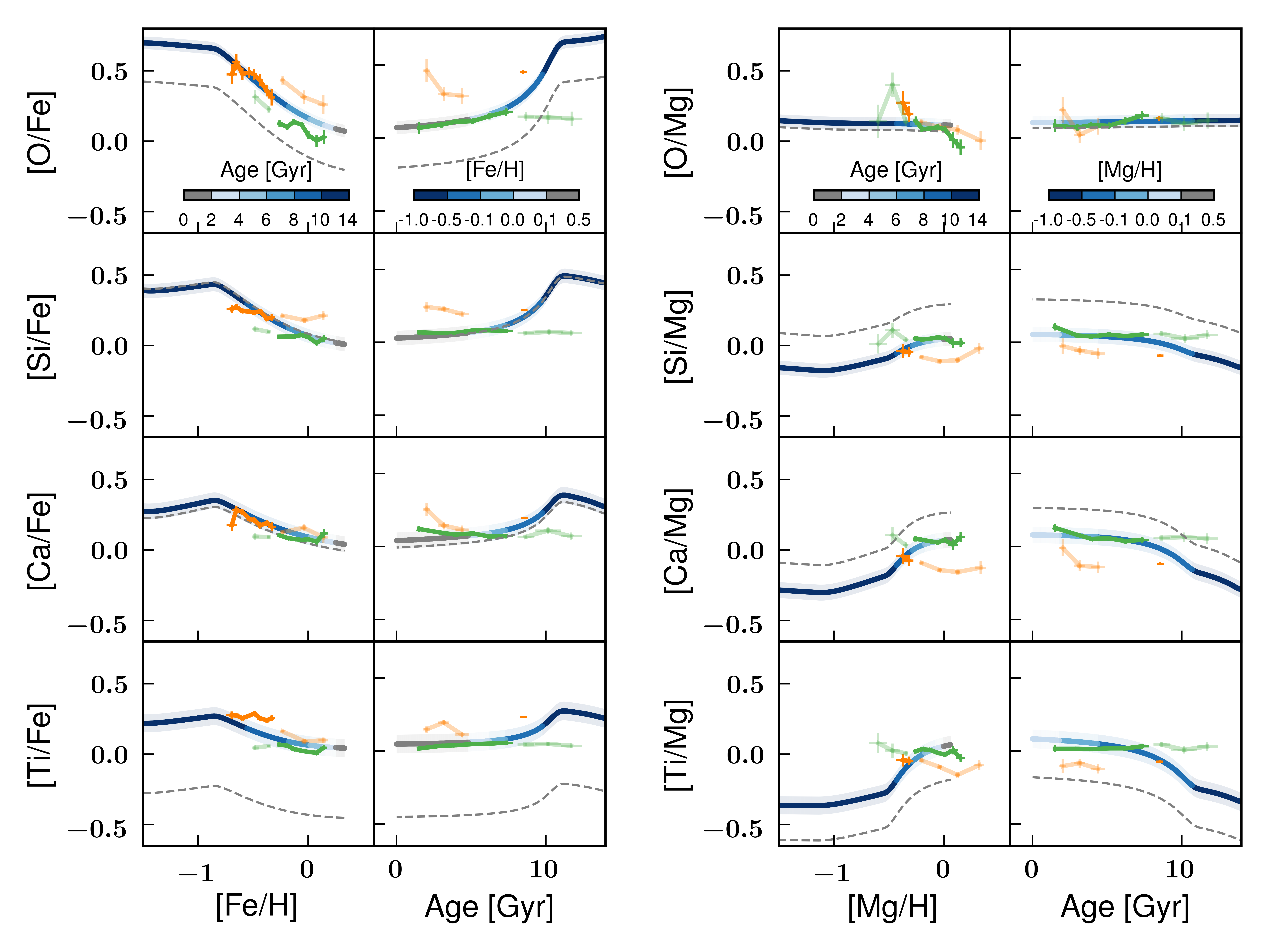}
    \caption{Abundance ratios and age-abundance patterns for $\alpha$
      elements, shown in both \fe{X} and \mg{X} space. The green and
      orange error bars and curves indicate the $2\sigma$
      uncertainties in the binned weighted mean of the low-$\alpha$
      and high-$\alpha$ stars. The blue segmented curves show the
      predictions from the Galactic chemical evolution model of
      \cite[K20]{kobayashi_karakas_lugaro2020}, displaced additively in X
      to agree with the median \fe{X} or \mg{X} of the observed stars (grey
      dashed curves indicate the models before this re-scaling) and
      colored by either \feh, [Mg/H], or age, according to the color bars. The transparency in the weighted means of the data emphasizes where the K20 models track the high-$\alpha$ stars ($\feh \lesssim -0.3$ and $\tau \gtrsim 5 \mathrm{Gyr}$) and the low-$\alpha$ stars ($\feh \gtrsim -0.3$ and $\tau \lesssim 8 \mathrm{Gyr}$). We note that not every star has GALAH measurements for every element, particularly at low metallicities.} 
    \label{fig:alpha}
\end{figure*}

\begin{figure*}
    \includegraphics[width=0.95\textwidth]{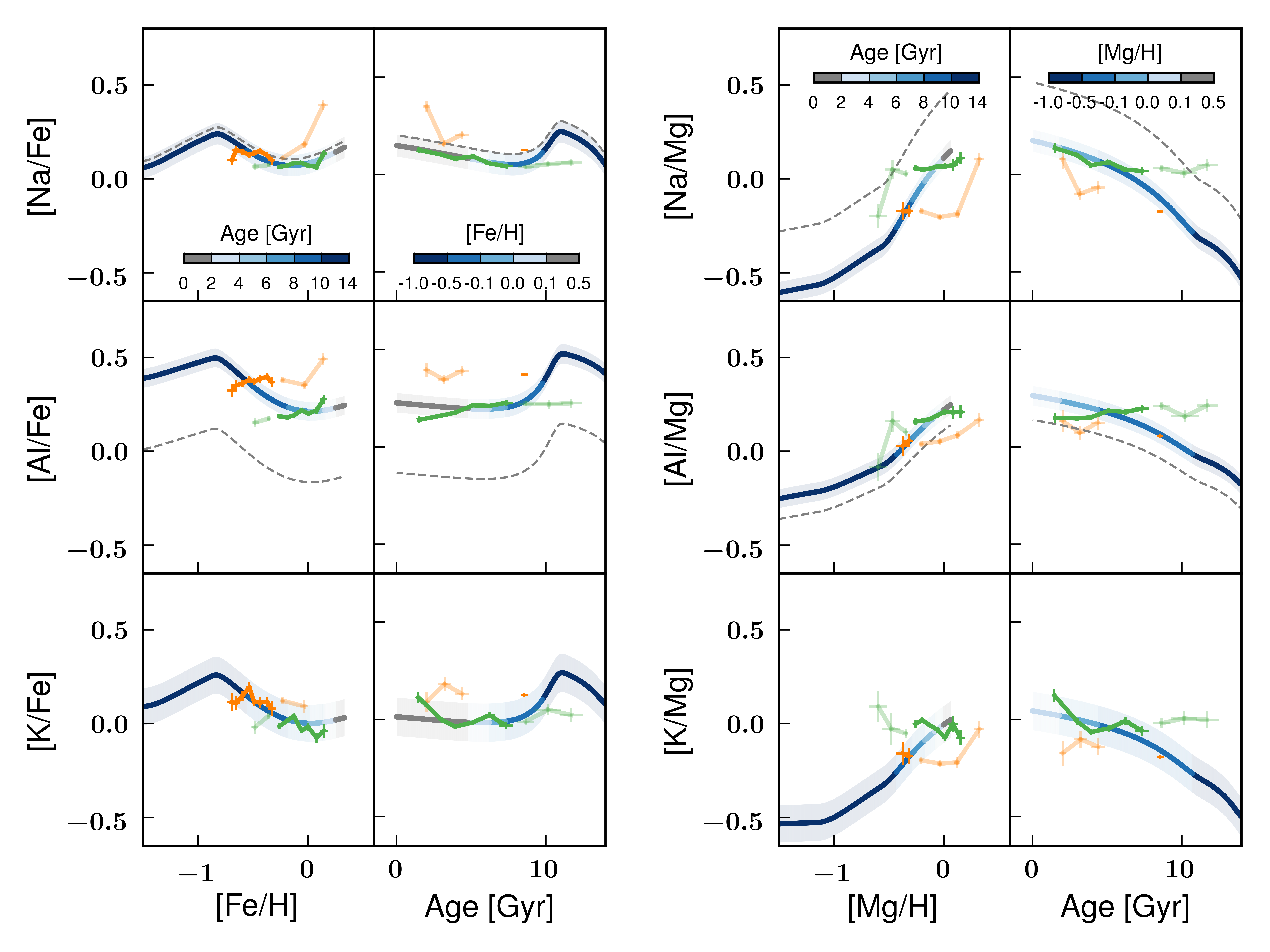}
    \caption{Same as Figure~\ref{fig:alpha}, but for light, odd-Z elements.} 
    \label{fig:lightodd}
\end{figure*}

We show in Figures~\ref{fig:alpha}-\ref{fig:mainr} abundance ratios
(\fe{X} versus \feh\ or \mg{X} versus \mgh) and age-abundance
patterns/enrichment histories
(\fe{X} versus stellar age or \mg{X} versus stellar age) for different
nucleosynthetic families of elements. A running weighted average of
the data are shown as colored error bars connected by lines, with
green indicating the low-$\alpha$ population and orange indicating the
high-$\alpha$ population. Not plotted are stars with flagged GALAH DR3
abundance measurements in \feh\ or in \fe{X}. As mentioned above, the
extent to which the low-$\alpha$ (green curves) pattern is above the
high-$\alpha$ (orange curves) pattern in \mg{X}-\mgh\ space is
generally indicative of nucleosynthetic production site.

Regarding how to compare the K20 models with the data in these
figures, we note that at young and intermediate ages, the K20 models may be best interpreted as
 a low-$\alpha$ population, while the models represent a high-$\alpha$
 population at old ages. Because K20 models are one-zone models, there
 is a one-to-one mapping of age to abundance, which is not necessarily
 the case in the data. To guide the eye, we therefore highlight in
 Figure~\ref{fig:alpha_fe} and subsequent figures where the data should
 be compared to the models: bold curves indicate solidly
 old/metal-poor high-$\alpha$ stars or young/metal-rich low-$\alpha$
 stars, which are comparable to the K20 modelled populations shown in blue, whereas light curves
 indicate apparent young high-$\alpha$ or old low-$\alpha$ populations
 not directly comparable to the K20 models. To evaluate the agreement of the models with the
 data for older, \highalpha\ stars, we make reference here and in what
 follows to a single weighted average of the high-$\alpha$ abundances (which can be seen as a
 single orange error bar in the following figures), since the width of the
 \highalpha\ age distribution is dominated by uncertainties, and
 has a central value of $\approx 9$\,Gyr.


\subsection{$\alpha$ elements: O, Mg, Si, Ca, Ti}
Looking at O in Figure~\ref{fig:alpha}, it is clear that, after a global correction, the
observed abundance ratios for $\feh > -1$ are in excellent agreement
with K20 model predictions. That the metallicity-dependence of O
enrichment agrees with observations is a built-in feature of the
models: K20 models are adjusted by tuning the total number of
supernovae to agree with the observed literature O abundance
metallicity dependence (K20). With age information in hand, however,
we can independently test the models. We see that the agreement is
good when looking at the low-$\alpha$ \mg{O} trend as a function of
time up to $\tau \sim 8$Gyr, tracking Mg production, as an $\alpha$
element would.  We see that the high-$\alpha$ \fe{O} enrichment
history is in tension with model predictions at 9Gyr (orange error bar
versus blue curve). Given the agreement of the high-$\alpha$
population \fe{O} as a function of \feh, the disagreement of \fe{O}
for the high-$\alpha$ population in age space suggests an offset in
the observed and predicted high-$\alpha$ ages. A natural solution
would be to appeal to $\alpha$-enhanced stellar model
opacities. Indeed, \cite{warfield+2021} have demonstrated the increase
in stellar opacities due to non-solar $\alpha$ abundances can increase
low-mass (old) stellar ages by $\approx 10\%$ by decreasing core
temperature and extending a red giant's main sequence lifetime. For
the majority of elements considered in what follows of \S\ref{sec:ageabundance}, an increase in the high-$\alpha$ ages of that magnitude would improve agreement between the data and models. The global offset required to match the O abundances at high metallicities (blue curves versus grey dashed curves), could be due to GALAH $\alpha$-element abundances O, Mg, and Si having residual offsets of 0.1dex, in the sense that giants have larger [$\alpha$/Fe] compared to dwarfs even after non-LTE corrections \citep{amarsi+2020}. 

Both Ca and Si in Figure~\ref{fig:alpha} show good agreement between the predicted and observed
enrichment history: the predicted enrichment history at ages $\tau
\lesssim 8$Gyr tracks the observed trend (green curve) in \mgh\  and
\feh\ space. The models also predict a \fe{Si} at 9Gyr consistent with
the observed abundances of the old high-$\alpha$ population (orange
error bar). 

We consider Ti an $\alpha$ element, based on the findings in GJW19 that its production seems to be dominated by CCSNe contributions. Indeed, both the low-$\alpha$ and high-$\alpha$ curves share a similar \mg{Ti} in Figure~\ref{fig:alpha}. At older ages, however, the observed high-$\alpha$ \fe{Ti} abundances are in tension with model predictions for 9Gyr, which could be improved via older ages from aforementioned $\alpha$-enhanced stellar model opacities. Note that there is a large zero-point offset between the raw model abundances and the observed abundances (the offset to bring the raw model abundances into agreement with observations is the difference between the grey dashed curves and the blue segmented curves), which is a generic feature of nucleosynthetic Ti yield predictions, and may be remedied by 2- or 3-dimensional supernovae models (K20).

\subsection{Light odd-Z elements: Na, Al, K}
\label{sec:oddz}
Odd-Z element production is thought to depend on progenitor metallicity because their assumed production during explosive nucleosynthesis in CCSNe depends crucially on the neutron excess prior to the supernova, which itself is dependent on CNO cycle efficiency and therefore intial metal content \citep[e.g.,][]{truran_arnett1971}. The prediction of nucleosynthetic models for these elements, therefore is that 1) they should follow a CCSNe enrichment history (either a decreasing \fe{X} with younger stellar ages or, equivalently, constant \mg{X} with stellar age), and 2) they should be less abundant with decreasing metallicity. In Figure~\ref{fig:lightodd}, we show the light, odd-Z elements' abundance ratios and age-abundance patterns to test these predictions.

\begin{figure*}
    \includegraphics[width=0.95\textwidth]{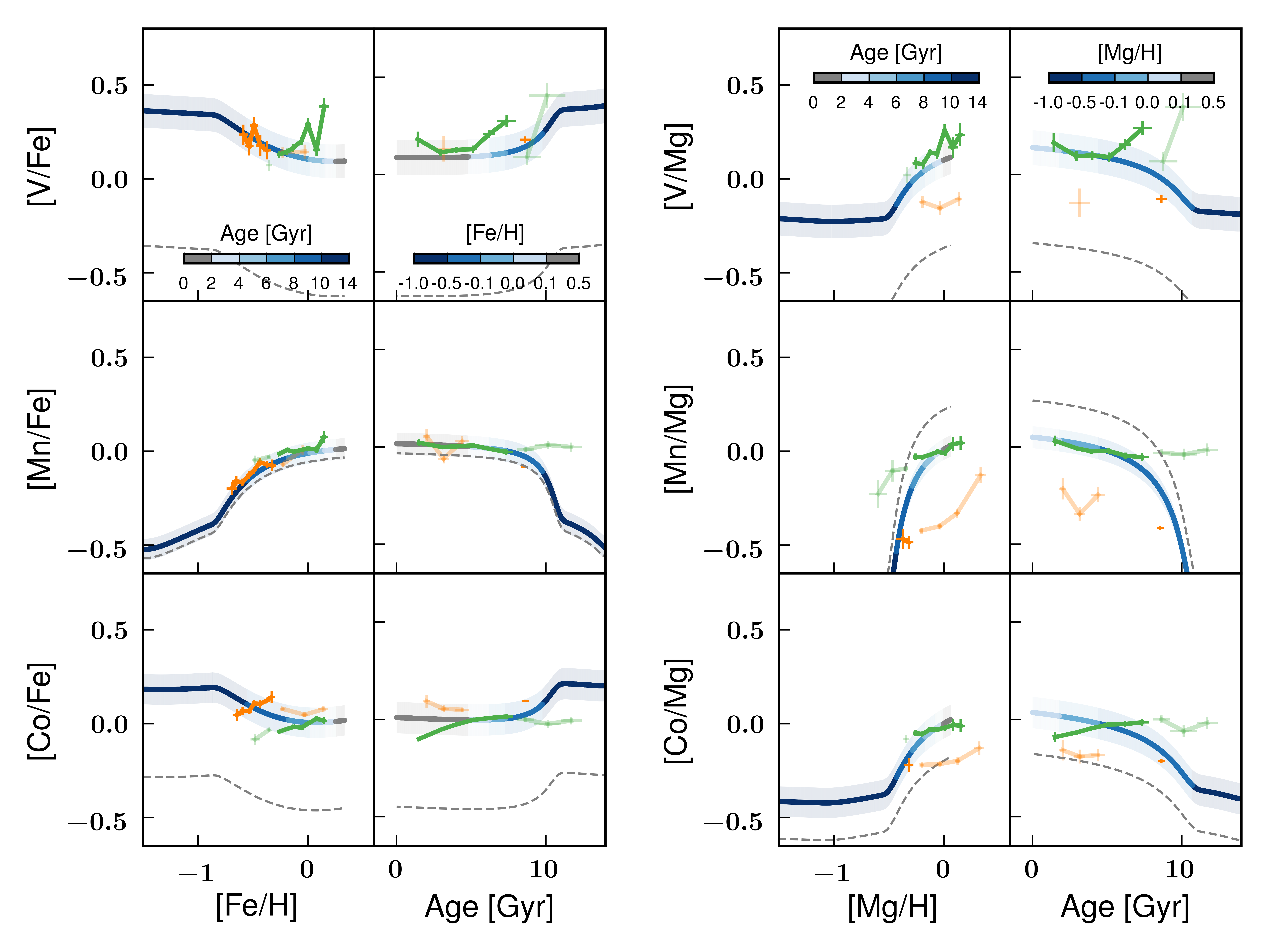}
    \caption{Same as Figure~\ref{fig:alpha}, but for odd-Z iron-peak elements.} 
    \label{fig:peakodd}
\end{figure*}

The DR3 GALAH \mg{Na} abundance ratios show a positive metallicity trend, consistent with findings from GJW19 using GALAH DR2, and broadly consistent with the predicted metallicity slope from K20 models. The enrichment history predictions appear to be consistent with observations, across all ages probed (keeping in mind the lack of resolution in age space for the \highalpha\ stars, which, to within uncertainties, are drawn from a single age of $\approx 9$Gyr).

The strong, negative metallicity gradient seen by GJW19 in \mg{K} is less pronounced with non-LTE corrections in GALAH DR3, and is in good agreement with K20 models in \feh\ space. The absolute abundances from K20 for K, however, are well below the observed value (the grey dashed curve is below the plotted region), and this offset may be alleviated by appealing to, e.g., rotating stellar models (K20 and references therein). The predicted abundances at old stellar age are consistent with those observed among high-$\alpha$ at 9Gyr.

The non-LTE GALAH DR3 corrections to Al reveal a strong metallicity trend with \mg{Al} not found in GALAH DR2 abundances, but corroborating the positive trend found in APOGEE abundances (GJW19). We confirm GJW19's interpretation of Al being produced largely in CCSNe production, given the relatively small separation between high- and low-$\alpha$ tracks (orange and green curves in \mgh\ space) compared to, e.g., Na. These observations are both consistent with theoretical predictions of significant, metallicity-dependent Al production during explosive C burning \citep{truran_arnett1971}. The observed and predicted enrichment histories are in disagreement. As with O, older high-$\alpha$ ages due to $\alpha$-enhanced stellar model opacities could improve agreement at old ages. This adjustment would also bring Na and K into even better agreement at old ages. The absolute yields are, as with K, severely under-predicted. This may very well be due to an over-prediction of the abundances on the observational side: even after non-LTE corrections, Al abundances for giants are larger than the dwarf abundances by 0.2dex \citep{amarsi+2020}.

\begin{figure*}
    \includegraphics[width=0.95\textwidth]{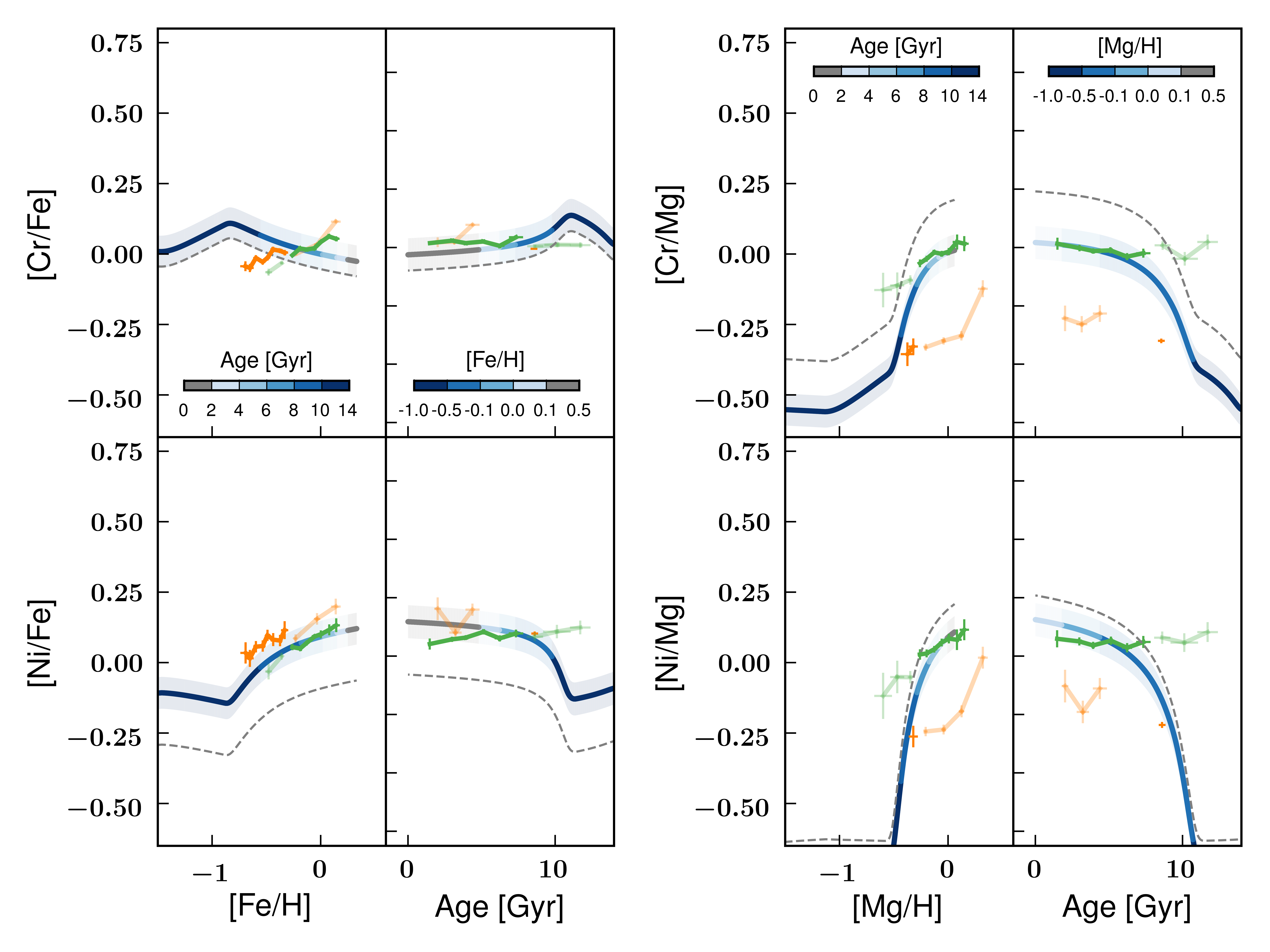}
    \caption{Same as Figure~\ref{fig:alpha}, but for even-Z iron-peak elements.} 
    \label{fig:peakeven}
\end{figure*}

\begin{figure*}
    \includegraphics[width=0.95\textwidth]{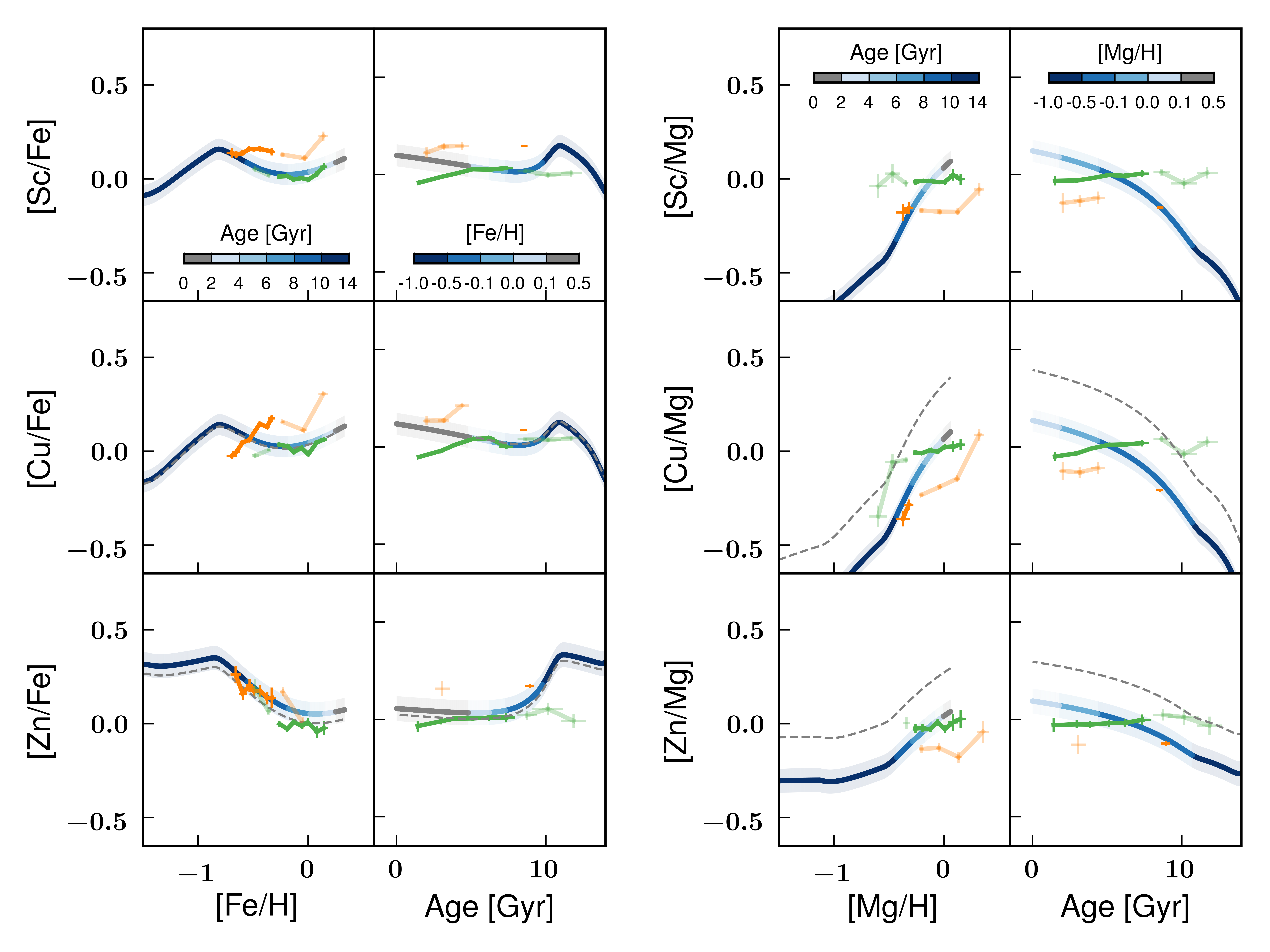}
    \caption{Same as Figure~\ref{fig:alpha}, but for iron-peak cliff elements.} 
    \label{fig:peakcliff}
\end{figure*}

\begin{figure*}
    \includegraphics[width=0.95\textwidth]{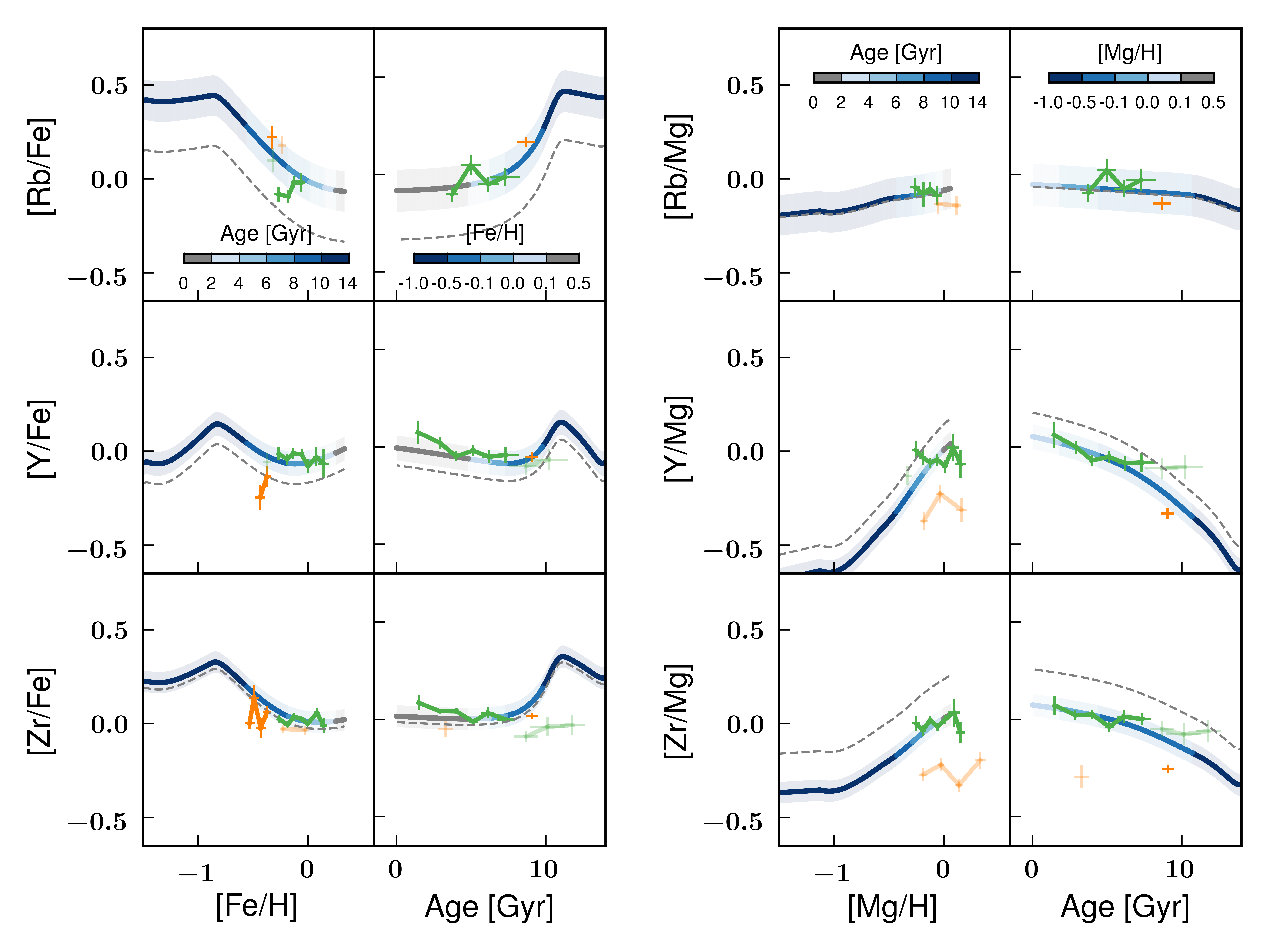}
    \caption{Same as Figure~\ref{fig:alpha}, but for weak s-process elements.} 
    \label{fig:firsts}
\end{figure*}

\subsection{Iron-peak elements}
Following GJW19's typology of iron-peak elements, we separate the elements just beyond iron as cliff elements, which seem to have distinct properties from other iron-peak elements. First, we consider the odd-Z iron-peak elements, then the even-Z elements, and, finally, iron-peak cliff elements.

\subsubsection{Odd-Z iron-peak elements: V, Mn, Co}
In this section, we discuss the odd-Z iron-peak element abundance
patterns and enrichment histories, as shown in
Figure~\ref{fig:peakodd}. First, we confirm with GALAH DR3 the metallicity trends at high metallicities in V and Mn abundances noted by W19 and GJW19 using APOGEE and GALAH DR2 abundances, respectively. This metallicity-dependent effect is most pronounced in Mn, and is in excellent agreement with the model predictions for the trend, which is the result of Mn production in deflagrations with the single-degenerate scenario \citep{kobayashi_leung_nomoto2020}. That the non-LTE Mn abundances from GALAH DR3 still show a metallicity dependence is in contrast to the decrease in the metallicity dependence going from LTE to non-LTE found in \cite{battistini_bensby2016}.

The observed low-$\alpha$ V pattern agrees well with the K20 predicted
\mg{V} enrichment history including at older ages, where the observed
high-$\alpha$ abundance at 9Gyr broadly agree with the predicted abundance. Nevertheless, the model abundances are uniformly vastly under-predicted compared to observations before a re-scaling is applied (grey dashed curves). This under-prediction could be remedied, however, by using yields from multi-dimensional supernovae yield predictions (K20).

The observed and predicted metallicity dependence for Mn are in very
good agreement. K20 models also reproduce well the enrichment history
of \mg{Mn} and \fe{Mn} in the low-$\alpha$ regime. For old,
high-$\alpha$ populations, however, both the \mg{Mn} and \fe{Mn}
enrichment histories could be improved by older high-$\alpha$ ages due
to $\alpha$-enhanced stellar model opacities (thereby shifting the
orange error bar at 9Gyr to older ages in the Mn
enrichment history panels of Fig.~\ref{fig:peakodd}).

The Co enrichment history agrees well in \mg{Co} space at old ages, though K20 predicts a too-fast enrichment in the younger low-$\alpha$ population (slope of blue segmented curve versus slope of green curve). As with V, the models significantly under-predict the global abundances for Co.

\subsubsection{Even-Z iron-peak elements: Cr, Ni}
To better reproduce the \fe{Cr} enrichment history and
\fe{Cr}-\feh\ ratios seen in Figure~\ref{fig:peakeven}, \fe{Cr} could
be made to be produced less overall, such as in the double-degenerate
scenario (the green dotted curve of Fig. 18 in
\citealt{kobayashi_karakas_lugaro2020}). Note, however, that such low
\fe{Cr} results in higher \fe{$\alpha$} and lower \fe{Mn} and \fe{Ni}
than observed. Otherwise, the observed enrichment history is flatter
than predicted in \feh\ space, but is in better agreement with the
models in \mgh\ space. The disagreement between observed and predicted
high-$\alpha$ \mg{Cr} cannot be redressed only with aforementioned
appeals to older high-$\alpha$ ages due to $\alpha$-enhanced stellar
model opacities, which would increase tension in high-$\alpha$
\fe{Cr}. Rather, this would need to be coupled with a significant
decrease in the production of Cr at early times.

Like Cr, the observed age-abundance pattern of Ni in \feh\ space seen
in Figure~\ref{fig:peakeven} is flatter than predicted. Though there
is broad agreement in the \fe{Ni} and \mg{Ni} high-$\alpha$ enrichment
history, it could be improved by an increase (as opposed to a decrease,
as with Cr) in Ni at early times combined with older high-$\alpha$ ages due to $\alpha$-enhanced stellar model opacities, as mentioned earlier. There is also an offset between the raw model abundances (grey dashed curves) and the observed abundances, though the offset is in the opposite direction to that of Cr. Note that the metallicity dependence is in good agreement with model predictions in \fe{Ni}-\feh\ space, in contrast to Cr. 

\subsubsection{Iron-peak cliff elements: Sc, Cu, Zn}
Looking at Figure~\ref{fig:peakcliff}, the observed and predicted \fe{Zn} v. \feh\ and age-abundance trend are in good agreement. The small separation in \mg{Zn} of the low- and high-$\alpha$ sequences corroborate the CCSNe-dominated production assumed in the K20 models and the interpretations of the Zn abundance ratios in GJW19 that Zn is mostly a CCSNe element.

The enrichment history predicted by the K20 models for Cu show strong
increases in both \mg{Cu} and \fe{Cu} for younger stellar age, which
is in disagreement with a slight trend in the other direction among
the low-$\alpha$ population (green curves in Fig.~\ref{fig:peakcliff}) in both \feh\ and \mgh\ space. Slightly higher high-$\alpha$ ages in the data would help reconcile the observed and predicted \fe{Cu}.

The Sc age-abundance patterns in Figure~\ref{fig:peakcliff} show the same behavior as Cu: the models
predict an age-dependent trend that is in the opposite direction to
that of the observed trends in \feh\ and \mgh\ space, and the observed
high-$\alpha$ population is offset in age compared to the models.

Taken together, the Cu and Sc trends are suggestive of different nucleosynthetic histories compared to Zn. The predicted increase in Cu and Sc yields is theoretically expected due to the metallicity dependence of Cu and Sc yields since both elements are odd-Z (see \S\ref{sec:oddz}). Indeed, the data do show this increase in \fe{Cu} and at least a flat trend in \fe{Sc} with \feh\ among the low-$\alpha$ population. The observed age trend (a flat or increasing abundance with increasing age among low-$\alpha$ stars) is therefore not straightforwardly related to metallicity-dependent yields, and is an interesting constraint on production of these elements; a similar enrichment history is also seen in the odd-Z element Al (see \S\ref{sec:oddz}).

\begin{figure*}
    \includegraphics[width=0.95\textwidth,trim=0 160 0 0, clip]{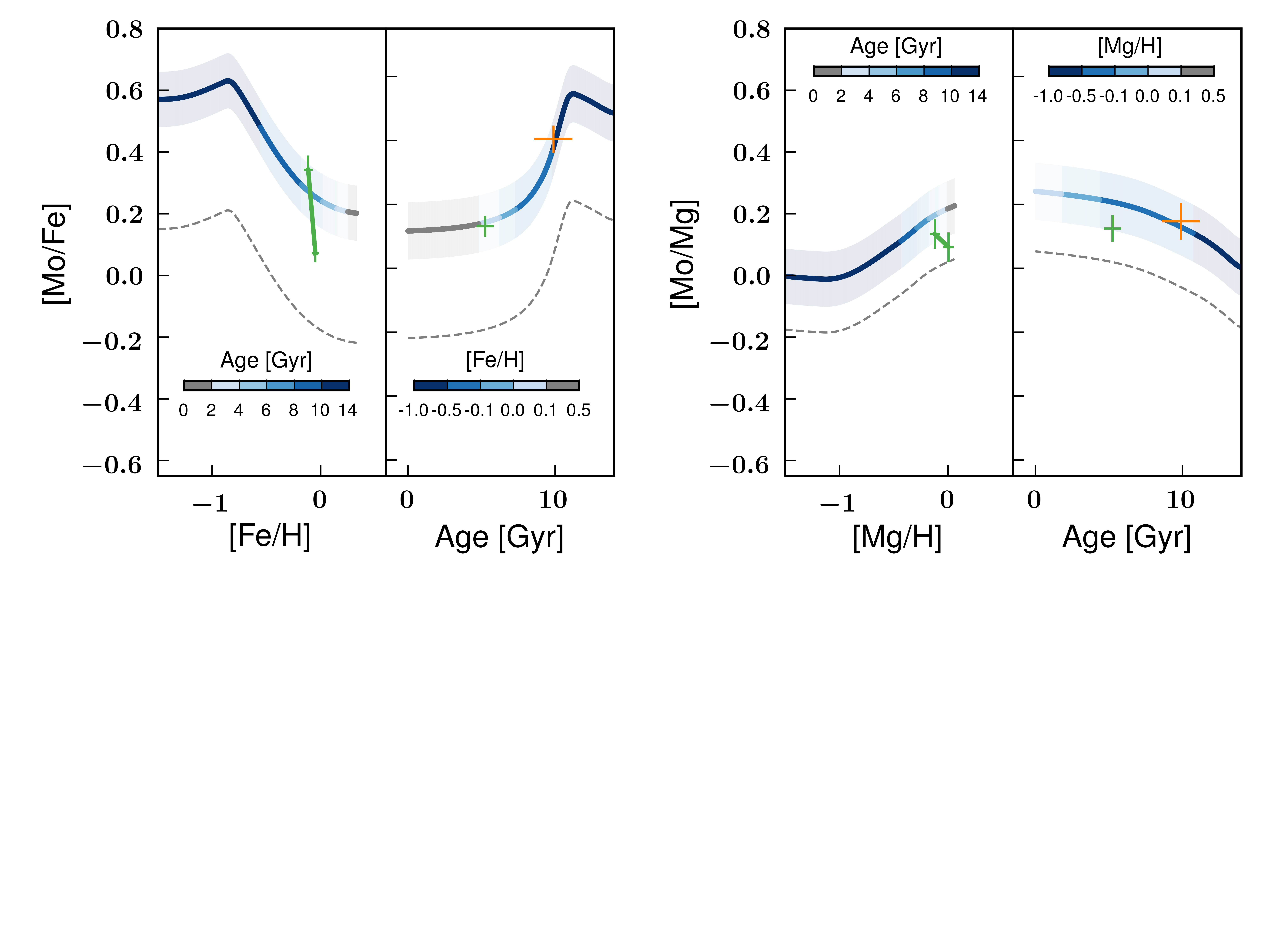}
    \caption{Same as Figure~\ref{fig:alpha}, but for the weak r-process element, Mo.}
    \label{fig:weakr}
\end{figure*}

\subsection{Neutron-capture elements}
Neutron-capture elements can be produced in one of two primary channels: s-process and r-process, which occur in neutron-poor and neutron-rich environments \citep[for a review, see][]{truran+2002}.

There is evidence of two different kinds of r-process production: a `weak' process that creates elements $A \lesssim 130-140$ \citep[e.g.,][]{honda+2004} and the main r-process for elements with $A \gtrsim 130-140$ \citep{truran+2002}. The main r-process production site has been proposed to be decompressing neutron-rich ejecta from a neutron star--neutron star (NS-NS) merger \citep{lattimer_schramm1974,lattimer+1977,rosswog+1999}. However, the delay-time distribution of NS-NS mergers is difficult to reconcile with that needed to reproduce observed r-process enrichment histories both at early and late times \citep[e.g.,][]{hotokezaka_beniamini_piran2018,haynes_kobayashi2019}. Other r-process channels involving neutrino-driven winds during neutron or magnetar birth may be plausible alternatives \citep[e.g.,][]{qian_woosley1996,hoffman_woosley_qian1997}.

Like the r-process, there seem to be a weak and main s-process. The weak s-process occurs in core He burning of $M > 25\msun$ stars \citep{peters1968,lamb+1977,raiteri+1993}, and works by way of neutron production from the $^{22}$Ne($\alpha$, n)$^{25}$Mg reaction to create free neutrons, which can then build elements up to $A \approx 90$ \citep{truran+2002}. The main s-process occurs during the AGB phase of low- and intermediate-mass stars, \citep[$M \sim 1-3\msun$;][]{schwarzschild_harm1967} acting through the $^{13}$C($\alpha$, n)$^{16}$O reaction, and forming elements with $A \gtrsim 90$.

With K2 GAP DR3 age estimates and GALAH DR3 abundances, we are in a position to test assumed production mechanisms of neutron-capture elements by comparisons to K20 models. 

\begin{figure*}
    \includegraphics[width=0.95\textwidth]{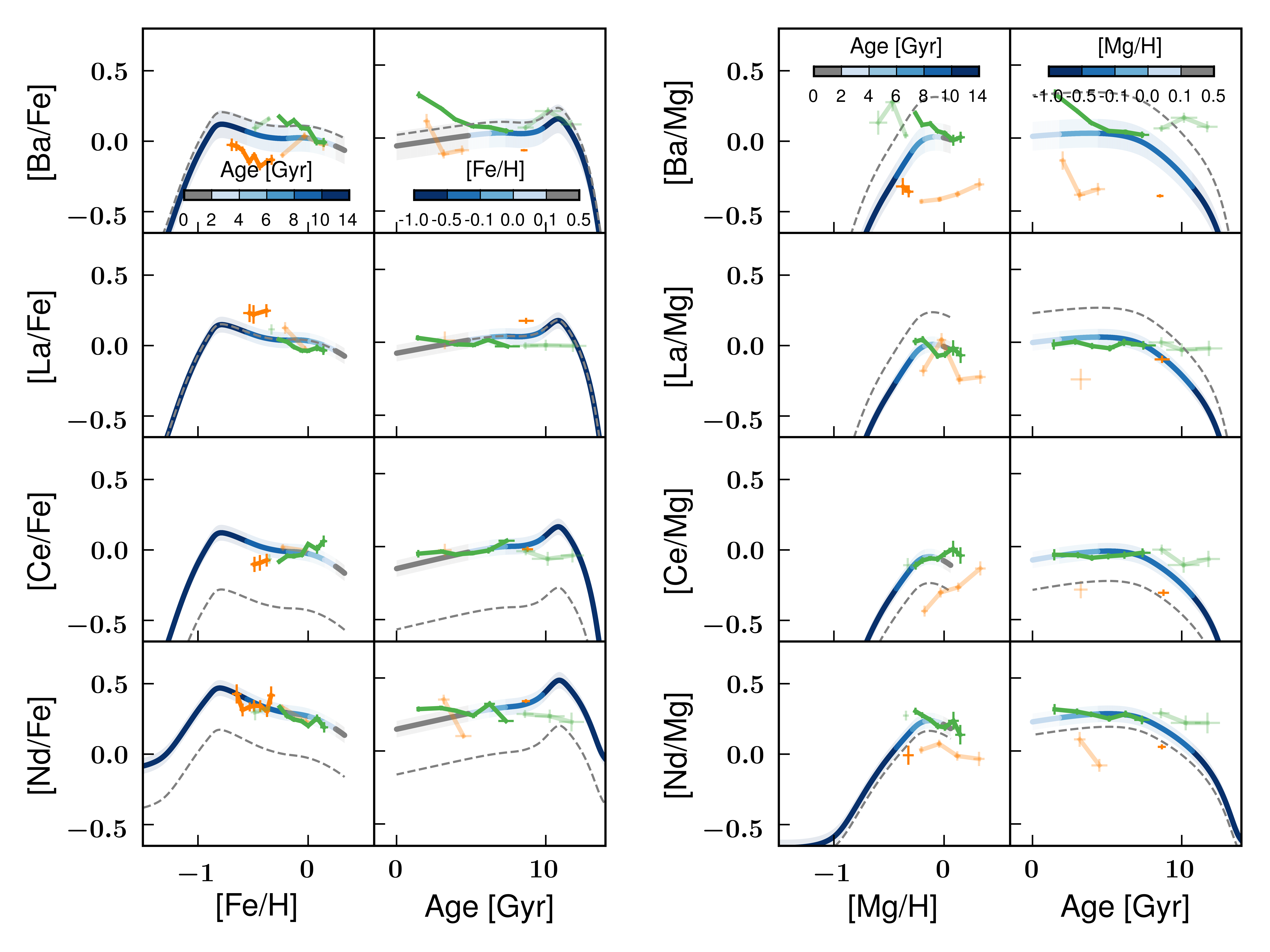}
    \caption{Same as Figure~\ref{fig:alpha}, but for main s-process elements.}
    \label{fig:mains}
\end{figure*}

Electron-capture supernovae (ECSNe) are included as a source of neutron-capture elements in K20 models, the effect of which is to form first-peak s-process elements Sr, Y, Zr, Mo, and Ru via nuclear equilibrium processes, as well as weak r-process production from Nd to In \citep{WJM2011}. ECSNe are assumed in K20 models to occur between the relatively narrow mass range of $\sim 8.8-9\msun$; increases of an order of magnitude in the electron-capture supernova rate have been assumed in the literature, and so may be a tunable parameter to increase bulk yields (see K20 and references therein). The largest contributors to r-process production in the K20 models, however, are magneto-rotational supernovae (MRSNe), which are theorized to be core-collapse supernovae of massive stars with large magnetic fields and/or strong rotation, which develop accretion discs and jets that can be conducive to r-process production \citep[e.g.,][]{symbalisty_schramm_wilson1985,cameron2003,nishimura_plus_2017}. There is also a NS-NS merger r-process contribution included in K20 models, though its contribution is subdominant compared to that of MRSNe.

\subsubsection{Weak s-process elements: Rb, Y, Zr}
Shown in Figure~\ref{fig:firsts} are the age-abundance ratios of elements thought to be formed through the `weak' s-process.\footnote{In detail, the K20 models predict that Y and Zr are in fact produced mostly in low- and intermediate-mass AGB stars as part of what we label here the main s-process.}

The agreement between the observed and predicted age-abundance patterns of \mg{Rb} and \fe{Rb} is very good, across both high- and low-$\alpha$ populations. 

Although there is an over-prediction in the abundances for the \highalpha\ stars at 9Gyr, the agreement between K20 models and data are also good for \mg{Zr} and \fe{Zr}, as well.

The K20 age-\mg{Y} pattern does not reach an equilibrium value, indicating a metallicity-dependence to the s-process production of Y. This metallicity-dependence is also borne out in the data, save for a zero-point offset in \mg{Y}. The predicted enrichment history is in good agreement in both \mg{Y} and \fe{Y} space. The agreement of the observed and predicted Y enrichment histories represents another endorsement of dating stars with Y abundances \citep[e.g.,][]{nissen2015}. 

\subsubsection{Weak r-process: Mo}
Although GALAH can measure Ru, the number of stars with good Ru measurements is small, and so we only consider Mo as representative of elements produced as part of the so-called `weak' r-process.

The K20 models under-predict Mo compared to GALAH (grey curve versus
error bars in Fig.~\ref{fig:weakr}), which is inconsistent with the inferred
over-production compared to high-resolution Mo abundance measurements
from the literature (K20). Nevertheless, the predicted history of Mo enrichment is consistent
with the observed \fe{Mo} and \mg{Mo} age-abundance patterns (blue curve compared to error bars).  

\begin{figure*}[!ht]
    \includegraphics[width=0.95\textwidth]{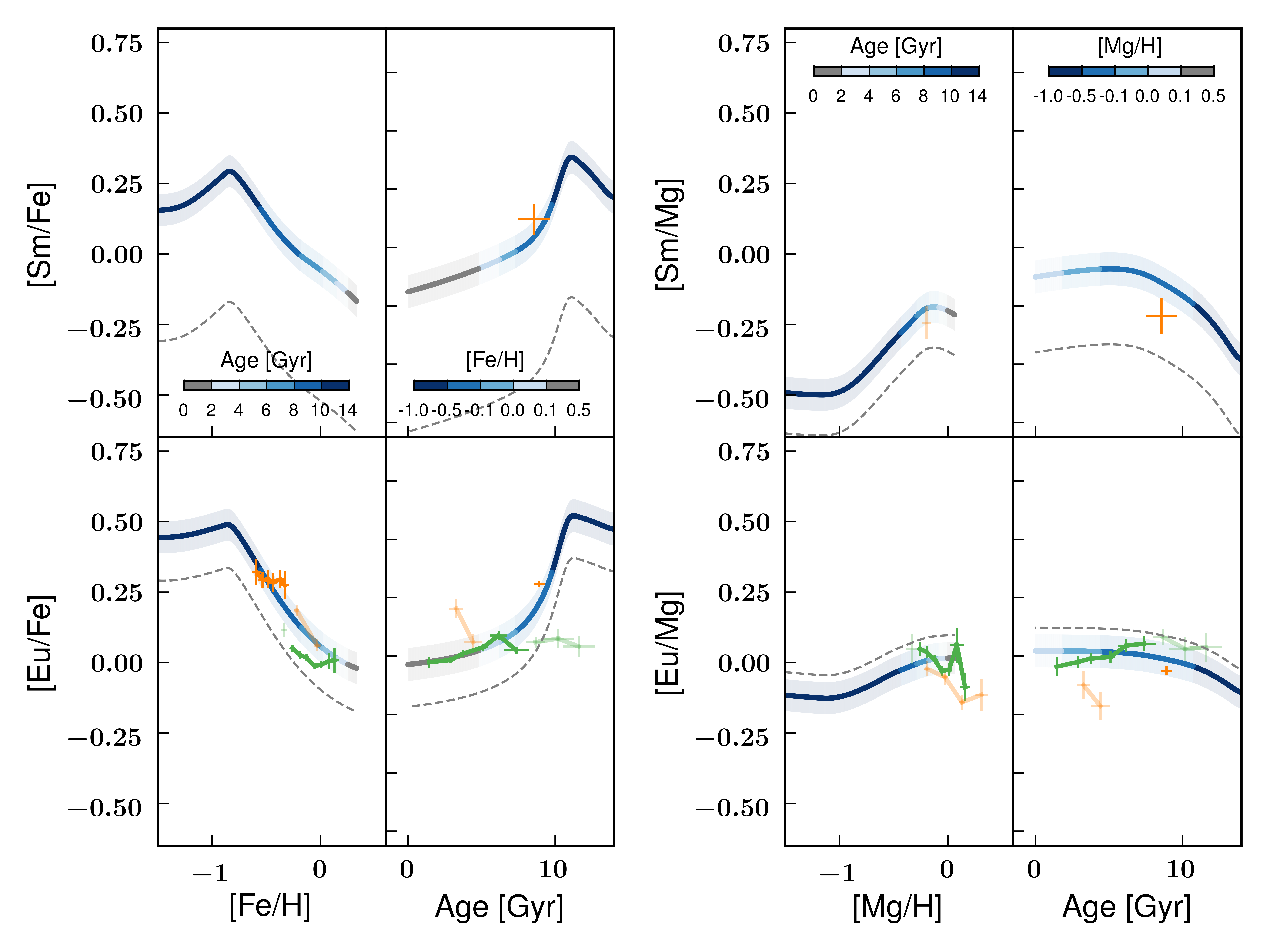}
    \caption{Same as Figure~\ref{fig:alpha}, but for main r-process elements.}
    \label{fig:mainr}
\end{figure*}

\subsubsection{Main s-process: Ba, La, Ce, Nd}
Ba, shown in Figure~\ref{fig:mains} is thought to be primarily produced by the s-process at the metallicities considered here ($ \feh > -2$; \citealt{gilroy+1988}; \citealt{arlandini+1999}; \citealt{burris+2000}). \mg{Ba} is predicted to reach a plateau in young stars according to K20 models, but the data disfavor a plateau and rather suggest a continually-increasing ratio with younger ages, like \mg{Y}.

 \cite{dorazi+2009} have observed a similar unexpected increase in \fe{Ba} at young stellar ages, based on open cluster measurements. They proposed that an increased production in low-mass stars would explain the observations, and which could possibly be related to enhanced mixing into the helium-burning shell thought to be the site of s-process production in low- and intermediate-mass AGB stars. Similar behavior has also been seen in more recent studies \citep{mishenina+2013b,magrini+2018,casamiquela+2021}. It is unlikely that this enrichment history is explicable by an astrophysical, metallicity-dependent yield, since \fe{Ba} decreases with increasing \feh, which demonstrates the advantage of analyzing nucleosynthetic yields with age information.

Our results therefore corroborate a mass-dependent Ba yield interpretation, though it is possible the GALAH DR3 Ba abundances themselves are responsible: a trend in Ba with stellar mass would mimic this effect. One candidate for such a systematic may be choice in the microturbulance parameter, given the sensitivity of one of the GALAH Ba lines, Ba II, to that parameter \citep{dobrovolskas+2012}. GALAH DR3 Ba abundances are calculated assuming that RGB stars with the same effective temperature, surface gravity, and $\feh$ have the same microturbulence velocity. That is not necessarily the case, and could lead to artificial shifts in the measured Ba abundance with, e.g., mass/age. Regarding the zero-point offset in the Ba abundances compared to the K20 models under no re-scaling (grey dashed lines), the RGB Ba abundances are systematically larger than the dwarf abundances by 0.2 dex. Were the RGB Ba abundances placed on the dwarf scale, then, there would be an even larger global offset between observations and models than shown here. On the model side, K20 noted offsets compared to other literature Ba abundances, noting that they could be remedied by imposing a smaller mixing region during AGB dredge-up. Based on our findings, the models may improve agreement with observations more specifically with a mass-dependent increase in the mixing region.

The predicted and observed enrichment histories of La, Ce, and Nd are
in good agreement, as seen in Figure~\ref{fig:mains}. The exception in this agreement is Ce among high-$\alpha$ stars, which is low compared to predictions, even after considering a $\approx 10\%$ increase in the observed ages, potentially indicating the need for less Ce production in K20 models at early times.

\subsubsection{Main r-process elements: Sm, Eu}
We show in Figure~\ref{fig:mainr} the abundances of the two main r-process elements available in GALAH DR3, Sm and Eu.

Given the relatively few stars in GALAH DR3 with measured Sm abundances, it is difficult to determine the precise agreement with K20 models as a function of age. It does, however, appear that the high-$\alpha$ abundances at old ages are broadly consistent with the predicted enrichment history, though could be improved more with older high-$\alpha$ ages from $\alpha$-enhanced stellar opacities.

Eu is mostly produced via the r-process \citep{arlandini+1999,battistini_bensby2016}, and, according to the K20 models, the primary site of r-process production is MRSNe (see Fig. 32 in K20), where the rate of MRSNe is chosen to be 3\% of massive CCSNe (hypernovae) with mass $M> 25\msun$ in order to reproduce the \fe{Eu}-[Fe/H] trend in the solar neighborhood.

The K20 models are in good agreement with both GALAH abundance ratios,
and the asteroseismic age-abundance patterns in \fe{Eu}, when
comparing models to low-$\alpha$ stars (green curves) at intermediate
and young ages ($\tau \lesssim 8\gyr$) and comparing models to
high-$\alpha$ stars (orange error bars at 9Gyr in the Eu enrichment history panels
of Fig.~\ref{fig:mainr}) at older ages. Consistent with studies of metal-poor systems with significant r-process enrichment \citep[e.g.,][]{barklem+2005,hansen+2018}, \cite{lin+2019} corroborated the short time-delay of Eu production sites using isochronal stellar ages of subgiants combined with GALAH DR2 abundances. In this context, the agreement of the observed K2 GAP DR3--GALAH DR3 Eu enrichment history with that of the MNSRe-dominated K20 Eu models gives further credence to a significant contribution to r-process elements from a prompt source --- e.g., late-time collapsar accretion disk outflows associated with MRSNe \citep{symbalisty_schramm_wilson1985,cameron2003,mosta+2013,vlasov_metzger_thompson2014,siegel+2019}.

\section{Conclusions}
\label{sec:conc}
The K2 GAP DR3 sample, as the largest asteroseismic sample published to date, and probing a range of Galactic environment, represents an important tool for Galactic archaeology and stellar physics. With 18821 total radius and mass coefficients for red giant branch (RGB) and red clump (RC)  stars delivered as part of this final data release, below are our main results.

\begin{enumerate}
\item We calibrated our asteroseismic values to be on the \Gaia\ parallactic scale. The radius and mass coefficients, $\kappar$ and $\kappam$, that are released in K2 GAP DR3 need only be multiplied by a temperature-dependent factor according to the user's preferred temperatures to yield radii and masses. The typical uncertainties in these coefficients are $2.9\%\  (\rm{stat.})\ \pm 0.1\%\  (\rm{syst.})\ $ \& $6.7\%\  (\rm{stat.})\ \pm 0.3\%\ (\rm{syst.})\ $ in $\kappar$ \& $\kappam$ for RGB stars and $4.7\%\  (\rm{stat.})\ \pm 0.3\%\ (\rm{syst.})\ $ \& $11\%\  (\rm{stat.})\ \pm 0.9\%\  (\rm{syst.})$ for RC stars. All of the stars with $\kappar$ and $\kappam$ are classified as red giant branch or red clump stars, according to a machine-learning approach.

\item Using injection tests, we estimate that our completeness in radius peaks for stars with $R \sim 10\rsun$, where our recovery rate is around $80\%$. There is a sharp decline in completeness at smaller radii, and a more gradual decline in completeness at larger radii. We estimate a nearly uniform completeness in mass space of $\sim 60\%$.

\item Injection tests suggest systematics of $1-3\%$ may arise due to the shorter time baseline of \Ktwo\ compared to \Kepler, and take the form of both zero-point biases and trends as a function of $\numax$, $\dnu$, and signal-to-noise ratio. These findings should be informative for future studies using short--time baseline TESS light curves, which would presumably suffer from similar, if not more severe, systematics.

\item We derived ages with typical
  precisions of 20\% for a subset of the K2 GAP DR3 sample based on
  GALAH metallicities and effective temperatures. In combination with GALAH abundances, we compared observed age-abundance patterns with those predicted by \cite{kobayashi_karakas_lugaro2020} as an independent check on the abundance evolution of low- and high-$\alpha$ stars. We corroborate recent inferences regarding the nucleosynthesis of $\alpha$, light, iron-peak, and neutron-capture elements based on abundance ratios alone \citep[e.g.,][]{griffith_johnson_weinberg2019}. Following similar indications from the \cite{lin+2019} analysis of GALAH DR2 subgiants with isochronal ages, we find evidence for significant production of Eu at early times, consistent with core-collapse supernovae as the predominant site of r-process production. Our findings also suggest mass-dependent Ba yields, in support of indications from \cite{dorazi+2009}.
\end{enumerate}

Studies of Galactic chemical evolution stand to benefit enormously
from a continued focus on considering age and not just stellar
abundances themselves, as we have shown here. Indeed, ages are of
crucial importance in interpreting chemo-kinematic relations
\citep{minchev+2019} --- particularly ages with the level of precision
reported here \citep[e.g.,][]{2014MNRAS.443.2452M}. As the largest
asteroseismic dataset in the literature, K2 GAP DR3 will prove useful
not only for Galactic studies, but also for testing
stellar models using the sample's evolutionary state classifications
coupled with its accurate and precise asteroseismic masses and
radii.

\begin{longrotatetable}
\begin{splitdeluxetable*}{lcccccccccccccBcccccccccccccl}
  \tablecaption{Derived asteroseismic $\numax$ and $\dnu$ values \label{tab:derived}}
  \tabletypesize{\footnotesize}
  \tablehead{EPIC ID & $\langle \numax' \rangle$ & $\sigma_{\langle \numax'
      \rangle}$ & $\varsigma_{\langle \numax'
      \rangle}$ & $N_{\nu\mathrm{max}}$ & $\langle\dnu'\rangle$ &
    $\sigma_{\langle\dnu'\rangle}$ & $\varsigma_{\langle\dnu'\rangle}$ & $N_{\dnu}$ & $X_{\mathrm{Sharma}}$ & $\sigma_{X{\mathrm{Sharma}}}$
    & $\langle\dnu\rangle$ & $\nu_{\mathrm{max,A2Z}}'$ &
    $\nu_{\mathrm{max,BAM}}'$ &
    $\nu_{\mathrm{max,BHM}}'$ & $\nu_{\mathrm{max,CAN}}'$ & $\nu_{\mathrm{max,COR}}'$ & $\nu_{\mathrm{max,SYD}}'$ & $\dnu_{\mathrm{A2Z}}'$  & $\dnu_{\mathrm{BAM}}'$ & $\dnu_{\mathrm{BHM}}'$ &
     $\dnu_{\mathrm{CAN}}'$ &
 $\dnu_{\mathrm{COR}}'$ &
     $\dnu_{\mathrm{SYD}}'$ & EPIC $T_{\rm{eff}}$ & $\sigma_{T}$ & EPIC [Fe/H] & $\sigma_{\rm{[Fe/H]}}$}
\startdata 
 & $\muhz$ & $\muhz$ & $\muhz$ & & $\muhz$ & $\muhz$ & $\muhz$ & & & & $\muhz$ & $\muhz$ & $\muhz$ & $\muhz$ & $\muhz$ & $\muhz$ & $\muhz$ & $\muhz$ & $\muhz$ & $\muhz$ & $\muhz$ & $\muhz$ & $\muhz$ & K & K & & \\  
210306475 &  28.041 &   1.274 &   1.495 &     5 &   3.495 &   0.112 &   0.176 &     3 &   1.026 &   0.008 &   3.405 &  28.575 &  30.053 &  26.808 &  27.801 &     --- &  29.351&     ---&   3.361&   3.568&     ---&     ---&   3.539 & 4797 & 134 & -0.266 & 0.300 \\
210307958 &  28.380 &   0.933 &   1.308 &     5 &   3.965 &   0.134 &   0.100 &     3 &   1.032 &   0.014 &   3.842 &  27.964 &  30.233 &     --- &  29.399 &  28.349 &  28.369&     ---&   3.852&     ---&   4.112&   3.925&     --- & 4750 & 138 & -0.359 & 0.260 \\
210314854 &  30.482 &   1.034 &   1.669 &     6 &   4.171 &   0.064 &   0.129 &     5 &   1.017 &   0.019 &   4.102 &  29.055 &  31.776 &  31.093 &  30.745 &  31.639 &  31.693&     ---&   4.201&   4.169&   4.119&   4.268&   4.112 & 4953 & 174 & -0.510 & 0.330 \\
210315825 &  58.454 &   0.341 &   1.990 &     6 &   6.093 &   0.059 &   0.096 &     5 &   1.025 &   0.011 &   5.944 &  59.632 &  59.091 &  59.795 &  59.671 &  58.958 &  59.540&     ---&   6.065&   6.175&   6.083&   6.123&   6.018 & 4827 & 180 & -0.298 & 0.300 \\
210318976 &  24.052 &   0.501 &   1.350 &     5 &   3.541 &   0.043 &   0.083 &     2 &   1.031 &   0.014 &   3.434 &  24.821 &  25.012 &     --- &  23.772 &  24.539 &  24.162&     ---&     ---&     ---&   3.568&   3.507&     --- & 4680 & 140 & -0.199 & 0.260 \\
\enddata
\tablecomments{Asteroseismic values re-scaled for scalar offsets among
  pipelines are denoted by a prime (the pipeline-specific solar
  reference scale factors are listed in Table~\ref{tab:refs}); mean $\numax$ and $\dnu$ values for each star across all pipelines are denoted by
  $\numaxmean$ and $\dnumean$; the standard deviation of these values for each star
  across all pipelines are denoted by $\sigma_{<\nu\mathrm{max}'>}$ and
  $\sigma_{<\dnu'>}$, and are the adopted uncertainties for K2 GAP DR3. $\varsigma_{\langle \numax'
      \rangle}$ and $\varsigma_{\langle\dnu'\rangle}$ are conservative estimates of statistical uncertainties based on reported pipeline statistical uncertainties. $\dnumean$ is adjusted using theoretically-motivated correction factors, $X_{\mathrm{Sharma}}$ \protect\citep{sharma+2016}, for use in asteroseismic scaling relations; an uncorrected version of $\dnumean$ for each star is provided, $\langle \dnu \rangle  = \dnumean/X_{\mathrm{Sharma}}$, should the user wish to compute custom $\dnu$ corrections. EPIC temperatures and metallicities are provided for this purpose, though these are relatively uncertain estimates of the true temperatures and metallicities (these uncertainties are also provided for convenience). The uncertainties in $X_{\mathrm{Sharma}}$,  $\sigma_{X{\mathrm{Sharma}}}$, are computed by perturbing the EPIC temperature and metallicities in a Monte Carlo procedure. Note that $\sigma_{X{\mathrm{Sharma}}}$ are not provided for EPIC ID 240289249 and EPIC ID 235193028, which have anomalously large EPIC temperature uncertainties. $\numaxmean$ values have an  evolutionary state--dependent correction applied to align asteroseismic radii with Gaia radii, per \S\ref{sec:calibratoin}. Pipeline-specific re-scaled values, $\numax'$ and
  $\dnu'$, are only provided for targets
  for which at least two pipelines returned concordant results, and otherwise
  have a blank entry; the numbers of
  pipelines returning valid results for $\numax$ or $\dnu$ are denoted by
  $N_{\nu \mathrm{max}}$ and $N_{\dnu}$. A2Z $\dnu'$ values are not provided, since A2Z $\dnu$ values do not contribute to $\dnumean$.
  See text for details. A full version of this table is available in the online journal.}
\end{splitdeluxetable*}
\end{longrotatetable}

\begin{longrotatetable}
\begin{deluxetable}{cccccccc}
\tablecaption{Solar reference value scale factors and solar reference values \label{tab:refs}}
  \tablehead{ & A2Z &  CAN &  COR &  SYD &  BAM & BHM & K2 GAP DR3 }
  \startdata
$X_{\numax,{\mathrm{\ RGB,\ APOKASC2}}}$ & 1.00230 $\pm$ 0.00002 & 1.00820 $\pm$ 0.00002 & 0.99890 $\pm$ 0.00002 & 1.00060 $\pm$ 0.00002 &     --- &     ---& \\  
$X_{\numax,\mathrm{\ RGB}}$ & 0.9991 $\pm$ 0.0006 & 0.9953 $\pm$ 0.0003 & 1.0000 $\pm$ 0.0003 & 0.9990 $\pm$ 0.0007 & 1.0027 $\pm$ 0.0003 & 1.0034 $\pm$ 0.0003& \\ 
 $\nu_{\mathrm{max,}\odot{\mathrm{,{RGB}}}}$ &   3095 $\pm$      2 $\muhz$ &   3125 $\pm$      1 $\muhz$ &   3050 $\pm$      1 $\muhz$ &   3087 $\pm$      2 $\muhz$ &   3102 $\pm$      1 $\muhz$ &   3060 $\pm$      1 $\muhz$ &   3076 $\muhz$\\  
$X_{\dnu,\mathrm{\ RGB,\ APOKASC2}}$ & 0.99930 $\pm$ 0.00001 & 1.00070 $\pm$ 0.00001 & 1.00510 $\pm$ 0.00001 & 0.99950 $\pm$ 0.00001 &     --- &     ---& \\  
$X_{\dnu,\mathrm{\ RGB}}$ &    --- & 1.0042 $\pm$ 0.0002 & 1.0004 $\pm$ 0.0004 & 1.0001 $\pm$ 0.0012 & 0.9969 $\pm$ 0.0007 & 0.9978 $\pm$ 0.0005& \\ 
 $\dnu_{\odot\mathrm{,RGB}}$ & --- & 135.48 $\pm$   0.03 $\muhz$ & 134.97 $\pm$   0.05 $\muhz$ & 135.1 $\pm$   0.2 $\muhz$ & 134.42 $\pm$   0.10 $\muhz$ & 134.62 $\pm$   0.06 $\muhz$ & 135.146 $\muhz$ \\  
$X_{\numax,\mathrm{\ RC,\ APOKASC2}}$ & 1.00350 $\pm$ 0.00003 & 1.00670 $\pm$ 0.00002 & 0.99090 $\pm$ 0.00002 & 1.00100 $\pm$ 0.00003 &     --- &     ---& \\  
$X_{\numax,\mathrm{\ RC}}$ & 0.9951 $\pm$ 0.0011 & 0.9935 $\pm$ 0.0006 & 0.9992 $\pm$ 0.0005 & 0.996 $\pm$ 0.001 & 1.0131 $\pm$ 0.0005 & 1.0024 $\pm$ 0.0006& \\ 
 $\nu_{\mathrm{max,}\odot\mathrm{,RC}}$ &   3082 $\pm$      4 $\muhz$ &   3120 $\pm$      2 $\muhz$ &   3048 $\pm$      2 $\muhz$ &   3077 $\pm$      4 $\muhz$ &   3134 $\pm$      1 $\muhz$ &   3057 $\pm$      2 $\muhz$ &   3076 $\muhz$ \\  
$X_{\dnu,\mathrm{\ RC,\ APOKASC2}}$ & 0.99650 $\pm$ 0.00003 & 1.01080 $\pm$ 0.00002 & 0.99600 $\pm$ 0.00001 & 1.00320 $\pm$ 0.00002 &     --- &     ---&\\  
$X_{\dnu,\mathrm{\ RC}}$ &   --- & 1.0066 $\pm$ 0.0005 & 1.0010 $\pm$ 0.0005 & 0.999 $\pm$ 0.002 & 0.993 $\pm$ 0.002 & 0.9971 $\pm$ 0.0007& \\ 
 $\dnu_{\odot\mathrm{,RC}}$ & --- & 135.81 $\pm$   0.07 $\muhz$ & 135.06 $\pm$   0.07 $\muhz$ & 134.9 $\pm$   0.3 $\muhz$ & 133.9 $\pm$   0.3 $\muhz$ & 134.53 $\pm$   0.10 $\muhz$ & 135.146 $\muhz$\\  
\enddata
  \tablecomments{Solar reference value scale factors and solar reference values (see \S\ref{sec:methods}), 
    compared to those computed for some of the same pipelines using a
    similar method with {\it Kepler} data (\citealt{pinsonneault+2018}; ``APOKASC-2''). The adopted solar reference values for K2 GAP DR3 are listed in the last column. A2Z $\dnu$ solar reference value scale factors and solar reference values are not provided, since A2Z $\dnu$ values do not contribute to $\dnumean$; see Table~\ref{tab:solar_refs} for the default A2Z $\dnusun$ value.}
\end{deluxetable}
\end{longrotatetable}

\begin{longrotatetable}
\begin{splitdeluxetable*}{lcccccccccccccBccccccccccccccl}
  \tablecaption{Radius and mass coefficients \label{tab:kappa}}
  \tabletypesize{\footnotesize}
  \tablehead{EPIC ID & $\kapparmean$ & $\sigma_{\kapparmean}$ & $\kappammean$ & $\sigma_{\kappammean}$ & $\kappa_{R,\mathrm{A2Z}}^{\prime}$ & $\kappa_{R,\mathrm{BAM}}^{\prime}$ & $\kappa_{R,\mathrm{BHM}}^{\prime}$ & $\kappa_{R,\mathrm{CAN}}^{\prime}$ & $\kappa_{R,\mathrm{COR}}^{\prime}$ & $\kappa_{R,\mathrm{SYD}}^{\prime}$ & $\sigma_{\kappa R',\mathrm{A2Z}}$ & $\sigma_{\kappa R',\mathrm{BAM}}$ &  $\sigma_{\kappa R',\mathrm{BHM}}$ & $\sigma_{\kappa R',\mathrm{CAN}}$ &  $\sigma_{\kappa R',\mathrm{COR}}$ &  $\sigma_{\kappa R',\mathrm{SYD}}$ & $\kappa_{M,\mathrm{A2Z}}^{\prime}$ &
     $\kappa_{M,\mathrm{BAM}}^{\prime}$ &  $\kappa_{M,\mathrm{BHM}}^{\prime}$ & $\kappa_{M,\mathrm{CAN}}^{\prime}$ & $\kappa_{M,\mathrm{COR}}^{\prime}$ & $\kappa_{M,\mathrm{SYD}}^{\prime}$ & $\sigma_{\kappa M',\mathrm{A2Z}}$ & $\sigma_{\kappa M',\mathrm{BAM}}$ & $\sigma_{\kappa M',\mathrm{BHM}}$ & $\sigma_{\kappa M',\mathrm{CAN}}$ & $\sigma_{\kappa M',\mathrm{COR}}$ & 
    $\sigma_{\kappa M',\mathrm{SYD}}$ }
  \startdata
  & $$ & $$ & $$ & $$ & $$ & $$ & $$ & $$ & $$ & $$ & $$ & $$ & $$
  & $$ & $$ & $$ \\
210306475  &  13.406 &   1.086 &   1.694 &   0.331 &  12.815 &  15.798 &  12.502 &     --- &     --- &  13.912 &   1.012 &   1.901 &   1.298 &     --- &     --- &   1.573&   1.515&   2.438&   1.362&     ---&     ---&   1.847 &   0.359 &   0.605 &   0.324 &     --- &     --- &   0.484\\
210307958  &  10.541 &   0.818 &   1.060 &   0.186 &     --- &  12.100 &     --- &  10.323 &  10.927 &     --- &     --- &   0.836 &     --- &   0.594 &   0.562 &     ---&     ---&   1.439&     ---&   1.019&   1.100&     --- &     --- &   0.213 &     --- &   0.142 &   0.134 &     ---\\
210314854  &  10.232 &   0.480 &   1.073 &   0.133 &  10.158 &  10.690 &  10.625 &  10.757 &  10.313 &  11.131 &   0.926 &   0.493 &   1.029 &   0.599 &   0.479 &   1.870&   0.968&   1.181&   1.141&   1.157&   1.094&   1.277 &   0.233 &   0.125 &   0.236 &   0.180 &   0.123 &   0.552\\
210315825  &   9.194 &   0.193 &   1.661 &   0.075 &  10.034 &   9.539 &   9.312 &   9.576 &   9.338 &   9.763 &   0.818 &   0.292 &   0.609 &   0.335 &   0.332 &   0.541&   1.938&   1.748&   1.686&   1.779&   1.671&   1.845 &   0.412 &   0.140 &   0.235 &   0.151 &   0.143 &   0.296\\
210318976  &  11.201 &   0.368 &   1.015 &   0.084 &  13.643 &     --- &     --- &  11.089 &  11.847 &     --- &   1.087 &     --- &     --- &   0.798 &   0.623 &     ---&   1.492&     ---&     ---&   0.950&   1.120&     --- &   0.347 &     --- &     --- &   0.175 &   0.139 &     ---\\
\enddata
\tablecomments{$\kapparmean$ and
    $\kappammean$, and their uncertainties, computed based on $\langle \dnu' \rangle$ and
$\langle \numax' \rangle$, according to Equations~\ref{eq:radius}~\&~\ref{eq:mass}. $\kapparmean$ and $\kappammean$ values have an evolutionary state--dependent correction to align asteroseismic radii with Gaia radii, per \S\ref{sec:calibratoin}. Pipeline-specific radius and mass coefficients, $\kappa_R'$ and $\kappa_M'$, are
    computed with pipeline-specific
    asteroseismic parameters, $\dnu'$ and
$\numax'$. See \S\ref{sec:methods} for details. A full version of this table is available in the online journal.}
\end{splitdeluxetable*}
\end{longrotatetable}

\begin{deluxetable*}{ccccc|cccc}
  \tablecaption{Median fractional uncertainties of Kepler and K2 asteroseismic quantities (in percent) \label{tab:unc}}
  \tabletypesize{\footnotesize}
  \tablehead{& & RGB or & RGB/AGB & & &  & RC &}
  \startdata
  & APOKASC-2 & Y18 & K2 GAP DR2 & K2 GAP DR3 & APOKASC-2 & Y18 & K2 GAP DR2 & K2 GAP DR3\\ \hline
  $\sigma_{\nu_{\rm{max}}}$ & 0.9 & 1.0 & 1.7 &    1.3      & 1.3 & 2.1 & 2.4 &    2.2        \\ \hline
  $\sigma_{\Delta{\nu}}$ &0.4 & 0.3 & 1.7   &   1.1       & 1.1 & 1.1 & 2.3 &       1.8      \\ \hline
  $\sigma_{\kappa_R}$ & 1.3 & 1.1 & 3.3 &          2.9        & 2.7 & 3.3 & 5.0 &       4.7\\ \hline
  $\sigma_{\kappa_M}$  & 3.4 & 3.1 & 7.7 &         6.7       & 6.2 & 8.4 & 10.5 &        11\\ \hline
\enddata
\tablecomments{``APOKASC-2" indicates median fractional uncertainties from the analysis of \cite{pinsonneault+2018}, while ``Y18" refers to the analysis of \cite{yu+2018}. \Ktwo\ GAP DR2 uncertainties are taken from Table 7 of \cite{zinn+2020}.}
\end{deluxetable*}

\begin{longrotatetable}
\begin{splitdeluxetable*}{lccccccccccBcccccccccccBccccccccccccl}
  \tablecaption{Raw asteroseismic $\numax$ and $\dnu$ values, with evolutionary states \label{tab:raw}}
  \tabletypesize{\footnotesize}
\tablehead{ID & EPIC ID & Campaign & Priority & Evo. State (EV) & A2Z EV & BAM EV & BHM EV & CAN EV & COR EV & SYD EV & $\nu_{\mathrm{max,A2Z}}$ &
  $\sigma_{\nu\mathrm{max,A2Z}}$ & $\dnu_{\mathrm{A2Z}}$ &
  $\sigma_{\dnu,\mathrm{A2Z}}$ & $\nu_{\mathrm{max,BAM}}$ &
  $\sigma_{\nu\mathrm{max,BAM}}$ & $\dnu_{\mathrm{BAM}}$ &
  $\sigma_{\dnu,\mathrm{BAM}}$ & $\nu_{\mathrm{max,BHM}}$ &
  $\sigma_{\nu\mathrm{max,BHM}}$ & $\dnu_{\mathrm{BHM}}$ &
  $\sigma_{\dnu,\mathrm{BHM}}$ & $\nu_{\mathrm{max,CAN}}$ &
  $\sigma_{\nu\mathrm{max,CAN}}$ & $\dnu_{\mathrm{CAN}}$ &
  $\sigma_{\dnu,\mathrm{CAN}}$ & $\nu_{\mathrm{max,COR}}$ &
  $\sigma_{\nu\mathrm{max,COR}}$ & $\dnu_{\mathrm{COR}}$ &
  $\sigma_{\dnu,\mathrm{COR}}$ & $\nu_{\mathrm{max,SYD}}$ &
  $\sigma_{\nu\mathrm{max,SYD}}$ & $\dnu_{\mathrm{SYD}}$ &
  $\sigma_{\dnu,\mathrm{SYD}}$}
\startdata
& & & & & & & & & & &$\muhz$ & $\muhz$ & $\muhz$ & $\muhz$ & $\muhz$ & $\muhz$ & $\muhz$
& $\muhz$ & $\muhz$ & $\muhz$ & $\muhz$ & $\muhz$  & $\muhz$ & $\muhz$
& $\muhz$ & $\muhz$ & $\muhz$ & $\muhz$ & $\muhz$ & $\muhz$ & $\muhz$
& $\muhz$ & $\muhz$ & $\muhz$ \\
210306475\_4 & 210306475 & 4 &   903 & RGB  & RGB & RGB & RC & --- & --- & RC &  28.550 & 2.25 &   3.620 &   0.010 &  30.135 & 0.808 &   3.277 &   0.197 &  26.900 & 1.4 &   3.470 &   0.160 &  27.670 & 1.27 &     --- &     --- &     --- & --- &     --- &     --- &  29.322 & 1.742 &   3.450 &   0.170\\
210307958\_4 & 210307958 & 4 &  2771 & RGB  & --- & RC & --- & RC & RGB & --- &  27.940 & 2.47 &   3.890 &     --- &  30.316 & 0.723 &   3.723 &   0.124 &     --- & --- &     --- &     --- &  29.260 & 1.03 &   4.004 &   0.094 &  28.350 & 0.83 &   3.805 &   0.083 &  28.341 & 1.478 &     --- &     ---\\
210314854\_4 & 210314854 & 4 &  1141 & RGB  & RGB & RC & RGB & RGB & RC & RC &  29.030 & 2.03 &   4.100 &   0.120 &  31.863 & 0.743 &   4.115 &   0.083 &  31.200 & 1.0 &   4.090 &   0.190 &  30.600 & 1.49 &   4.066 &   0.056 &  31.640 & 0.89 &   4.204 &   0.079 &  31.662 & 3.853 &   4.033 &   0.238\\
210315825\_4 & 210315825 & 4 &  1651 & RGB  & RGB & RGB & RGB & RGB & RGB & RGB &  59.580 & 3.63 &   5.910 &   0.160 &  59.253 & 1.379 &   5.899 &   0.060 &  60.000 & 1.3 &   6.010 &   0.190 &  59.390 & 1.28 &   5.960 &   0.084 &  58.960 & 1.26 &   5.975 &   0.087 &  59.482 & 3.086 &   5.873 &   0.058\\
210318976\_4 & 210318976 & 4 &   988 & RGB  & RGB & --- & --- & RGB & RGB & --- &  24.800 & 1.88 &   3.270 &   0.040 &  25.080 & 0.841 &     --- &     --- &  21.700 & 1.1 &   3.080 &   0.100 &  23.660 & 1.22 &   3.482 &   0.090 &  24.540 & 0.73 &   3.402 &   0.076 &  24.138 & 2.070 &     --- &     ---\\
\enddata
\tablecomments{`Raw' asteroseismic parameters returned by
  a given pipeline, along with their uncertainties, without the re-scaling described in \S\ref{sec:methods}
  applied. Evolutionary states are also given for stars with both $\numaxmean$ and $\dnumean$ values (EV) as well for individual pipeline values (A2Z EV, BAM EV, etc.); see text for details. If classified, a
  star's evolutionary state is assigned as either ``RGB'', ``RGB/AGB", or ``RC''. ``Priority" refers to the K2 GAP target priority for a given \Ktwo\ campaign, which is discussed in \S\ref{sec:data} (a smaller numerical value corresponds to higher priority); serendipitous targets do not have a populated priority entry. `ID' is a unique combination of the EPIC ID and from which campaign the measurements come (some stars were observed in multiple campaigns). A full version of this table is available in the online journal.}
\end{splitdeluxetable*}
\end{longrotatetable}

\begin{deluxetable*}{cccccccc}
  \tablecaption{Numbers of stars with raw asteroseismic values ($\numax$, $\dnu$), re-scaled asteroseismic values ($\numax'$, $\dnu'$), and radius~\&~mass coefficients ($\kappa_R'$, $\kappa_M'$), as a function of pipeline and campaign \label{tab:numbers}}
  \tabletypesize{\footnotesize}
  \tablehead{ & &  $\numax$  &  $\numax'$ & $\dnu$ & $\dnu'$ & $\kappa_R'$ & $\kappa_M'$ }
  \startdata
C1 & A2Z & 672 & 541 & 672 &   0 & 541 & 541 \\ \hline 
C2 & A2Z & 2326 & 993 & 1932 &   0 & 833 & 833 \\ \hline 
C3 & A2Z & 1418 & 834 & 1042 &   0 & 636 & 636 \\ \hline 
C4 & A2Z & 1966 & 1272 & 1536 &   0 & 1116 & 1116 \\ \hline 
C5 & A2Z & 3088 & 2088 & 2398 &   0 & 1761 & 1761 \\ \hline 
C6 & A2Z & 1086 & 1311 & 1086 &   0 & 1215 & 1215 \\ \hline 
C7 & A2Z & 993 & 835 & 293 &   0 & 224 & 224 \\ \hline 
C8 & A2Z & 1254 & 718 & 959 &   0 & 581 & 581 \\ \hline 
C10 & A2Z & 1660 & 832 & 1213 &   0 & 629 & 629 \\ \hline 
C11 & A2Z & 1359 & 670 & 1058 &   0 & 540 & 540 \\ \hline 
C12 & A2Z & 1717 & 866 & 1280 &   0 & 678 & 678 \\ \hline 
C13 & A2Z & 2393 & 1578 & 1924 &   0 & 1304 & 1304 \\ \hline 
C14 & A2Z & 1571 & 799 & 1138 &   0 & 621 & 621 \\ \hline 
C15 & A2Z & 3777 & 2598 & 2906 &   0 & 2055 & 2055 \\ \hline 
C16 & A2Z & 2685 & 1621 & 2025 &   0 & 1388 & 1388 \\ \hline 
C17 & A2Z & 1913 & 1173 & 1458 &   0 & 1016 & 1016 \\ \hline 
C18 & A2Z & 423 & 230 & 323 &   0 & 221 & 221 \\ \hline 
Total & A2Z & 30301 & 17291 & 23243 &   0 & 13827 & 13827 \\ \hline 
C1 & BAM & 948 & 698 & 757 & 457 & 457 & 457 \\ \hline 
C2 & BAM & 2591 & 1030 & 361 & 264 & 264 & 264 \\ \hline 
C3 & BAM & 1288 & 791 & 493 & 434 & 434 & 434 \\ \hline 
C4 & BAM & 2478 & 1282 & 844 & 751 & 751 & 751 \\ \hline 
C5 & BAM & 3001 & 2066 & 1158 & 1242 & 1242 & 1242 \\ \hline 
C6 & BAM & 2529 & 1626 & 955 & 1005 & 1005 & 1005 \\ \hline 
C7 & BAM & 2315 & 1202 & 677 & 587 & 587 & 587 \\ \hline 
C8 & BAM & 1107 & 719 & 426 & 385 & 385 & 385 \\ \hline 
C10 & BAM & 1568 & 852 & 428 & 348 & 348 & 348 \\ \hline 
C11 & BAM & 1339 & 647 & 275 & 229 & 229 & 229 \\ \hline 
C12 & BAM & 1603 & 878 & 471 & 419 & 419 & 419 \\ \hline 
C13 & BAM & 2262 & 1547 & 817 & 734 & 734 & 734 \\ \hline 
C14 & BAM & 1304 & 810 & 479 & 433 & 433 & 433 \\ \hline 
C15 & BAM & 3526 & 2589 & 1367 & 1261 & 1261 & 1261 \\ \hline 
C16 & BAM & 2269 & 1645 & 2269 & 1400 & 1400 & 1400 \\ \hline 
C17 & BAM & 1713 & 1209 & 609 & 763 & 763 & 763 \\ \hline 
C18 & BAM & 408 & 227 & 108 & 149 & 149 & 149 \\ \hline 
Total & BAM & 32249 & 18115 & 12494 & 9641 & 9641 & 9641 \\ \hline 
C1 & BHM & 1030 & 670 & 1030 & 592 & 592 & 592 \\ \hline 
C2 & BHM & 1818 & 1009 & 1551 & 933 & 933 & 933 \\ \hline 
C3 & BHM & 1191 & 842 & 1086 & 775 & 775 & 775 \\ \hline 
C4 & BHM & 1984 & 1251 & 1529 & 1126 & 1126 & 1126 \\ \hline 
C5 & BHM & 2750 & 2120 & 2550 & 1933 & 1933 & 1933 \\ \hline 
C6 & BHM & 2275 & 1606 & 1702 & 1370 & 1370 & 1370 \\ \hline 
C7 & BHM & 1803 & 1186 & 1238 & 989 & 989 & 989 \\ \hline 
C8 & BHM & 1094 & 722 & 869 & 656 & 656 & 656 \\ \hline 
C10 & BHM & 1590 & 871 & 1081 & 723 & 723 & 723 \\ \hline 
C11 & BHM & 1012 & 612 & 807 & 546 & 546 & 546 \\ \hline 
C12 & BHM & 1416 & 867 & 1053 & 749 & 749 & 749 \\ \hline 
C13 & BHM & 2112 & 1548 & 1764 & 1403 & 1403 & 1403 \\ \hline 
C14 & BHM & 1261 & 817 & 1003 & 739 & 739 & 739 \\ \hline 
C15 & BHM & 3216 & 2532 & 2730 & 2299 & 2299 & 2299 \\ \hline 
C16 & BHM & 2198 & 1648 & 1731 & 1400 & 1400 & 1400 \\ \hline 
C17 & BHM & 1697 & 1226 & 1307 & 1069 & 1069 & 1069 \\ \hline 
C18 & BHM & 344 & 249 & 266 & 218 & 218 & 218 \\ \hline 
Total & BHM & 28791 & 18049 & 23297 & 16014 & 16014 & 16014 \\ \hline 
C1 & CAN & 1105 & 778 & 582 & 482 & 482 & 482 \\ \hline 
C2 & CAN & 1609 & 948 & 616 & 503 & 503 & 503 \\ \hline 
C3 & CAN & 1116 & 816 & 494 & 433 & 433 & 433 \\ \hline 
C4 & CAN & 1897 & 1204 & 968 & 793 & 793 & 793 \\ \hline 
C5 & CAN & 2530 & 2030 & 1559 & 1458 & 1458 & 1458 \\ \hline 
C6 & CAN & 1956 & 1514 & 1455 & 1273 & 1273 & 1273 \\ \hline 
C7 & CAN & 1564 & 1083 & 1048 & 879 & 879 & 879 \\ \hline 
C8 & CAN & 935 & 713 & 785 & 606 & 606 & 606 \\ \hline 
C10 & CAN & 1234 & 840 & 1093 & 679 & 679 & 679 \\ \hline 
C11 & CAN & 1000 & 612 & 874 & 515 & 515 & 515 \\ \hline 
C12 & CAN & 1181 & 854 & 1104 & 753 & 753 & 753 \\ \hline 
C13 & CAN & 1975 & 1548 & 1851 & 1326 & 1326 & 1326 \\ \hline 
C14 & CAN & 958 & 755 & 854 & 677 & 677 & 677 \\ \hline 
C15 & CAN & 3032 & 2497 & 2817 & 2250 & 2250 & 2250 \\ \hline 
C16 & CAN & 1814 & 1581 & 1633 & 1336 & 1336 & 1336 \\ \hline 
C17 & CAN & 780 & 1039 & 733 & 884 & 884 & 884 \\ \hline 
C18 & CAN & 297 & 235 & 260 & 191 & 191 & 191 \\ \hline 
Total & CAN & 24983 & 17386 & 18726 & 13650 & 13650 & 13650 \\ \hline 
C1 & COR & 777 & 681 & 777 & 610 & 610 & 610 \\ \hline 
C2 & COR & 1635 & 960 & 1635 & 930 & 930 & 930 \\ \hline 
C3 & COR & 1022 & 757 & 1022 & 699 & 699 & 699 \\ \hline 
C4 & COR & 1803 & 1233 & 1803 & 1177 & 1177 & 1177 \\ \hline 
C5 & COR & 2526 & 1953 & 2526 & 1774 & 1774 & 1774 \\ \hline 
C6 & COR & 1443 & 1404 & 1443 & 1286 & 1286 & 1286 \\ \hline 
C7 & COR & 1561 & 1162 & 1561 & 1127 & 1127 & 1127 \\ \hline 
C8 & COR & 955 & 681 & 955 & 617 & 617 & 617 \\ \hline 
C10 & COR & 1449 & 824 & 1449 & 737 & 737 & 737 \\ \hline 
C11 & COR & 1048 & 613 & 1048 & 570 & 570 & 570 \\ \hline 
C12 & COR & 1309 & 819 & 1309 & 740 & 740 & 740 \\ \hline 
C13 & COR & 1879 & 1470 & 1879 & 1389 & 1389 & 1389 \\ \hline 
C14 & COR & 1034 & 755 & 1034 & 699 & 699 & 699 \\ \hline 
C15 & COR & 2944 & 2432 & 2944 & 2342 & 2342 & 2342 \\ \hline 
C16 & COR & 1833 & 1490 & 1833 & 1323 & 1323 & 1323 \\ \hline 
C17 & COR & 1368 & 1080 & 1368 & 977 & 977 & 977 \\ \hline 
C18 & COR & 265 & 201 & 265 & 164 & 164 & 164 \\ \hline 
Total & COR & 24851 & 17019 & 24851 & 15850 & 15850 & 15850 \\ \hline 
C1 & SYD & 558 & 490 & 472 & 418 & 418 & 418 \\ \hline 
C2 & SYD & 1290 & 684 & 486 & 397 & 397 & 397 \\ \hline 
C3 & SYD & 1066 & 752 & 668 & 588 & 588 & 588 \\ \hline 
C4 & SYD & 2151 & 1270 & 735 & 670 & 670 & 670 \\ \hline 
C5 & SYD & 2627 & 2001 & 1627 & 1515 & 1515 & 1515 \\ \hline 
C6 & SYD & 2232 & 1472 & 782 & 743 & 743 & 743 \\ \hline 
C7 & SYD & 1678 & 1088 & 505 & 481 & 481 & 481 \\ \hline 
C8 & SYD & 937 & 681 & 588 & 518 & 518 & 518 \\ \hline 
C10 & SYD & 1138 & 715 & 466 & 385 & 385 & 385 \\ \hline 
C11 & SYD & 1583 & 699 & 459 & 378 & 378 & 378 \\ \hline 
C12 & SYD & 1097 & 751 & 584 & 511 & 511 & 511 \\ \hline 
C13 & SYD & 2007 & 1519 & 1230 & 1103 & 1103 & 1103 \\ \hline 
C14 & SYD & 1079 & 768 & 672 & 586 & 586 & 586 \\ \hline 
C15 & SYD & 3128 & 2492 & 1990 & 1819 & 1819 & 1819 \\ \hline 
C16 & SYD & 1886 & 1598 & 1229 & 1184 & 1184 & 1184 \\ \hline 
C17 & SYD & 598 & 1091 & 325 & 629 & 629 & 629 \\ \hline 
C18 & SYD & 266 & 232 & 176 & 183 & 183 & 183 \\ \hline 
Total & SYD & 25321 & 16688 & 12994 & 11080 & 11080 & 11080 \\ \hline 
\enddata
\end{deluxetable*}

\begin{deluxetable*}{ccccccccccc}
  \tablecaption{K2 GAP DR3 ages with GALAH spectroscopy \label{tab:age}}
  \tabletypesize{\footnotesize}
  \tablehead{EPIC ID & \texttt{sobject\_id} & $\tau$ & $\sigma_{\tau}$ & $\feh$ & $\sigma_{\feh}$ & $\fe{Mg}$ & $\sigma_{\fe{Mg}}$ & $\teff$ & $\sigma_{\teff}$ & $\alpha_{\mathrm{hi}}$}
   \startdata
 & & Gyr & Gyr &  &  &  &  & K & K &   \\ \hline
220387110 & 161007003801220 &  7.7 &  1.8 & -0.2 &  0.1 &  0.2 & 0.1 & 4691 &   91 & --- \\
220352927 & 161007003801158 & 11.1 &  1.5 & -1.3 &  0.1 &  0.1 & 0.1 & 4883 &  124 & --- \\
220420379 & 161007003801285 &  5.2 &  2.9 & -0.6 &  0.1 &  0.3 & 0.2 & 4705 &  136 & --- \\
220329169 & 161007003801110 &  4.1 &  0.7 & -0.3 &  0.1 &  0.1 & 0.1 & 4864 &   95 & --- \\
220425435 & 161007003801301 &  9.2 &  2.7 & -1.6 &  0.2 &  0.1 & 0.2 & 5060 &  174 & --- \\
220377647 & 161007003801390 &  4.2 &  0.9 & -0.4 &  0.1 &  0.2 & 0.1 & 4995 &  116 & --- \\
220382480 & 161007003801378 & 10.8 &  1.7 & -0.7 &  0.2 &  0.2 & 0.2 & 5085 &  190 & --- \\
220392564 & 161007003801360 &  6.6 &  2.5 & -0.9 &  0.1 &  0.3 & 0.2 & 4791 &  137 & --- \\
220408286 & 161007003801353 &  6.0 &  4.0 & -0.2 &  0.1 &  0.1 & 0.1 & 4770 &  111 & --- \\
220272081 & 161006004401209 &  9.1 &  3.0 & -0.6 &  0.1 &  0.4 & 0.1 & 4559 &   87 & 1 \\
\enddata
\tablecomments{Ages and GALAH metallicities, \fe{Mg}, and effective
  temperatures for the subset of the K2 GAP DR3 sample with GALAH
  data. \texttt{sobject\_id} is the GALAH observation ID, which may be used to
  cross-match with GALAH catalogues. $\alpha_{\mathrm{hi}}$ is 1 (0) if the star has GALAH abundances indicative of a high-$\alpha$ (low-$\alpha$) star at $2\sigma$ confidence; if the classification is ambiguous, the entry is blank (see text for details).  A full version of this table is available in the online journal.}
\end{deluxetable*}

\acknowledgments
We would like to thank the anonymous referee whose comments significantly
improved the manuscript. JCZ is supported by an NSF Astronomy and Astrophysics Postdoctoral Fellowship under award AST-2001869. JCZ and MHP acknowledge support from NASA grants
80NSSC18K0391 and NNX17AJ40G. YE and CJ acknowledge the support of the UK Science and Technology Facilities Council (STFC). SM acknowledges support from the Spanish Ministry of Science and Innovation with the Ramon y Cajal fellowship number RYC-2015-17697 and the grant number PID2019-107187GB-I00.  RAG acknowledges funding received from the PLATO CNES grant. CK acknowledges funding from the UK Science and Technology Facility 
Council (STFC) through grants ST/M000958/1 and ST/R000905/1, and 
ST/V000632/1.

Funding for the Stellar Astrophysics Centre (SAC) is provided by The Danish National Research Foundation (Grant agreement no. DNRF106). 

The {\it K2} Galactic Archaeology Program is supported by the National Aeronautics and Space Administration under Grant NNX16AJ17G issued through the {\it K2} Guest Observer Program. This publication makes use of data products from the Two Micron All Sky Survey, which is a joint project of the University of Massachusetts and the Infrared Processing and Analysis Center/California Institute of Technology, funded by the National Aeronautics and Space Administration and the National Science Foundation.

This paper includes data collected by the Kepler mission. Funding for the Kepler mission is provided by the NASA Science Mission directorate.

Parts of this research were supported by the Australian Research Council Centre of Excellence for All Sky Astrophysics in 3 Dimensions (ASTRO 3D), through project number CE170100013.

This research was partially conducted during the Exostar19 program at the Kavli Institute for Theoretical Physics at UC Santa Barbara, which was supported in part by the National Science Foundation under grant No. NSF PHY-1748958. 

Based in part on data obtained at Siding Spring Observatory via GALAH. We acknowledge the traditional owners of the land on which the AAT stands, the Gamilaraay people, and pay our respects to elders past and present.

This work has made use of data from the European Space Agency (ESA)
mission
{\it Gaia} (\url{https://www.cosmos.esa.int/gaia}), processed by the
{\it Gaia}
Data Processing and Analysis Consortium (DPAC,
\url{https://www.cosmos.esa.int/web/gaia/dpac/consortium}). Funding
for the DPAC
has been provided by national institutions, in particular the
institutions
participating in the {\it Gaia} Multilateral Agreement.

Funding for the Sloan Digital Sky Survey IV has been provided by the Alfred P. Sloan Foundation, the U.S. Department of Energy Office of Science, and the Participating Institutions. SDSS-IV acknowledges
support and resources from the Center for High-Performance Computing at
the University of Utah. The SDSS web site is www.sdss.org.

\software{asfgrid \citep{asfgrid}, corner \citep{corner}, emcee \citep{foreman-mackey+2013}, NumPy \citep{numpy}, pandas \citep{pandas}, Matplotlib \citep{matplotlib}, IPython \citep{ipython}, SciPy \citep{scipy}}

\bibliography{paper/bib.bib}

\end{document}